\documentclass[12pt]{iopart}

\usepackage{graphicx}
\usepackage{epstopdf} % When an eps file, say fig_xx.eps, is included for the first time, one needs to pdflatex from terminal in order to have the pdf file 'fig_xx-eps-converted-to.pdf' generated: 
\usepackage{floatrow}
\usepackage{subfig,amssymb}

\usepackage{caption}

\begin{document}

% \title[Author guidelines for IOP Publishing journals in  \LaTeXe]{How to prepare and submit an article for 
% publication in an IOP Publishing journal using \LaTeXe}
\title[{\color{black}Edge} States in the Climate System]{{\color{black}Edge} States in the Climate System: Exploring Global Instabilities and Critical Transitions}

%\author{Content \& Services Team}
\author{Valerio Lucarini$^{1,2,3,4}$ \& Tam\'as B\'odai$^{1,2}$}

%\address{IOP Publishing, Temple Circus, Temple Way, Bristol BS1 6HG, UK}
\address{$^{1}$ Department of Mathematics and Statistics, University of Reading, Reading, UK}
\address{$^{2}$ Centre for the Mathematics of the Planet Earth, University of Reading, Reading, UK}
\address{$^{3}$ Centre for Environmental Policy, Imperial College, London, UK}
\address{$^{4}$ CEN, Meteorological Institute, University of Hamburg, Hamburg, Germany}
%\ead{submissions@iop.org}
\ead{v.lucarini@reading.ac.uk \& t.bodai@reading.ac.uk}
\vspace{10pt}
    \begin{center}
\today
\end{center}
\vspace{10pt}

\begin{abstract}
Multistability is a ubiquitous feature in systems of geophysical relevance and provides key challenges for our ability to predict a system's response to perturbations. Near critical transitions small causes can lead to large effects and - for all practical purposes - irreversible changes in the properties of the system.  As is well known, the Earth climate is multistable: present astronomical and astrophysical conditions support two stable regimes, the warm climate we live in, and a snowball climate, characterized by global glaciation. We first provide an overview of methods and ideas relevant for studying the climate response to forcings and focus on the properties of critical transitions in the context of both stochastic and deterministic dynamics, and assess strengths and weaknesses of simplified approaches to the problem. Following an idea developed by Eckhardt and collaborators for the investigation of multistable turbulent fluid dynamical systems, we study the global instability giving rise to the snowball/warm multistability in the climate system by identifying the climatic edge state, a saddle embedded in the boundary between the two basins of attraction of the stable climates. The edge state attracts initial conditions belonging to such a boundary and, while being defined by the deterministic dynamics, is the gate facilitating noise-induced transitions between competing attractors. We use a simplified yet Earth-like intermediate complexity climate model constructed by coupling a primitive equations model of the atmosphere with a simple diffusive ocean. We refer to the climatic edge states as Melancholia states and provide an extensive  analysis of their features. {\color{black}We study their dynamics, their symmetry properties, and we follow }a complex set of bifurcations. We find situations where the Melancholia state has chaotic dynamics. {\color{black}In these cases, we have that the basin boundary between the two basins of attraction is a strange geometric set with a nearly zero codimension, and relate this feature to the time scale separation between  instabilities occurring on weather and climatic time scales.} We also discover a new stable climatic state {that is \color{black}similar to a Melancholia state and is} characterized by non-trivial symmetry properties. 
\end{abstract}

%\maketitle

\section{Introductory Remarks}\label{intro}

Providing a comprehensive framework able to bring together the understanding of climate dynamics, able to encompass its variability and its responses to forcings, is one of the grand challenges of contemporary science \cite{ghil_topics_1987,Chekroun2011,LBHRPW14,LLR16}. While classical approaches to the definition of climate sensitivity are based on concepts borrowed from the analysis of systems at or near equilibrium, a more general point of view is indeed needed \cite{Ghil:2013,Ghil2015}. The issue has relevance both for the understanding of a) the history of our planet, the relationship between climatic conditions and the development of terrestrial life across the geological epochs, and the pressing issue of anthropogenic climate change; and b) for the possibility of having an encompassing view of planetary circulations, a much coveted goal in view of the extraordinary achievements of contemporary planetary science. Many planets are being discovered by the day, and the quest for discovering atmospheres able to support life (as we know it) is ongoing{\color{black}; see the NASA exoplanet archive (\texttt{http://exoplanetarchive.ipac.caltech.edu}) for some useful information at this regard.}

The climate is a complex system featuring variability of a vast range of spatial and temporal scales, resulting from instabilities fundamentally fuelled by the inhomogeneous absorption of solar radiation, and from the corresponding fluxes of mass, chemical species, and energy that, acting as negative feedbacks, allow for the establishment of steady state conditions. Interestingly, one can interpret the presence of organized motions of the geophysical fluids as a result of mechanical work, transforming available potential energy into kinetic energy, which is eventually dissipated by friction \cite{Lor67,Peixoto:1992}. Altogether, the climate can be seen as a thermal engine able to transform heat into mechanical energy with a given efficiency, and featuring many different irreversible processes that make it non-ideal \cite{Kleidon05,Lucarini09PRE,LBHRPW14}. An overarching theory of climate dynamics, able to encompass instabilities, re-equilibrating feedbacks, and large scale dynamical structures is still far from being achieved, despite lots of progress in the last decades. 

Nonequilibrium  systems like the climate can be essentially characterized as being in contact with at least two thermostats with different temperatures (or, \textit{e.g.} chemical potentials) \cite{G14}. Nonequilibrium systems exhibit an extremely complex phenomenology: while a lot is known for (near)equilibrium systems, our understanding of nonequilibrium systems is comparatively poor and limited, in spite of the wealth of phenomena occurring out of equilibrium. 

Obviously, much of the interest on the understanding of the response of the climate system to forcings comes from the evidence of the complex interplay between climate conditions and life. Since the early 1990s, the Intergovernmental Panel on Climate Change (IPCC) has started collecting and analyzing in a systematic way the scientific literature on the topic of climate dynamics, looking at the distant and near paleoclimatological conditions, but with a special focus on global warming and on the socio-economic impacts of anthropogenic climate change \cite{IPCC01,IPCC07,IPCC13}. An enormous scientific effort is aimed at improving our understanding of climate change, and climate-related investigations account for some of the most extensive data collection campaigns and computational exercises across all disciplines.

The response of the climate system to perturbations can be qualitatively divided in two different yet coexisting forms. In some cases, such a response is smooth with respect to the perturbation parameters. In other cases, the smoothness is lost and abrupt changes can result from small perturbations. {\color{black}In some previous works, we have  studied using a statistical mechanical approach  the regime of smooth response \cite{LBHRPW14,RLL16,LLR16,Grits_Luca2016} and the conditions under which such smoothness is lost \cite{Tantet2015b}. In other previous works, we have, instead, used thermodynamical concepts to better understand  under which conditions small perturbations can lead to qualitative changes in the properties of the climate system \cite{Luchyst,Lucarini2013a,Boschi}. In this work we wish to further advance the understanding of the global instabilities of the climate system by taking advantage of some powerful concepts and  methods borrowed from the theory of dynamical systems.}

In Sect. \ref{response} we review the two scenarios of smooth vs. nonsmooth climatic response to perturbations, in order to provide a meaningful context for the results contained in this paper and for the motivations behind our work. {\color{black}We introduce the concept of \textit{critical transitions} (\textit{tipping points}) in geosciences \cite{Lenton2008,Scheffer12}, and provide basic concepts and models for understanding the multistability of the Earth climate coming from the coexistence of the so-called snowball and warm climate states in a wide range of boundary conditions \cite{Budyko,Sellers,Ghil,Hoffman98,saltzman_dynamical,Pierrehumbert}}.

In Sect. \ref{edge2} we present an overview of some methods discussed in the literature for approaching the problem of investigating the critical transitions. We briefly summarize the rationale behind using stochastic models for studying climate dynamics  \cite{hasselmann_stochastic_1976,arnold_hasselmanns_2001,saltzman_dynamical} and discuss how large deviations theory \cite{T09} can be relevant, in the context of the Freidlin-Wentzell theory \cite{FW98} for explaining the transitions of the system between coexisting regimes \cite{Bouchet2014,Laurie2015}. Following the work of Eckhardt and collaborators in the context of the fluid dynamics of turbulent flows, and sticking to the paradigm of deterministic dynamics, we  introduce the concept of \textit{edge state}, the unstable saddle separating different regimes of motions and discuss how it is possible to find it using the edge tracking algorithm  \cite{PhysRevLett.96.174101,Schneider13022009,Vollmer09}. We also briefly discuss the application we have previously presented \cite{BLL2014} on a simple geophysical model \cite{Ghil}.% for studying the multistability of the Earth climate coming from the coexistence of the so-called snowball and warm climate states in a wide range of boundary conditions \cite{Hoffman98,Pierrehumbert}.  

In Sect. \ref{model} we introduce PUMA-GS, a new simple yet potentially relevant climate model constructed by coupling a modified version of the model introduced in \cite{Ghil} with PUMA \cite{puma}, a simple open-source primitive equations model of the atmospheric circulation. We then clarify how using a suitably adapted version of the edge tracking algorithm it is possible to identify the {\color{black}edge states of the system}, \textit{i.e.} unstable climatic states living on the boundary between the basins of attraction of the snowball and warm climate states\footnote{We have decided to use for the edge states of the climate system the name of \textit{Melancholia states}, as a homage to the 2011 movie \textit{Melancholia} directed by L. von Trier, where a masterful and accurate portrait of the challenges of prediction in the vicinity of critical transitions is given. The movie itself has provided a fundamental inspiration for this investigation. Just to give a - possibly unwanted - spoiler to the reader who has not yet watched the movie: things go terribly wrong. See \texttt{http://www.melancholiathemovie.com}.}. 

In Sect. \ref{results} we  {\color{black}study the geometrical properties of the boundary between competing basins of attraction, along the lines of \cite{Grebogi83,Kaplan84,McDonald85,Grebogi87,Vollmer09}}, the properties of the climatic edge states, investigating whether they correspond to fixed points, periodic points or chaotic saddles featuring a limited horizon of predictability, 
and the bifurcations between such regimes, including a process of symmetry breaking. {\color{black}We also show how the methodology proposed here allows for identifying a new stable climate regime that would hardly be discoverable through standard direct numerical integration.}

In Sect. \ref{concl} we summarize the main ideas presented in this paper and present our conclusions and perspectives for future work. 

\section{Response of the Climate System to Perturbations}\label{response}

\subsection{Response Theory and Climate Change}

In (near-)equilibrium systems, we have formidable tools for relating forced and free fluctuations of a system. Kubo \cite{K57,kubo_statistical_1988} constructed a mathematically nonrigorous but physically extremely powerful theory able to predict how a system whose statistics is described by the canonical ensemble responds to external perturbations. Kubo's response theory also includes the fluctuation-dissipation theorem, which basically shows that the linear response of a system to forcings can be derived from its natural fluctuation through a suitably defined linear operator. This result has been extremely influential in many areas of the physical and chemical sciences. See a comprehensive review in \cite{marconi2008} and  a recent extension to the nonlinear case in \cite{LC12}.

The lack of mathematical rigour of the response theory was later addressed by Ruelle \cite{R97,ruelle_smooth_1999,ruelle_review_2009}, who showed that if one considers {Axiom A} dynamical systems, and limits oneself to considering sufficiently smooth observables, it is possible to prove the differentiability of the invariant measure of the system with respect to small parameters describing the change in the dynamics, and to provide explicit formulas for computing the change in the measure. Axiom A systems are chaotic and uniformly hyperbolic on the attractor, where the asymptotic dynamics takes place. 

Moreover, modern methods of spectral theory have provided different and elegant proofs of Ruelle's results. The response theory can be developed by comparing the Perron-Frobenius transfer operator \cite{B00} of the unperturbed and of {\color{black} the} perturbed system. The spectrum of the transfer operator describes, among other things, how an initial probability distribution converges to - in the case of mixing systems - a unique invariant measure. In this way, one can study directly the change in the invariant measure resulting from the the applied perturbation \cite{BL07}. {\color{black}Such a point of view has also allowed the extension of Ruelle's results to the study of the response for smooth observables in more general dynamical systems \cite{B08,B14b} or of non-smooth observables in Axiom A systems \cite{BKL16}.}

Axiom A systems are indeed far from being typical dynamical systems,
but, according to the \textit{chaotic hypothesis} of Gallavotti and Cohen \cite{GC95}, they
can be taken as effective models for  chaotic physical systems with many
degrees of
freedom. Specifically, this means that when looking at macroscopic observables in \textit{sufficiently} chaotic (to be intended in a qualitative sense) high-dimensional  systems, it is extremely hard to distinguish their properties from those of an Axiom A system, including some degree of structural stability. One can interpret the chaotic hypothesis as the possibility of constructing robust physical properties for the system under investigation. Therefore, providing results for Axiom A systems can be thought of as being of rather general physical relevance. 

{Axiom A} systems corresponding to equilibrium physical systems possess an invariant measure
that is absolutely continuous with respect to Lebesgue because the phase space does not
contract nor expand, as the flow is nondivergent. Indeed, in this case the fluctuation-dissipation theorem applies. 

Axiom A systems corresponding to nonequilibrium conditions feature a phase space that contracts on the average, so that the asymptotic dynamics takes place on a strange attractor with dimension smaller than the dimension of the phase space. As a result of the singularity of the invariant measure, the usual form of the fluctuation dissipation theorem
does not hold. Ruelle \cite{R97,ruelle_smooth_1999,ruelle_review_2009} provides a mathematical explanation of this property, while a physical interpretation is given in \textit{e.g.} \cite{lucarini08,L09,lucarini2011}. This implies that for nonequilibrium systems there is no obvious relationship between free and forced fluctuations of the system, as already suggested by Lorenz \cite{lorenz79}.  

Various proposals have been put forward to avoid the problem of the lack of smoothness of the invariant measure in nonequilibrium statistical mechanical systems. Some authors consider adding some stochastic forcing on top of the deterministic dynamics, in order to take into account the effect of unresolved scales \cite{AM07a}, having in mind the Mori-Zwanzig theory  \cite{zwanzig_memory_1961,mori_transport_1965}. %Interestingly, one can use the Ruelle response theory to compute explicitly the effect of small scale, fast degree on freedom on the macroscopic one and, indeed, confirm that the coupling between the two leads to reintroducing smoothness in the projected invariant measure of the slow variables of the system \cite{WL12,WL13}; see a mathematically more rigorous  treatment in \cite{CLWa,CLWb}. 

Interestingly, while on one side there have been positive examples of applications of the fluctuation-dissipation theorem for studying the climate, it is clear that the quality of the obtained response operator depends substantially on the chosen observable  \cite{gritsun_climate_2007,GBM,cooper_climate_2011}.  The difficulties in constructing \textit{ab-initio} the response operator using Ruelle's explicit formulas come from the extremely different mathematical properties of the contributions coming from the unstable and stable manifold \cite{AM07a} plus from the possible presence, in the case of high-dimensional chaos, of rather diverse time scales. Algorithms based on adjoint methods seem to  partially ease these issues \cite{EHL04,W13}. A possible way to sort out the problem of estimating the response operator relies on projecting the perturbation flow on the covariant Lyapunov vectors, which allow for constructing a covariant base spanning the stable and unstable manifolds \cite{GPTCLP07}, or exploiting the reconstruction of the invariant measure obtained using unstable periodic orbits \cite{svita88,CE91}. {\color{black}In fact, through a combined use of the formalism of covariant Lyapunov vectors and of unstable periodic orbits, it has been recently possible  to elucidate interesting mathematical and physical aspects  behind the violation of the fluctuation-dissipation theorem in a simple atmospheric model \cite{Grits_Luca2016}.}

 Convincingly good results in terms of climate prediction performed using the linear response theory have instead been obtained by bypassing the problem of constructing the response operator and using, instead, the formal properties of the Green's function \cite{lucarini2011,LBHRPW14,RLL16}. Note that the response theory is able to predict not only the response of globally averaged quantities, but also the spatial patterns of the change \cite{LLR16}. %Tests in simple models have emphasized that also the nonlinear response theory is extremely solid and amenable of numerical verification \cite{L9}.  

\subsection{Where the Response Theory Fails: Critical Transitions of the Climate System}

The response theory is based on a perturbative expansion of the invariant measure, so its intrinsic limitation comes from the fact that the radius within which the expansion is valid could be extremely small. Of course, one might think of covering an extended range of parametric changes by repeating the expansion in neighbouring intervals. {\color{black}However, }the procedure will fail in the vicinity of critical transitions, which correspond to qualitative change in the dynamics of the system resulting from a crisis of the attractor. 

Bifurcations involving non-chaotic attractors have been very accurately studied, {\color{black} usually taking the point of view of analysing how the existence and the stability properties of invariant sets of the considered system change when one parameter is changed \cite{GuckHolmes83}. The resulting bifurcations can  be classified according to universal classes \cite{Arnold}, which has proved of immense relevance for studying a large {\color{black}variety} of physical systems, and indeed successfully in the context of climate dynamics \cite{ghil_topics_1987,dijkstra2013}. 

We remark that  Ashwin et al. \cite{Ashwin12} have recently emphasized the need for widening the usual treatment of bifurcations in dynamical systems in order to accommodate for more general scenarios of critical transitions. In addition to the usual bifurcation scenario described above (\textit{B-tipping}), they have studied the transitions between competing basins of attraction depending on the presence of a finite rate of change of one or more parameters of the system (\textit{R-tipping}){\color{black},} or on the presence of stochastic forcing of finite amplitude (\textit{N-tipping}). See in \cite{Lucarini05,Lucarini07} an earlier discussion of the role of time-dependent forcings in the context of another geophysically relevant critical transition, \textit{i.e.}, the collapse of the large scale oceanic circulation \cite{rahmstorf}.}   

The situation is much more complex in the case of high-dimensional {\color{black}and/or chaotic} attractors, where critical transitions are associated to a large variety of mathematical mechanisms, including boundary crises, where an attractor is destroyed by contact with its basin; interior crises, where the shape of the attractor undergoes a sudden change; synchronization and symmetry breaking (and restoring) processes, whereby different subsystems exhibit (or lose) special temporal and/or spatial correlations; and many others; see comprehensive treatments of these fascinating issues in, e.g., \cite{Ott2002}{\color{black}, and a survey of recent advances on synchronization phenomena (which have been initially studied in the simpler context of non-chaotic attractors) in \cite{Pikovsky2001}}.

{\color{black}For reasons that will be clear in Sec. \ref{edge2},} in the rest of the paper our main (but not exclusive) focus will be on the boundary crises, which are accompanied by the disappearance of an attractor. The vicinity of such critical transitions is accompanied{, \color{black} on the one hand,} by a rough dependence of the system's properties on the parameter to be tuned, as a result of the Ruelle-Pollicott resonances, and{\color{black}, on the other hand,} by the presence of a slow decay of correlations (\textit{critical slowing-down}) of the system's observables, with the decorrelation times diverging when the control parameter reaches its critical value \cite{chekroun2014,Tantet2015a,Tantet2015b}. Looking at the Ruelle response formulas, it is clear that the lack of a slow decay of correlations leads to a lack of convergence in the estimate of the Green's function, even though the relationship between the presence of sufficiently fast decay of correlations and the validity of response theory can be clarified more easily using spectral theory \cite{butterley2007}.

The climate system is characterized by such critical transitions, which in the geophysical literature are usually dubbed as \textit{tipping points}. {\color{black}Relevant examples of tipping points include the dieback of the Amazon forest, the shut-down of the thermohaline circulation of the Atlantic ocean, the methane release resulting from the melting of the permafrost, and the collapse of the atmospheric circulation regime associated to Indian monsoon; see \cite{Lenton2008,Scheffer12} for a useful perspective.} Obviously, not all tipping points are equally \textit{critical}, {\color{black}both in terms of global climatic impacts, and, more technically, in the sense that the lack of smoothness in the response might be practically detectable only when looking at specific observables. As an example, one can expect that non-smoothness in the response associated to the dieback of the Amazon forest will be easier to detect when looking at climatic fields in South America rather than in Siberia or Australia.} Understanding the qualitative and quantitative properties of climatic tipping points is another great challenge of climate science. 

\subsection{Snowball and Warm Climate States}

Probably the most impressive example of critical transition in the Earth system can be found when considering that the current astronomical and astrophysical conditions support at least two possible steady climate states, one being the one we are experiencing, and the other one characterized by much lower temperatures, {\color{black}almost total absence of water vapour in the atmosphere}, and global ice-cover, referred to as \textit{snowball} Earth. Models of different levels of complexity support such a statement, and, more importantly, geological evidence suggests that during the Neoproterozoic (the period spanning from about 1000 million to about 540 million years ago), the Earth suffered two of its most severe periods of glaciation \cite{Hoffman98,Hoffman,HoffmanSchrag} and entered into a snowball climate state.  

The onset and decay of snowball conditions are associated to changes in the value of the solar constant and in the composition of the atmosphere, with the ensuing variation of the greenhouse effect. For a fixed atmospheric composition one finds a range of values of the solar constant $S^*$ where {\color{black}the system features} multistability: the warm climate {\color{black}can exist} for all values of $S^*>S^*_{W\rightarrow SB}$ and the snowball state {\color{black}can exist} for all values of $S^*<S^*_{SB\rightarrow W}${\color{black}, with $S^*_{W\rightarrow SB}<S^*_{SB\rightarrow W}$.}  The critical value  $S^*=S^*_{W\rightarrow SB}$ ($S^*=S^*_{SB\rightarrow W}$) is such that if we are in the warm (snowball) state and slowly  reduce (increase) the value of $S^*$, for $S^*=S^*_{W\rightarrow SB}$ ($S^*=S^*_{SB\rightarrow W}$) the stability of the warm (snowball) climate is lost and the system tips to the snowball (warm) state. 

The critical transitions associated with the boundary crisis happen in conditions where the stabilizing negative feedbacks{\color{black}, to be described later,} are {\color{black}overcome by }the positive ice-albedo feedback. {\color{black}The positive feedback works as follows: colder (warmer) temperatures lead to an increase (a decrease) in the ice cover, which, by increasing (decreasing) the albedo, leads in turn to  lower (higher) temperatures \cite{saltzman_dynamical}. When the ice-albedo feedback dominates, the temperature then decreases (increases) substantially until a new balance sets in, corresponding to the cold (warm) stable state realized after the critical transition.}

{\color{black}We wish to remark that some theoretical investigations and observational evidence point at the possible existence of a third possible stable climate state, the so called \textit{slushball} Earth. Such a state is characterized by cold conditions (yet not as extreme as in the snowball), where the equatorial belt is ice-free and characterized by vigorous oceanic and atmpsoheric circulations and a non-trivial hydrological cycle; see, e.g., \cite{Lewis07,Abbott,Rose2015}. In the rest of the paper, we will only comment briefly on this - indeed relevant - geophysical problem.}

A simple mathematical formulation of the physical processes responsible for the snowball-warm multistability of the Earth system can be given by energy balance models (EBMs), which provide a simplified yet relevant representation of the energy exchanges in the climate system \cite{north81,saltzman_dynamical}. Reducing the Earth to a 0-dimensional (0-D) system (in physical space) where the only relevant quantity is the global temperature indicator $T$ (which can be heuristically interpreted as a globally averaged effective temperature), one can construct an EBM by modelling the energy budget as follows \cite{saltzman_dynamical}:
\begin{equation}
C\frac{\mathrm{d}}{\mathrm{d}t}{T}(t)= I(1-\alpha(T))-O(T)\rightarrow\frac{\mathrm{d}}{\mathrm{d}t}{T}(t)=-\frac{\mathrm{d}}{\mathrm{d}T}V(T),\label{0DEBM}
\end{equation}
where $t$ is time, $C$ is an effective average heat capacity per unit area, $I=S^*/4$ is the average incoming solar radiation per unit area, $S^*$ is the solar constant  (the factor 4 emerges looking at the geometry of the Earth-Sun system, see \cite{saltzman_dynamical}), $\alpha$ is the albedo, which is a non-increasing function of $T$, and $O$ is the outgoing radiation per unit area. $O(T)$ is a monotonically increasing function of $T$,{\color{black} i.e. an increasing surface temperature leads to an increase in the outgoing radiation, which is the basic mechanism of the Boltzmann (negative) radiative feedback.}

\begin{figure} [ht]
    \begin{center}
	%\scalebox{0.5}{\includegraphics{bif_ta_3}} 
	\includegraphics[trim=1cm 1cm 1cm 1cm, clip=true, angle=0,width=0.33\textwidth]{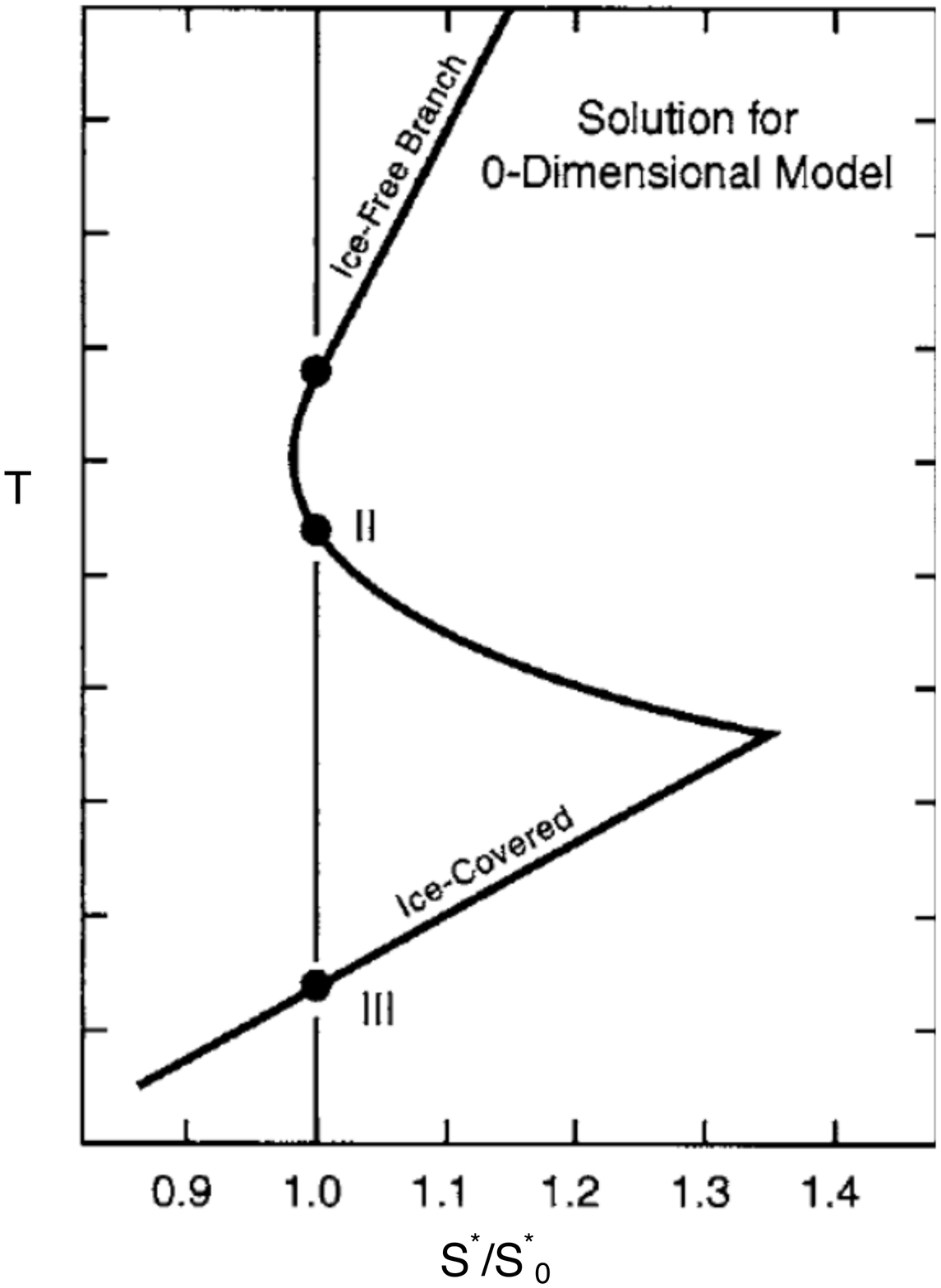}\\a)\\
	\includegraphics[trim=0cm 0cm 0cm 0cm, clip=true, angle=270,width=0.50\textwidth]{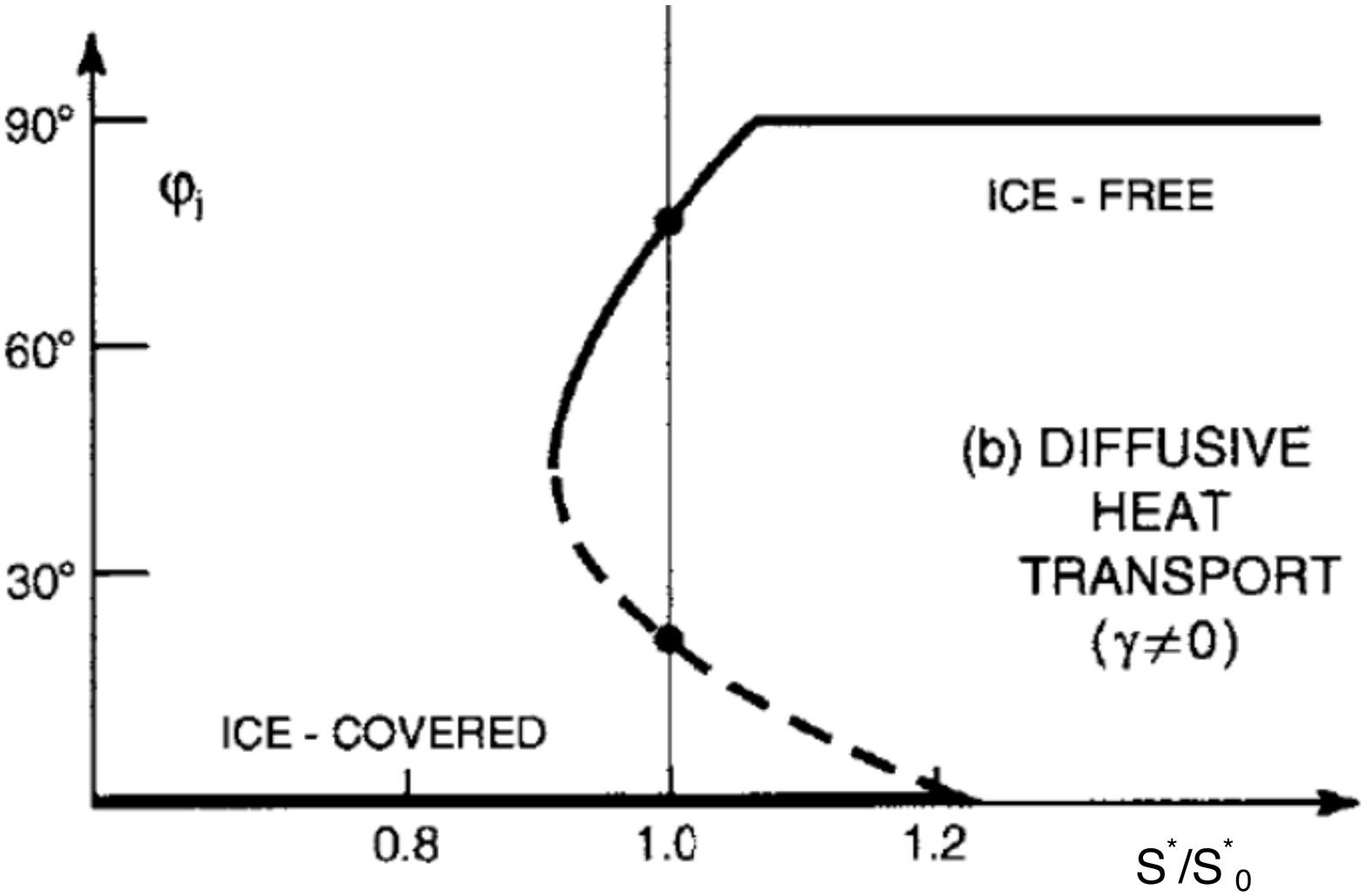}\\b)
        \caption{\label{EBMsnowball}         Multistability of the climate system corresponding to the coexistence of warm and snowball conditions. a) Bifurcation diagram obtained from the 0-D EBM as given in Eq. \ref{0DEBM}. The y-axis represents the temperature $T$,  the control parameter in the abscissa is $S^*/S^*_0$, {\color{black}where $S^*=S_0^*\approx1360$ $Wm^{-2}$ is the present-day value of $S^*$}. b) Bifurcation diagram obtained using a 1-D EBM as described in Eq. \ref{1DEBM}, where $\gamma$ controls the strength of the diffusion operator. The y-axis shows the ice-line latitude $\varphi_i$, the control parameter in the abscissa is $S^*/S^*_0$. Adapted from \cite{saltzman_dynamical}.}
    \end{center}
\end{figure}

By suitable (and reasonable) choices of $\alpha$ and $O$, one indeed finds bistability as a function of $S^*$; see the discussion in \cite{saltzman_dynamical} and Fig. \ref{EBMsnowball}a, where the two stable (warm and snowball) solutions are separated by the unstable solution $II$. 

\begin{figure}[ht] %[t!]
    \begin{center}
	%\scalebox{0.5}{\includegraphics{bif_ta_3}} 
	\includegraphics[width=0.5\textwidth]{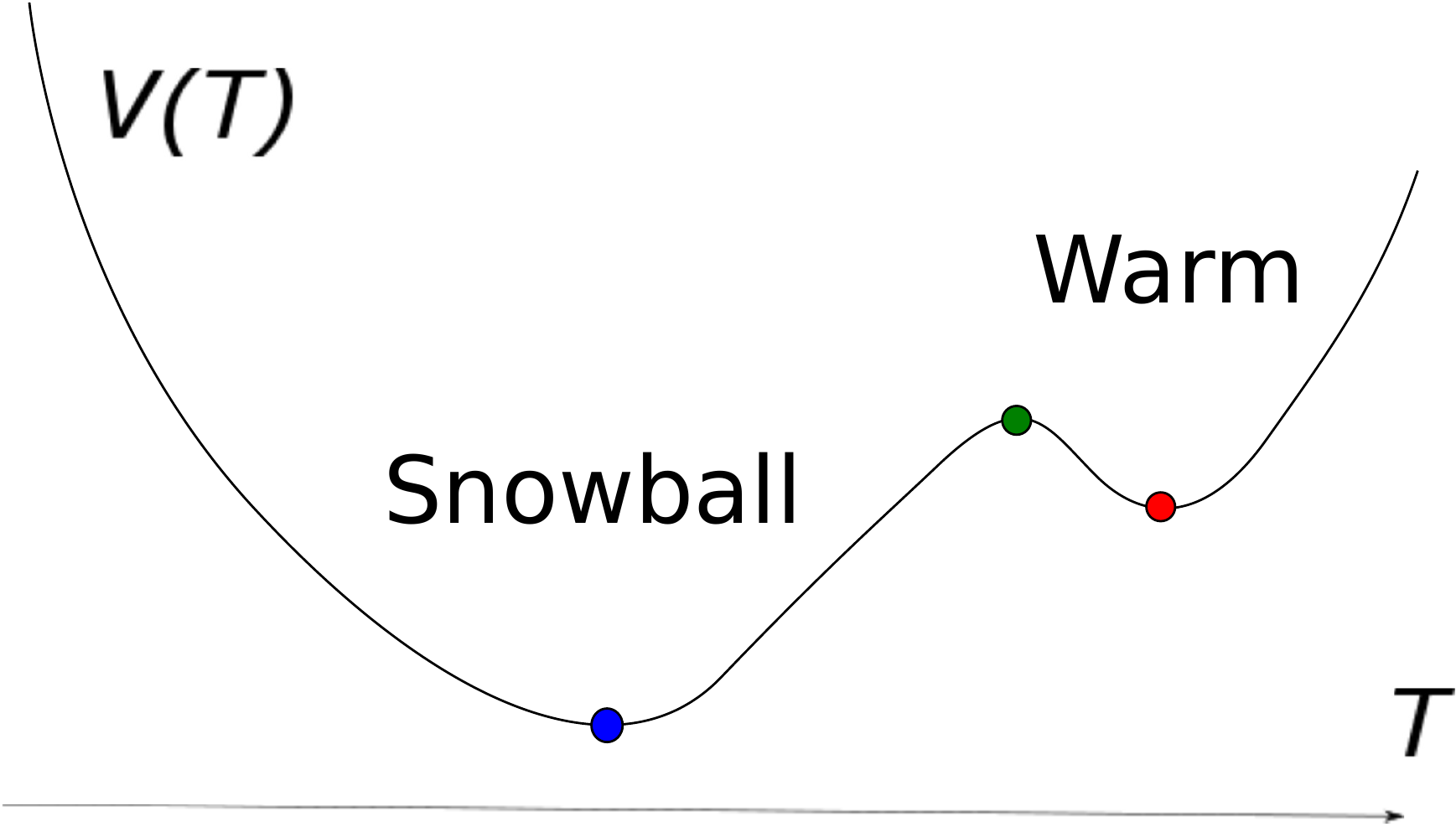}
        \caption{\label{fig:double_well} 
        Scalar double well potential function $V(T)$; the {\color{black}warm} and the {\color{blue}snowball} states correspond to the two attractors of the system, separated by the {\color{green}saddle} point. The noise triggers the transition from one basin of attraction to the other with a mean exit {\color{black}time} described by Eq. \ref{kramers}.}
    \end{center}
\end{figure}

Note that the time derivative of the temperature in Eq. \ref{0DEBM} can be written as minus the derivative of a potential $V(T)$, and in the case of bistability the two local minima of $V$ correspond to the stable solutions, and the local maximum of $V$ corresponds to the unstable solution, see Fig. \ref{fig:double_well} for a qualitative description.

EBMs can be extended in such a way as to include a latitudinal dependence of the Earth's temperature, as in the case of the celebrated models by Budyko \cite{Budyko}, Sellers \cite{Sellers}, and Ghil  \cite{Ghil}. The time evolution of the temperature for one-dimensional EBMs (1D-EBMs) can be written in terms of the following partial differential equation:
\begin{equation}
C(\phi)\partial_tT(t,\phi)= I(\phi)(1-\alpha(\phi,T))-O(T)-D[T,\phi],\label{1DEBM}
\end{equation}
where, as opposed to Eq. \ref{0DEBM}, the heat capacity $C$, the incoming radiation $I$, and the albedo $\alpha$ are explicitly dependent on the latitude $\phi$, and we have to include in the energy budget the divergence of energy transport performed by the geophysical fluids, represented as the diffusion operator $D[T,\phi]$. {\color{black} It is important to note that the diffusion operator describes in a very simple yet efficient way the action of the negative feedbacks regarding the meridional temperature gradient, through the meridional heat transport facilitated by atmospheric and oceanic motions. On weather time scales (few days), the feedback is powered by the  baroclinic instability: the stronger the large scale meridional temperature gradient, the higher the intensity of the midlatitude atmospheric eddies and of their related counter-gradient heat transport \cite{Peixoto:1992,Holton}. On climatic time scales (several years) the ocean acts in a similar way (though by  different specific physical mechanisms), by dampening the large scale temperature differences through the transport of heat from warm to cold regions \cite{Pierrehumbert}.}

With a suitable choice for the parameters controlling the terms responsible for the energy budget in Eq. \ref{1DEBM}, we obtain a range of values of $S^*$ where bistability is found; see Fig. \ref{EBMsnowball}b, where the unstable solution sits in between the two stable solutions. \footnote{It is historically remarkable to note that, after the Budyko and Sellers models were introduced, it became apparent that a \textit{Nuclear Winter}, by reducing the incoming solar radiation as a result of increased albedo due to a dramatic increases in the particulate matter in the atmosphere, could potentially trigger an even greater disaster for life on Earth than the nuclear war itself. This has been influential in reducing the size of the nuclear arsenals at the end of the Cold War. It is comforting to note that, somewhat ironically, the two contributions came almost simultaneously from scientists belonging to the two  opposing geopolitical blocks. }

\begin{figure} [ht]
    \begin{center}
	%\scalebox{0.5}{\includegraphics{bif_ta_3}} 
	a)\includegraphics[angle=270,width=0.5\textwidth]{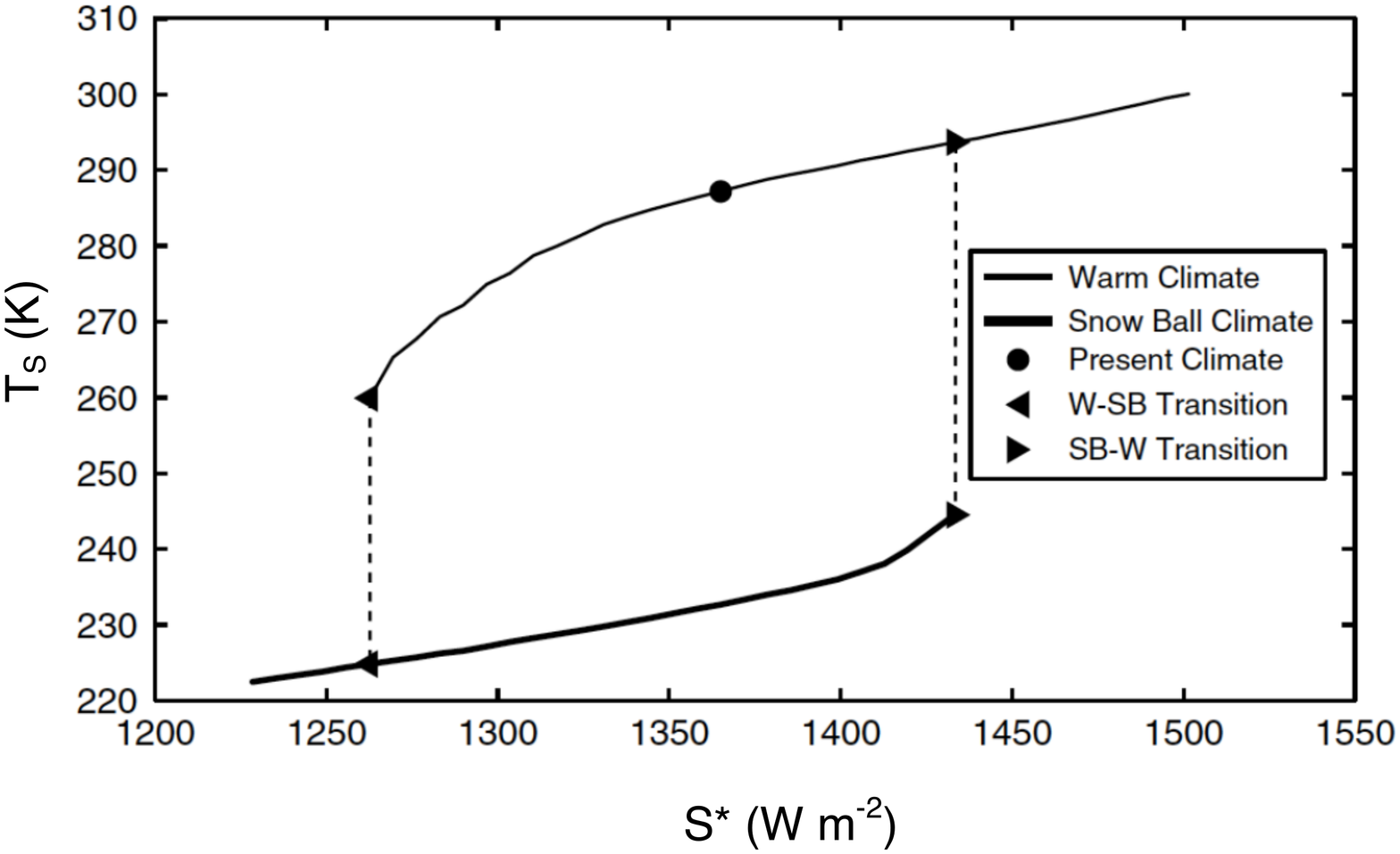}b)\includegraphics[angle=270,width=0.5\textwidth]{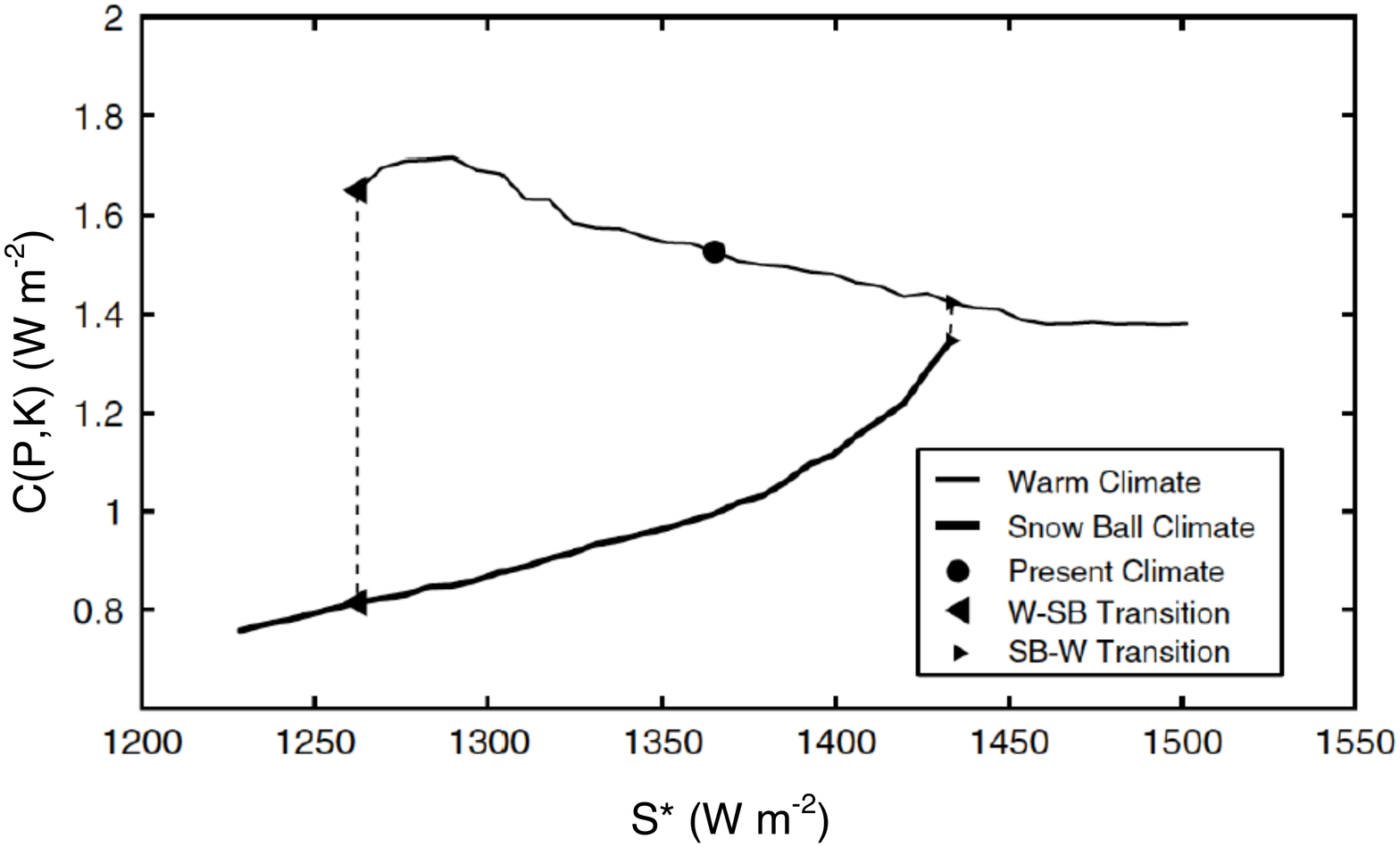}\\
	c)\includegraphics[trim=2cm 2cm 2cm 2cm, clip=true, angle=270,width=0.5\textwidth]{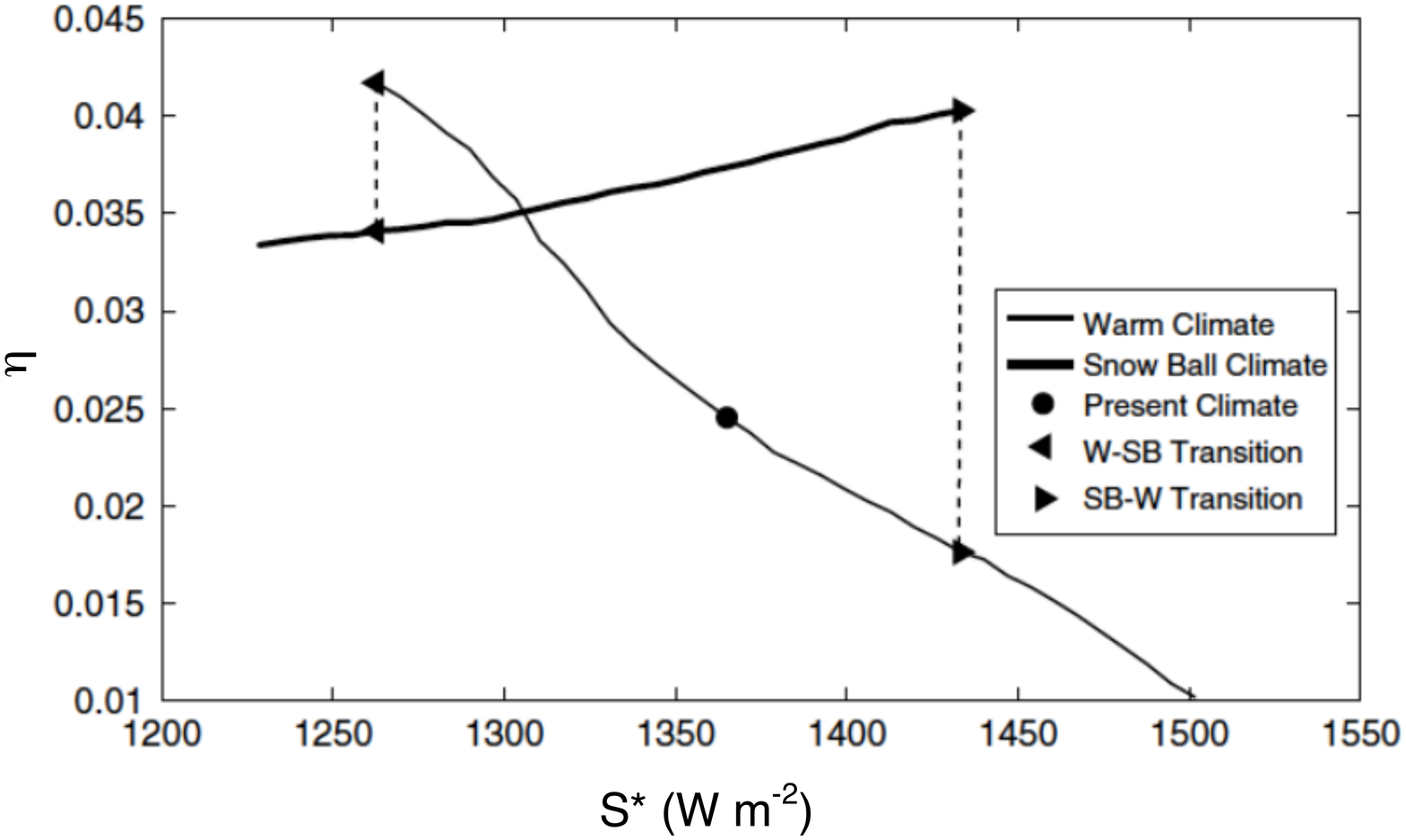}
        \caption{\label{plasimsnowball} 
        Multistability of the climate system corresponding to the coexistence of warm and snowball conditions. a) Globally averaged surface temperature $T_S$. b) Average intensity of the Lorenz energy cycle $C(P,K)$. c) Efficiency $\eta$ of the climate system. W-SB (SB-W) indicates the occurrence of transitions from warm to snowball conditions (from snowball to warm conditions). {\color{black}These results are obtained using the open source climate model PlaSim \cite{plasim}}. Adapted from \cite{Luchyst}.}
    \end{center}
\end{figure}

{\color{black}The existence of multistability for the Earth's climatic conditions is a robust feature across a modelling hierarchy, ranging from the simple EBMs described above to global climate models: see a detailed analysis of acting feedbacks as well as glaciation/deglaciation scenarios and mechanisms in \cite{Pierrehumbert}. Global climate models feature a range of bistability $[S^*_{W\rightarrow SB},\quad S^*_{SB\rightarrow W}]$ that includes the present value of the solar constant $S^*_0$, see Fig. \ref{plasimsnowball} a) for an example. While in simple EBMs the stable solutions correspond to fixed points in the phase space, in global climate models, the attracting solutions correspond to  states featuring, in general, chaotic behaviour, where the invariant measure is supported on a strange attractor \cite{Luchyst,LBHRPW14}. The  temperature profiles of the stable climates obtained  with the 1D-EBM given in Eq. \ref{1DEBM} {\color{black}with a suitable choice of the value of the parameters} are in broad agreement with what obtained with complex general circulation models, see \cite{BLL2014}.} 

For a given value of $S^*$ within the range of bistability, the warm climate is characterized by a very active atmosphere whose variability is sustained by the presence of a {\color{black}large meridional temperature gradient, which supports the existence of eddies as result of the baroclinic conversion between potential and kinetic energy}, whereas the snowball climate has a much lower level of atmospheric variability, with a correspondingly low meridional temperature gradient. Figure \ref{plasimsnowball} b) shows that the intensity of the Lorenz energy cycle {\color{black}$C(P,K)$, given by the average rate of conversion of potential into kinetic energy, which, at steady state, }is also equal to the average rate of dissipation of kinetic energy due to friction \cite{Lor67,Peixoto:1992,Lucarini09PRE}, is substantially higher in the warm than in the snowball state. See \cite{Pierrehumbert} for a comprehensive discussion of the atmospheric circulation in the two regimes. 

Thermodynamics provides additional insight to the critical transitions. One can define a Carnot-like climate \textit{efficiency} $\eta=(\Theta^+-\Theta^-)/\Theta^+$ in terms of the average temperatures inside the system where  heating ($\Theta^+$) and cooling ($\Theta^{\color{black}-}$) {\color{black}processes} take place {\color{black}on the average}, respectively \cite{Lucarini09PRE}. The efficiency is higher when there is stronger correlation between temperature fluctuations and heating rates. The tipping point is accompanied by a sudden decrease in the efficiency $\eta$, which can be interpreted as a result of the system getting closer to thermodynamic equilibrium, see Fig. \ref{plasimsnowball}c) and discussion in \cite{LBHRPW14,Luchyst,Lucarini2013a,Boschi}.
\footnote{A larger (smaller) concentration of greenhouse gases leads to a shift of the range of multistability towards lower (higher) values of $S^*$. The presence of gigantic concentration of $CO_2$ is deemed responsible for the exit of the Earth from the snowball state experienced in the Neoproterozoic \cite{Pierrehumbert}. {\color{black}Note also that the range of multistability is altered when one considers variations in the rotation rate of the planet, with multistability being eventually lost and a unique climate emerging when very slow rotation rates are considered \cite{Boschi,Lucarini2013a}}.}

\section{Critical Transitions and Edge States}\label{edge2}

%The examples discussed above look at the bifurcation of the system by studying how the stability properties of the solutions change when a parameter is varied. Ashwin et al. \cite{Ashwin12} have recently emphasized the need for widening the usual treatment of bifurcations in dynamical systems in order to accommodate transitions between competing basins of attraction strongly dependent on the presence of a finite rate of change of one or more parameters of the system or on the presence of stochastic forcing of finite amplitude. 
\subsection{Energy Landscapes, Large Deviations, and Transitions}
Note that, far from the critical transitions indicated in Fig. \ref{plasimsnowball}, it is possible to jump from the warm attractor to the snowball attractor and \textit{vice versa} by perturbing the system through suitably defined time-dependent deterministic and/or stochastic forcing.  The transition from one basin of attraction to another one  is most easily obtained through a Dirac's $\delta$-like perturbation applied to the orbit (this might correspond, in physical terms, to the impact of an asteroid on our planet). 

Nonetheless, arguably the most interesting scenario of forcing is given by the presence of stochastic perturbations. Climate science has  indeed been one of the first areas of natural sciences where the paradigm of stochastically perturbed dynamical systems has been extensively used, following the Hasselmann programme \cite{hasselmann_stochastic_1976}; see also later discussions in  \cite{arnold_hasselmanns_2001,saltzman_dynamical,MC11}. The basic idea is the following: if one considers separately slow and fast modes of variability of the climate system, in the limit of infinite time scale separation, the impact of the fast modes on the slow modes can be essentially treated as a stochastic correction to the deterministic dynamics of the slow modes. This point of view is closely related to the mathematical theory of the \textit{averaging method} for performing the coarse-graining of the dynamics of multiscale dynamical systems \cite{arnold1988,K92}.

The Freidlin-Wentzell theory \cite{FW98} provides a comprehensive framework in multistable systems for studying the probability of transitions outside one of the basins of attraction triggered by stochastic forcing. The Freidlin-Wentzell theory is most easily understood in the case of a system whose dynamics takes place in an energy landscape (plus additive white noise), such that:
\begin{equation}
\mathrm{d}{y}=-\nabla_y V(y)\mathrm{d}t +\epsilon \mathrm{d}W\label{ito}
\end{equation}
where $y\in\mathbb{R}^n$, $V:\mathbb{R}^n\rightarrow \mathbb{R}$ is sufficiently smooth, and $\mathrm{d}W$ is a vector of $n$ independent increments of Brownian motions ({\color{black}the It\^o or Stratonovich conventions are equivalent, in this case}), and $\epsilon$ determines the strength of the noise. As is well known, one can associate to the stochastic differential equation given by Eq. \ref{ito} a Fokker-Planck equation describing the evolution of the probability density function (pdf) of  an ensemble of trajectories obeying the stochastic differential equation as follows:
\begin{equation}
\frac{\partial p(y,t)}{\partial t}=\nabla_y\cdot (\nabla_y V(y) p(y,t))  + \frac{\epsilon^2}{2} \Delta p(y,t) \label{fp}
\end{equation}
where, under suitable conditions of integrability and in the weak noise limit $\epsilon\rightarrow 0$, the stationary solution corresponding to the invariant measure is given by $\lim_{t\rightarrow \infty} p(y,t)=p(y)$, where at leading order $p(y)\propto \exp[-2V(y)/\epsilon^2]$. The local minima of the potential $V$ ({\color{black}fixed-point }attractors in the case of deterministic dynamics with $\epsilon=0$) correspond to the local maxima of $p$, so that, \textit{e.g.}, a double-well potential corresponds to a bimodal pdf. In the limit of weak noise, trajectories starting near a local minimum of $V$ typically a wait long time before moving to the neighborhood of a different local minimum of $V$,  and the transitions take place most likely through the lowest energy saddle linking the initial basin of attraction to any other basin. 

A comprehensive treatment of the critical transitions accounting for the effect of noise can be found in \cite{Kuehn11}. Let's now consider the simple case of $y\in\mathbb{R}$ and focus again on the problem of the transitions between the snowball and warm climates. If one includes stochastic forcing in the form of additive white noise on the right hand side of Eq. \ref{0DEBM}, we have 
\begin{equation}
{\mathrm{d}}{T}(t)=-\frac{\mathrm{d}}{\mathrm{d}T}V(T)dt+\epsilon \mathrm{d}W \label{0DEBMb}
\end{equation}
where $\mathrm{d}W$ is the increment of a Brownian motion and $\epsilon>0$. In the limit of weak noise, the stationary distribution can be expressed at leading order as \begin{equation}
p(T)\propto\exp[-2V(T)/\epsilon^2],\label{pdf}.
\end{equation}
{\color{black}Using large deviations theory, }one can derive that the mean exit time corresponding to the transition from the basin of attraction of the stable solution $A$ to the basin of attraction of the stable solution $B$  through the unstable saddle $U$ can be written {\color{black}in first approximation as}: 
\begin{equation}
\mathcal{\tau}_{A\rightarrow B} \propto \exp{[2(V(U)-V(A))/\epsilon^2]},\label{kramers}
\end{equation}
{\color{black}while the constant in Eq. \ref{kramers} is specified by the celebrated Kramer's escape rate formula \cite{K40,HTB90}}. See Fig. \ref{fig:double_well}, where the states $A$ and $B$ can be though of as corresponding to the warm and snowball state, and the saddle $U$ coincides with the boundary of the two basins of attraction. %It is worth remembering that a great stimulation of 

%\begin{figure}[ht] %[t!]
%    \begin{center}
%	%\scalebox{0.5}{\includegraphics{bif_ta_3}} 
%	\includegraphics[width=0.5\textwidth]{double_well}
%        \caption{\label{fig:double_well} 
%        Scalar double well potential function $V(T)$; the {\color{black}warm} and the {\color{blue}snowball} states correspond to the two attractors of the system, separated by the {\color{green}saddle} point. The noise triggers the transition from one basin of attraction to the other with a mean exit team  described by Eq. \ref{kramers}.}
%    \end{center}
%\end{figure}

%Similar formulas apply in higher dimensional cases, where the dominant contribution to the probability of transition comes from lowest saddle connecting the two basins of attraction.  
In the geophysical literature, tipping points are mostly studied by using Eq. \ref{pdf} to construct an equivalent one-dimensional pseudo-potential from the pdf derived starting from the time series of a selected climate observable. Subsequently, Kramer's theory is used to estimate the probability of transition from one basin of attraction to another one within a given time frame \cite{Lenton2008,Scheffer12}. Nonetheless, inconsistencies in this method emerge from the fact that the operation of projecting the dynamics onto only one observable is very likely to cause errors in the evaluation of the time scales of the dynamics, thus breaking the simple and powerful relations existing between invariant density and escape probability from a local mimimum of the effective potential, see the discussion in \cite{LFW12}. 

When considering dynamical systems featuring multiple attractors and undergoing gaussian stochastic forcing, large deviations theory \cite{T09} provides methods for constructing the so-called instantons, which are the most likely noise-driven trajectories leading to the transitions from one basin of attraction to the other one. An instanton is constructed as a minimizer of the related Freidlin-Wentzell action and can be interpreted as the \textit{most efficient} way to exit a local potential minimum. See \cite{Grafke2015} for an introduction in the context of fluid dynamics.  

Some geophysical fluid dynamical systems feature extremely rare transitions between different regimes of motions. These are rare excursions between regions of the attractor of the system that take place only on ultralong time scales, so that, performing an operation of coarse graining to the dynamics, one can describe them as noise-induced transitions between separate attracting sets for the deterministic part of the dynamics \cite{Bouchet2014,Laurie2015}. 

The instantons method allows for treating more general cases than those of systems whose dynamics is governed by an energy landscape. What we find a bit unsatisfactory in this otherwise extremely powerful framework of noise-driven transitions is the fact that one is left with the question of what is the mathematical nature and the physical justification of the noise. Note that, given the lack of time-scale separation in the climate variability \cite{Peixoto:1992}, it is a bit of a stretch to use standard arguments to justify the presence of white noise as resulting from fast atmospheric motions. In other terms, despite the huge insight of the Hasselmann program mentioned above, climate dynamics is  in fact not the best setting for using the averaging method.  As discussed in \cite{WL12,WL13} in the spirit of  \cite{zwanzig_memory_1961,mori_transport_1965} and confirmed in the mathematically more rigorous treatment in \cite{CLW15a,CLW15b}, when no scale separation is present between the scales of motions we want to resolve and those we want to parameterize, the effect of the latter on the former entails considering time-correlated noise and non-markovian effects. 

As a final note we wish to remark that, even in the context of systems of the form given in Eq \ref{ito}, if one assumes that the noise possesses instead slow decay of correlations, such as L\'evy processes,  the standard Freidlin-Wentzell picture is dramatically altered, with an entirely different dynamical scenario to be considered. While in the case of {\color{black}white} noise the transition from one basin of attraction to the other occurs in the very unlikely case that many subsequent stochastic perturbations \textit{conjure} to push the orbit across the saddle, in the case of L\'evy noise one or few large kicks do the job \cite{Dit99a,IP06}, so that the expression for mean exit time is rather different from what reported in Eq. \ref{kramers}.  This scenario seems to be relevant in a climate context as well \cite{Dit99b,Dit10,GIHP11}. 

\subsection{Edge States}

Following a classical point of view within statistical mechanics, we would like to be able to understand how {\color{black} noise-induced} transitions between separate climatic attractors may take place {\color{black} by gaining insight on the underlying purely deterministic  evolution laws}. In this regard, we take advantage of the \textit{edge tracking method} recently developed by Eckhardt and collaborators for studying fluid dynamical systems possessing multiple (quasi-)steady states \cite{PhysRevLett.96.174101,PhysRevE.78.037301,Schneider13022009}. 

The plane Couette flow or the pipe flow, in the case of sufficiently high Reynolds numbers, feature the coexistence of a (quasi-)attractor corresponding to the turbulent state and of an attracting fixed point corresponding to laminar flow. We remark that the turbulent state cannot be properly described as corresponding to a separate attractor, because turbulence is transient - even if with exceedingly long life time, which grows superexponentially fast with the Reynolds number \cite{Grossmann2000,Hof2006,Hof2008,PhysRevE.78.037301}. But, using arguments of time-scale separation, we can say that the scenario is (almost) indistinguishable from that of two separate attractors, where the turbulent state is steady, so we will use an abuse of terminology in describing these results, and refer to the turbulent state as corresponding to a proper attractor, possessing its own basin of attraction. 

The basic goal is to understand what lies in-between the two attractors in order to a) have a global view on the properties of the phase space of the system, beyond the paradigm of looking at steady states; and b) understanding which combinations of external forcings and internal processes might more easily lead to transitions between the steady states. 

Clearly, in the case of a bistable system, there must be a separatrix, a repelling set which acts as boundary between the two basins of attraction. The natural question one might ask is: what is the evolution of an orbit starting nearby or on such a separatrix? In the case of an energy landscape, such a basin boundary is the \textit{mountain crest} separating the two basins of attraction, and the evolution law would lead an orbit starting from a point on the boundary to the nearby {\color{black}local minimum \textit{restricted} to the subset of the phase space} corresponding to the basin boundary), where the evolution would stop. {\color{black}Clearly, the end point is in fact a saddle, because there is an (additional) unstable direction}. Orbits departing near the basin boundary but not exactly on it, would end up in the {\color{black}attracting} set corresponding to the basin they start from. 

Eckhardt and collaborators showed that trajectories initialized on the boundary between the two basins of attraction converge to a unstable saddle, which is a relative attractor with respect to the basin boundary (but obviously a repelling set when viewed globally). The scenario has been confirmed in the case of truly  separate attractors in a toy model discussed in \cite{Vollmer09}.

\begin{figure} [ht]
    \begin{center}
	%\scalebox{0.5}{\includegraphics{bif_ta_3}} 
	\includegraphics[trim=1.5cm 9cm 4cm 4cm, clip=true, angle=0,width=0.4\textwidth]{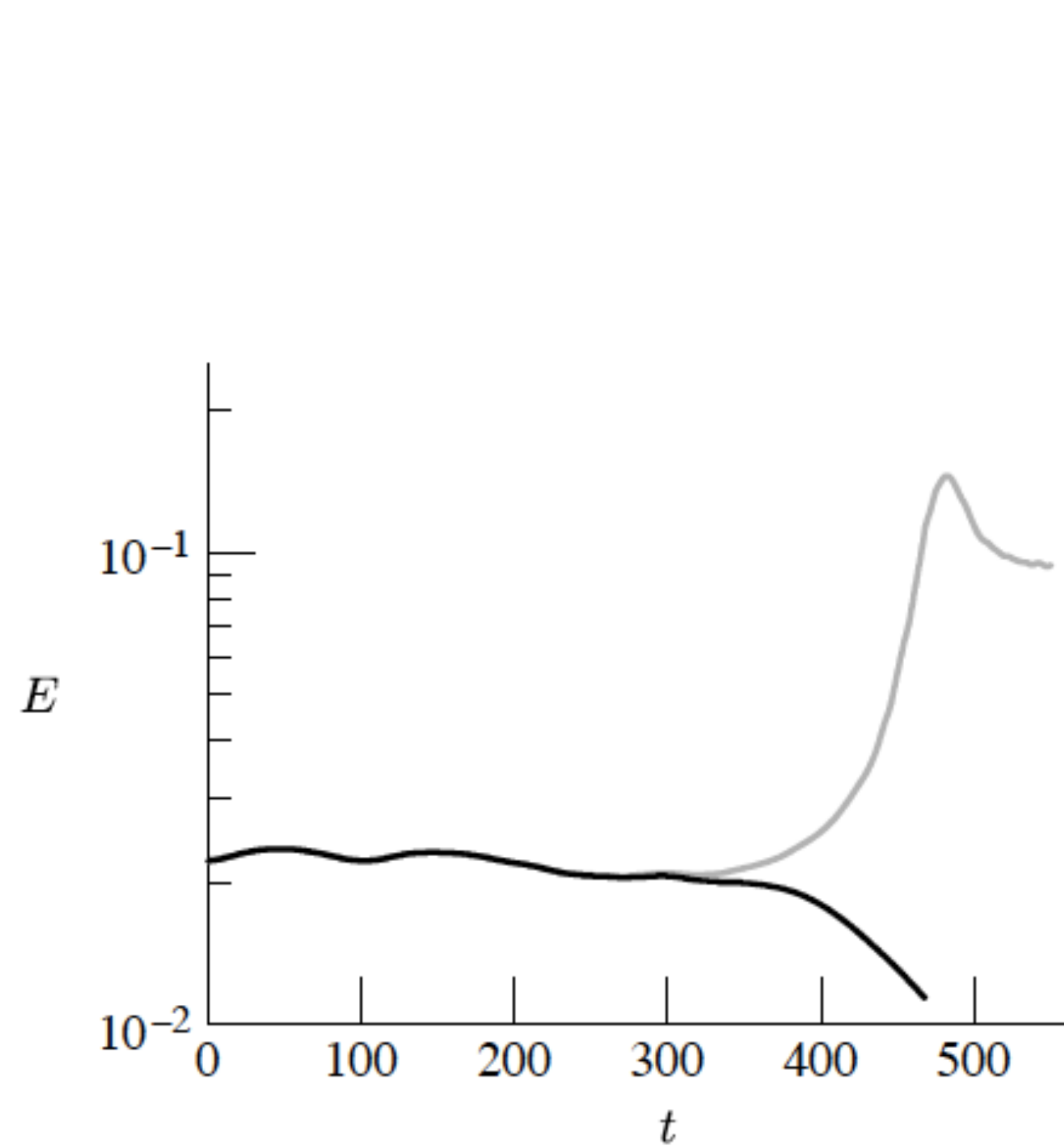}\\a)\\
\includegraphics[trim=1.5cm 1cm 1cm 1cm, clip=true, angle=270,width=0.4\textwidth]{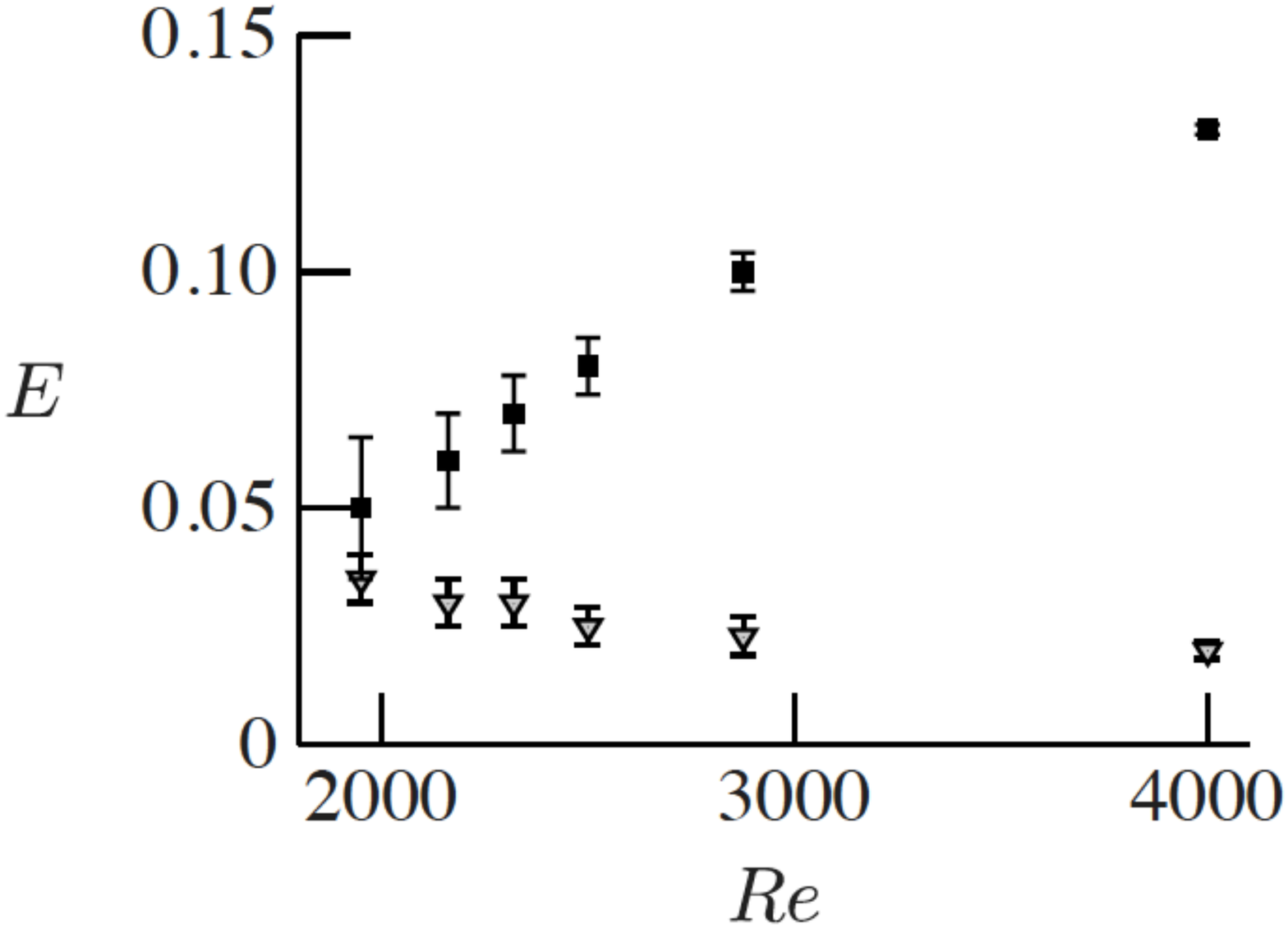}\\b)
        \caption{\label{edge} 
        Edge state in the pipe flow. a) Time evolution of the energy of the turbulent flow for two trajectories starting near the edge state but belonging to two different basins of attraction: one trajectory collapses to zero (laminar state), {\color{black} and} another one increases its turbulent kinetic energy until the quasi-steady turbulent state is reached. b) Turbulent kinetic energy of the steady turbulent state (square dots) and of the edge state (triangles) for various values of the Reynolds number. The turbulent energy of the other steady state, the laminar state, is zero by definition.  Note that we are using an abuse of language in defining the turbulent state as corresponding to a true attractor (it is a very long transient, instead, see text). Adapted from \cite{Schneider13022009}.}
    \end{center}
\end{figure}

Such an unstable saddle, the so-called \textit{edge state},  has  been found to be either a fixed point, or a periodic orbit, or, in some cases, a chaotic solution taking place on  a strange geometrical set. The relevance of the {edge state} lies in the fact that it is crucial for understanding the global properties of the system and that it provides the gate between the two attractors: trajectories resulting from (weak) external forcings that are successful in achieving the critical transition have to pass nearby the edge state.  

Of course, given its (extreme) instability, one cannot find the edge state by direct forward numerical integration; instead, one can find it by a suitable algorithmic procedure named \textit{edge tracking}. The basic idea is to find by bisections (and long forward integrations) two nearby points in the phase space that are on the two sides of the separatrix, follow their forward evolution for a given time (typically short), and then repeat the bisection  in order to reduce the divergence between the two trajectories due to the global instability connected with the bistability of the system. 

The bisection is obtained by interpolation to {\color{black}the} midpoint (more general constructions of convex combinations are possible) between the initial conditions belonging to the two sides of the basin boundary, checking, using a long run, whether the new initial condition leads to one or to the other attracting set, and repeating until two nearby (according to a prescribed criterion) initial conditions leading to different asymptotic dynamics are obtained. One retains short segments of the pair of control trajectories belonging to these most tightly bracketing initial conditions. The length of these segments are limited by a requirement on the maximal separation of the control trajectories. After a transient, any of the segments obtained iterating this procedure can be taken to represent the edge state, approximated by many discontinuous segments of {\color{black}the} trajectory.

A key mathematical fact to be considered here is that the basins of attraction are invariant sets, which ensures that the procedure is well-defined. % and it is useful to assume that the basin boundary, while possibly folded, is \textit{sufficiently} regular, which makes it possible for our finite-precision method to apply well. 
It is a crucial - in terms of robustness and efficiency - aspect of the procedure to choose a useful observable able to tell us unambiguously and soon enough whether the trajectory is ending up on one or on the other attractor, see Fig. \ref{edge}a. An efficient observable has a relatively small variability in the transient runs compared with the actual difference between the expectation values computed on the two separate attractors.  

{\color{black}Additionally, it seems quite natural to investigate the geometrical properties of the boundary separating the the basins of attraction, as it is not a-priori clear whether one should expect a manifold of co-dimension one or a more complex geometrical object. We will come back to this aspect in Sect. \ref{boundarygeometry}.} 

In the case of the problems studied by Eckhardt and collaborators, the natural choice is to use the total turbulent kinetic energy of the flow. In the limit of infinitely small separation between the two bracketing trajectories, one actually obtains the transient behaviour leading to the the edge state, and, then, is able to reconstruct the dynamics on the edge state. While this procedure is rather intuitive, one might reasonably expect lots of technical difficulties related to the need of controlling strong instabilities in a high-dimensional dynamical system. Eckhardt and collaborators have convincingly shown the robustness of the edge tracking method \cite{PhysRevLett.96.174101,Schneider13022009,Vollmer09}.

\begin{figure}[ht]
    \begin{center}
	%\scalebox{0.5}{\includegraphics{bif_ta_3}} 
	\includegraphics[trim=2cm 3cm 1cm 3cm, clip=true,angle=270,width=.7\textwidth]{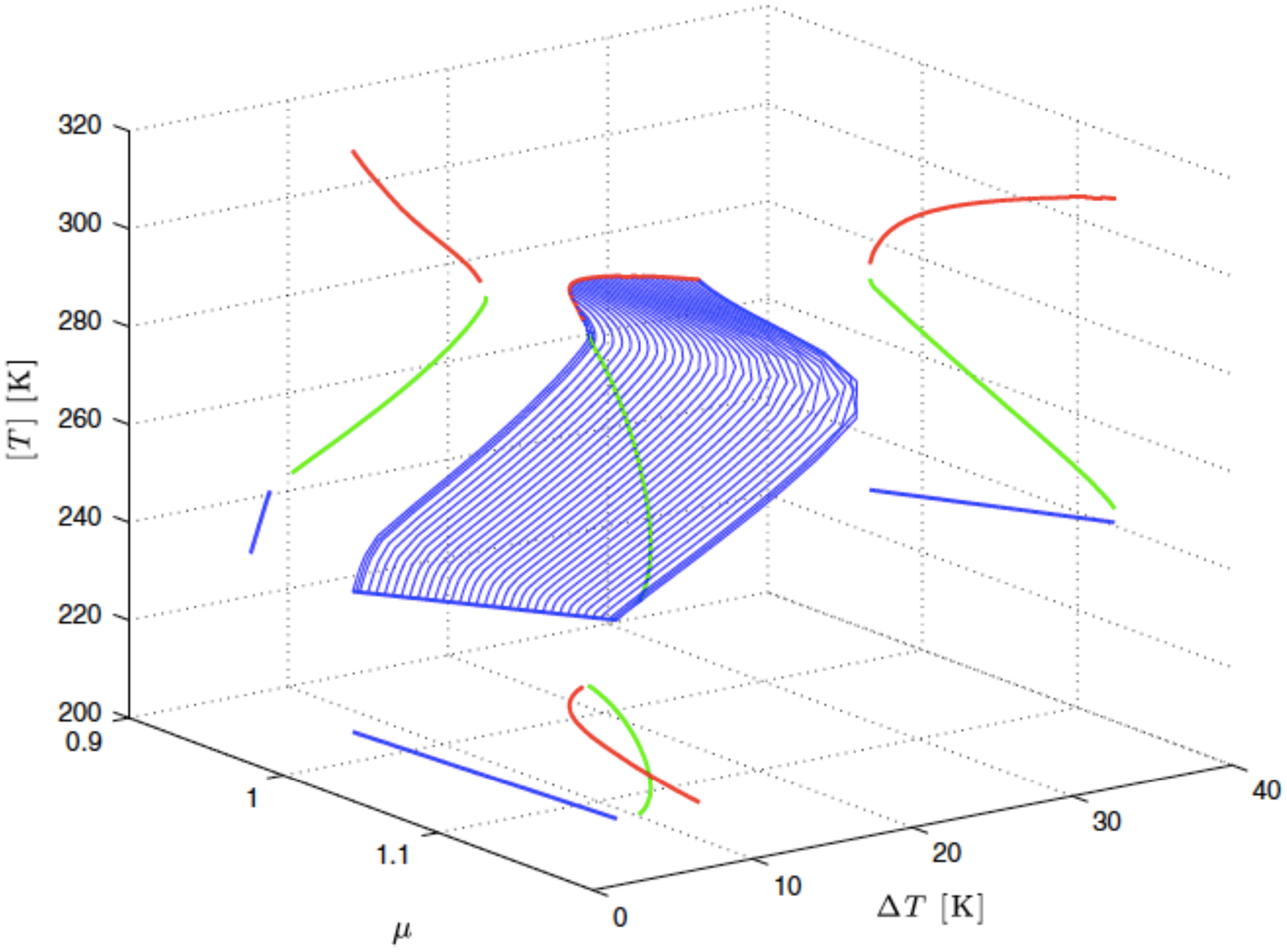}
        \caption{Bifurcation diagram of the Ghil \cite{Ghil} model defining the constitutive relations between the normalized intensity of the incoming radiation ($\mu=S^*/S^*_0=1$ corresponds to the current conditions), the globally averaged temperature $[T]$ and the low-to-high latitudes temperature difference $\Delta T$. The thick red, green, and blue lines correspond to the warm, edge, and snowball state, respectively. The thin blue lines depict heteroclinic  trajectories - the instantons. Adapted from \cite{BLL2014}.}\label{bodai14}
    \end{center}
\end{figure}

In a previous paper \cite{BLL2014}  we have introduced in the geophysical community  the edge tracking method to provide a fresh outlook on the properties of the Ghil-Sellers (GS) 1D-EBM) \cite{Ghil}, which can be written in the form of Eq. \ref{1DEBM}. Ghil \cite{Ghil} solved the appropriate boundary value problem and found  for a wide range of values of the solar constant three coexisting stationary solutions, and  carried out a stability analysis able to identify the stable warm and snowball climates, plus an unstable state. Such a state was identified by Ghil as the saddle point of a suitably constructed potential, and we proved that it coincides with the edge state found using the tracking procedure, see Fig. \ref{bodai14}. Our investigation made {\color{black}it} apparent that the most efficient (but definitely not unique) observable to be used for characterizing during the edge tracking whether the trajectory ends up in the snowball or warm state is the globally averaged surface temperature, because warming and cooling are mainly controlled by the anomalies in radiation emission and changes in albedo, which are both to a good extent controlled by changes in the globally averaged temperature. We also studied the physics of the three states, relating their instabilities  to relevant macroscopic thermodynamical properties such as large scale temperature gradients and entropy production, using the conceptual framework introduced in \cite{Luchyst,Lucarini2013a,Boschi}. 

\subsection*{Remark}
Looking at the top right projection of the three-dimensional curve portrayed in Fig. \ref{bodai14}, one sees that the edge state is characterized by the fact that lower average temperatures correspond to higher values of  solar constant. One can interpret this feature as representative of the fact that the edge state has, effectively, a \textit{negative} heat capacity, and is not thermodynamically stable. Making a \textit{Gedankenexperiment} where the Earth system in the edge state is put in an isolated box with another body emitting and absorbing radiation and looking at the dynamics of fluctuations should clarify this point. 

Interestingly, a qualitatively similar property can be found when looking at the case of turbulent flows analysed by Eckhardt and co. \cite{Schneider13022009} and portrayed in Fig. \ref{edge} b): in the edge state, larger values of the Reynolds number, which is representative of the applied pressure gradient, correspond to lower values of the turbulent kinetic energy of the flow. This can be interpreted as corresponding to conditions of mechanical instability, where the viscosity is \textit{negative}.

\subsection{Melancholia States in a Climate Model}

In this paper we want to show how the concept of the edge state can be used for characterizing  globally the dynamical properties of a severely simplified yet Earth-like new climate model featuring multistable properties in a realistic range of values of the solar constant. As mentioned in Sect. \ref{intro}, we propose to call the climatic edge states \textit{Melancholia states}. 

\begin{figure}[ht]
    \begin{center}
	%\scalebox{0.5}{\includegraphics{bif_ta_3}} 
	\includegraphics[trim=0cm 0cm 0cm 0cm, clip=true,angle=0,width=.7\textwidth]{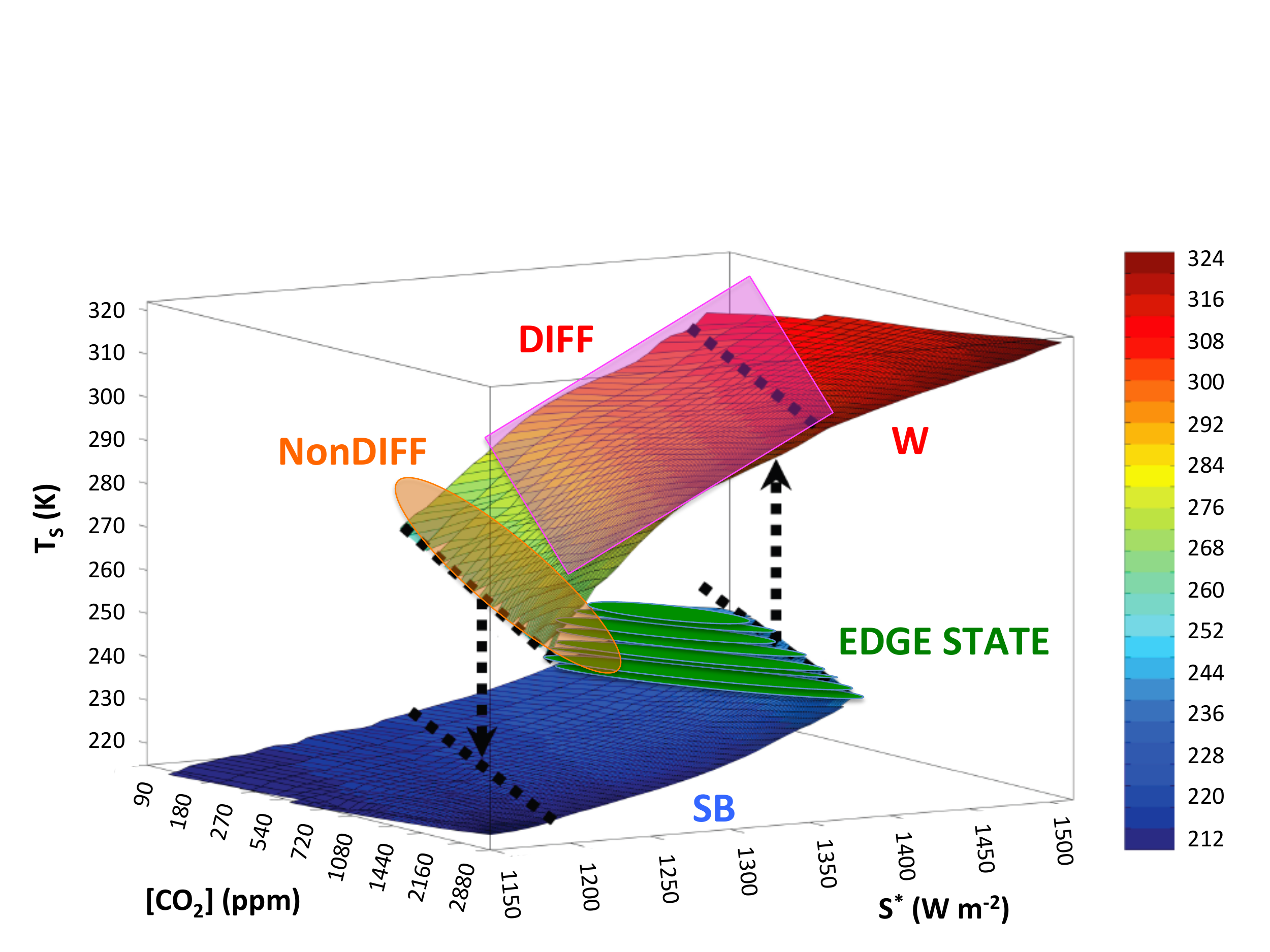}%{boschiadapted_rev.pdf}
        \caption{In \cite{Luchyst,LucACP,Lucarini2013a,Boschi} we have studied the macroscopic thermodynamic properties of the climate system in the {\color{black}W} state  and {\color{blue}SB} state. In \cite{RLL16,LLR16} we have studied the differentiability of the invariant measure in the W state around present climate conditions (purple shading, see DIFF). In \cite{Tantet2015b} we have studied the loss of differentiability of the invariant measure near the $W\rightarrow SB$ tipping point (orange shading, see NonDIFF). All of these analyses have been performed using the  climate model PlaSim \cite{plasim}. The goal of this paper is to charactacterize what is in-between the W and SB states, \textit{i.e.}, constructing and studying the edge states (green shading, see {\color{green}EDGE STATE}). For technical reasons, we use a simplified yet Earth-like climate model, PUMA-GS. The 3D surface plot is adapted from \cite{Lucarini2013a}.}\label{boschietal}
    \end{center}
\end{figure}

Figure \ref{boschietal} clarifies  how this work relates to previous analyses we have performed on the thermodynamics and statistical mechanics of climate. While most of the studies we consider here have been performed using the open source climate model PlaSim \cite{plasim}, in this specific investigation, given the complexity of the procedure of edge tracking, we resort to using a simpler model, later described in detail in Sect. \ref{model}.

The climate model PUMA-GS is constructed by coupling PUMA, an open source three-dimensional primitive equations spectral model of the atmosphere \cite{puma}, with the surface described by a slightly modified version of the reaction-diffusion GS model \cite{Ghil}, extended symmetrically along the longitudinal direction, thus making it possible to couple it with the atmosphere aloft.  The coupling is realized by imposing that the relaxation temperature profile that provides the baroclinic forcing to the atmospheric component is enslaved, through a simple representation of  radiative convective adjustment, to the surface temperature field of the GS model, which basically plays the role of the ocean. This provides a natural and coherent separation between fast variables (atmosphere) and slow variables (ocean). The ocean exchanges energy with the atmosphere aloft as well as featuring absorption of shortwave radiation and emission of longwave radiation. The model features multistability for reasonable values of its parameters and is described in detail later in the paper. This newly introduced model sits in terms of complexity between PUMA \cite{puma} and PlaSim \cite{plasim}. 

{\color{black} PlaSim is comparatively simple with respect to the state-of-the-art climate models considered in the compilation of the latest IPCC report \cite{IPCC13}, but nonetheless includes a (simplified) treatment via parameterizations of important physical processes such as convection, sea ice formation, clouds formation, precipitation in liquid and solid forms, boundary layer processes, which are entirely absent in PUMA-GS, plus a much more advanced treatment of radiative processes. Additionally, PlaSim allows for a flexible configuration of the land surface mask and of the orography, with an ensuing large variety of configurations for the ocean. Nonetheless, the ocean component of the model, similarly to PUMA-GS, has merely the role of horizontally transporting heat via diffusion, while not featuring any description of actual ocean currents. }

The challenge we address is threefold:
\begin{itemize}
\item We want to show that the edge tracking algorithm works also in the context of a system characterized by multiscale dynamics, a variety of positive and negative feedbacks, inhomogeneous physical and mathematical properties, and featuring intense fluxes across its subdomains. These features potentially result in  a time-dependent edge state, possibly characterized by nontrivial dynamics and nontrivial geometrical properties, and that cannot be found as a solution to a boundary value problem. This suggests an increased potential and scope for the methodology originally proposed by Eckhardt and collaborators; 
\item We want to specifically use the edge tracking algorithm for identifying the edge states of the climate system separating the warm climate states from the snowball states. We want to study what lies in-between the two curves corresponding to the stable climates, portrayed in, \textit{e.g.}, Fig. \ref{plasimsnowball}a, and derive the long term statistics of the climatic edge states, mirroring what has been shown in the case of edge states in turbulent flows in, \textit{e.g.}, Fig. \ref{edge}.    
\item As we observe that edge states are relative (in the reduced space of the basin boundary) attractors,  we want to complement the characterization of their climatology with the investigation of their dynamical properties, and in particular understand whether they feature trivial, quasi-periodic, or chaotic dynamics, in order to classify the properties of the weather of such special states. {\color{black}}Note that in this way we separate the {\color{black} slow} global climatic instability (driven by the ice-albedo feedback) leading to the presence of multiple steady states, from the {\color{black} fast} baroclinic instability \cite{Holton} responsible for the variability of the weather {\color{black} and acting as negative feedback}. 
\end{itemize}

\section{The Climate Model PUMA-GS: Formulation and Edge-Tracking}\label{model}
\subsection{Model Formulation}
The atmospheric component of the PUMA-GS model is provided by PUMA~\cite{puma}, which consists of a dynamical core: the dry hydrostatic primitive equations on the sphere (mapped laterally by the latitude $\phi$ and longitude $\lambda$), solved by a spectral transform method (only linear terms are evaluated in the spectral domain, nonlinear terms are evaluated in grid point space). %(a more concise description of PUMA can be found in \cite{Lunkeit:2001,PhysRevE.87.052113}):
The equations for the prognostic state variables, the vertical component (with respect to the local surface) of the absolute vorticity $\zeta=\xi+2\nu\Omega_E$ (where $\xi$ is the vertical component of the relative vorticity, $\nu=\sin\phi$, and $\Omega_E=2\pi$/day is the angular frequency of the Earth rotation) the (horizontal) divergence $D$, the (atmospheric) temperature $T_a=\bar{T_a}+T_a'$ (separated into a time-independent arbitrary reference part $\bar{T_a}$ and anomalies $T_a'$) and the logarithmic pressure (normalized by the surface pressure $p_s$) $\sigma=\ln p/p_s$, read as follows:
\begin{eqnarray}\label{eq:puma}
  \partial_t\zeta &= s^2\partial_{\lambda}F_v - \partial_{\nu}F_u - \tau^{-1}_f\xi - K\nabla^8\xi,\label{eq:puma_zeta} \\
  \partial_t D    &= s^2\partial_{\lambda}F_u + \partial_{\nu}F_v - \nabla^2[s^2(U^2+V^2)/2+\Phi+T_a\ln p_s]\label{eq:puma_D} \\
  &- \tau^{-1}_fD - K\nabla^8D, \nonumber \\
  \partial_t T_a' &= s^2\partial_{\lambda}(UT_a') - \partial_{\nu}(VT_a') + DT_a' - \dot{\sigma}\partial_{\sigma}T_a \label{eq:puma_T} \\
  &+ \kappa T_a\omega/p + \tau_c^{-1}(T_R(T_s) - T_a) - K\nabla^8T_a', \nonumber \\
  \partial_t\ln p_s &= -s^2\partial_{\lambda}\ln p_s - V\partial_{\nu}\ln p_s - D - \partial_{\sigma}\dot{\sigma},\label{eq:puma_p} \\
  \partial_{\ln\sigma}\Phi &= -T_a,\label{eq:puma_state}
\end{eqnarray}
where $s^2=1/(1-\nu^2)$, $F_u = V\zeta - \dot{\sigma}\partial_{\sigma}U - T_a'\partial_{\lambda}\ln p_s$, $F_v = -U\zeta - \dot{\sigma}\partial_{\sigma}V - T_a's^{-2}\partial_{\nu}\ln p_s$, $U=u\cos\phi$, $V=v\cos\phi$, $u$, $v$ being respectively the horizontal and vertical wind velocity components, and $\Phi$ is the geopotential height. %, $\sigma=\ln p/p_s$.
Equations \ref{eq:puma_zeta},\ref{eq:puma_D}, and \ref{eq:puma_p} express the conservation of momentum, Eq. \ref{eq:puma_T} expresses the conservation of energy, and Eq. \ref{eq:puma_state} is the equation of state.

A number of simple parametrizations are adopted in order to improve the realism and the stability of the model. Firstly, the hyperdiffusion operator $K\nabla^8$  is added to the equations of vorticity, divergence and temperature, to represent {\em subgrid-scale} eddies. Secondly, {\em large-scale} dissipation of vorticity and divergence is facilitated by Rayleigh friction of time scale $\tau_f$. Thirdly, the physics of diabatic heating due to radiative heat transport is parametrized by Newtonian cooling: the temperature field is relaxed (with a time scale $\tau_c$) towards a reference or \textit{restoration} temperature field $T_R$, which can be considered a radiative-convective equilibrium solution. We adopt the following simple expression for the restoration temperature~\cite{puma}:
\begin{eqnarray}\label{eq:res_temp}
 T_R = (T_R)_{tp} + \sqrt{[L(z_{tp} - z(\sigma))/2]^2 + S^2} + L(z_{tp} - z(\sigma))/2,\label{eq:res_temp_1} \\
 (T_R)_{tp} = [T_s] - \bar{L}z_{tp}, \label{eq:res_temp_2} \\
 L(\lambda,\phi) = \partial_zT_R=(T_s(\lambda,\phi) - (T_R)_{tp})/z_{tp}. \label{eq:res_temp_3}
\end{eqnarray}
where $(T_R)_{tp}$ and $z_{tp}$ are the temperature and height of the tropopause, respectively, $L$ ($\bar{L}$) is the (average) lapse rate, $[T_s]$ is the globally averaged surface temperature, and $z(\sigma)$ is determined by an iterative procedure~\cite{puma}. The above expressions indicate that the restoration temperature profile is \textit{anchored} to the surface temperature $T_s$. However, as Eq. \ref{eq:res_temp_2} indicates, $T_R$ at any one point on the sphere is determined by not only the local (dynamical) surface temperature, but also the global average $[T_s]$.

The surface temperature is taken to be governed by the 2D version of the GS EBM~\cite{Ghil}, which is in the form of Eq. \ref{1DEBM}, while the precise expression can be found in~\cite{BLL2014}. {\color{black}Since land masses are absent in this configuration, we are in fact dealing with an aquaplanet climate model.} Two changes are introduced to the GS EBM evolution equations with respect to our previous study \cite{BLL2014}:
\begin{itemize}
\item we introduce a longitudinal component in the model (even if diffusion takes places only along the meridional direction); adding a second dimension is important because we introduce the following atmosphere-to-surface coupling term on the right-hand-side of the equation for the field of the surface temperature: $k_3(T_a(\sigma=1)-T_s)$, where $T_a(\sigma=1)$ depends on latitude, longitude, and time;
\item {\color{black}we change the value of the original meridional diffusion coefficient, in order to make sure it represents the transport of the ocean only: in the original model the diffusion mimics the effects of both atmospheric and oceanic heat transports, but in PUMA-GS we have now a dynamical model for the atmosphere. We choose to reduce the value of the diffusion coefficient by a factor of 4 in order to make sure that in conditions resembling the present-day climate the ocean heat transport is few times weaker than the atmospheric one, in agreeement with observations and models outputs \cite{Peixoto:1992,LucariniRagone,LBHRPW14}}.
\end{itemize}
We note that $T_a(\sigma=1)$ is obtained by linear extrapolation, according to $T_a(\sigma=1) \approx T_a(\sigma=0.9) + \eta(T_s - T_R(\sigma=0.9))$, $0<\eta<1$. With $\eta=1$ the coupling term %in eq. (\ref{eq:gsebm}) 
is $k_3(T_a(\sigma=1)-T_s) \approx k_3(T_a(\sigma=0.9)-T_R(\sigma=0.9))$. Generally $\overline{T_a(\sigma=1)-T_s}\neq 0$ (laterally inhomogeneous heating), but $\overline{[T_a(\sigma=1)]}=\overline{[T_s]} $, where the overbar denotes averaging with respect to time. 

For our setup we chose: $K^{-1} = 0.25$ days, $\tau_c=30$ days, $\tau_f=1$ day. $\bar{L} = 0.0065$ K/m, $z_{tp} = 12000$ m. We also adopt a coarse resolution of T21 (i.e., the series of spherical harmonics are triangular-truncated at total wave number 21). This implies the optimal number of Gaussian grid points: $N_{lon}= 2N_{lat}=64$; and a number $N_{lev}=10$ of levels are taken. No orography is defined, \textit{i.e.}, we have a zonally-symmetric configuration, empirical functions of the GS EBM, like $C(\phi)$, depend on the latitude only. The equations are integrated numerically using a $\Delta t = 1$ [hour] time step size.

\begin{figure}[ht] %[t!]
    \begin{center}
	%\scalebox{0.5}{\includegraphics{bif_ta_3}} 
	\includegraphics[width=0.5\textwidth]{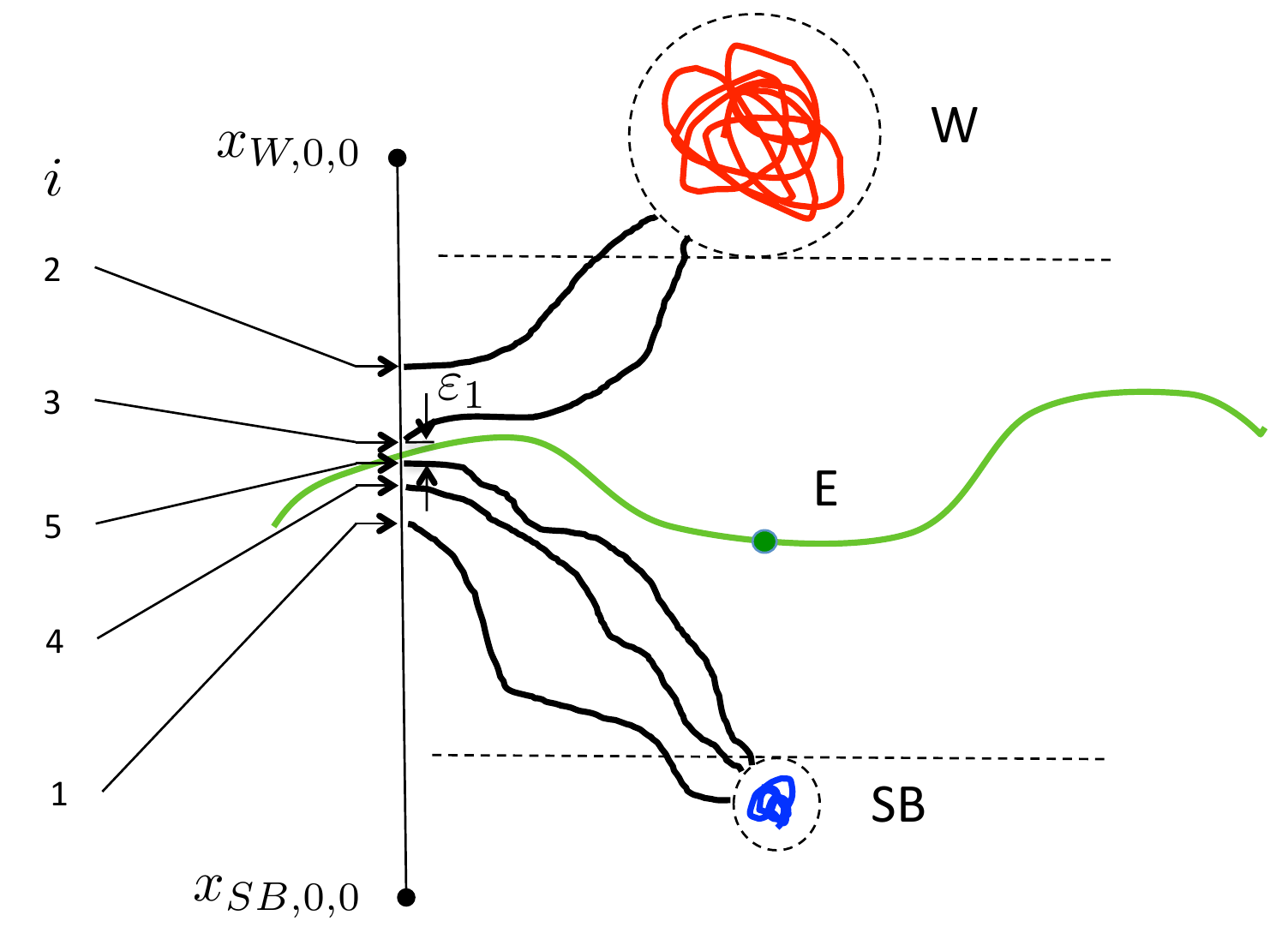}
        \caption{\label{fig:edge_tr_init} 
{\color{black}Scheme of the} initialization of the edge tracking procedure. By subsequent bisections, we define two initial conditions near the basin boundary and near to each other such that they belong to the two different basins of attraction. The bisection is guided by testing whether our guessed initial conditions get close enough to either attractor. The snowball attractor SB, the warm attractor W and the edge state E are depicted. Details in the text. }
    \end{center}
\end{figure}

With such a setup we find bistability in a range of the relative solar strength $\mu=S^*/S^*_0$: the warm-to-cold (cold-to-warm) tipping point is found at $\mu_{W\rightarrow SB}=S^*_{W\rightarrow SB}/S^0_0\approx 0.97$ ($\mu_{SB\rightarrow W}=S^*_{SB\rightarrow W}/S^0_0\approx 1.055$). We take 18 equally spaced ($\Delta\mu=0.005$) sample values in order to study how the properties of the system change when $\mu$ is varied.

\begin{figure}[ht]
    \begin{center}
	%\scalebox{0.5}{\includegraphics{bif_ta_3}} 
	\includegraphics[angle=0,width=0.7\textwidth]{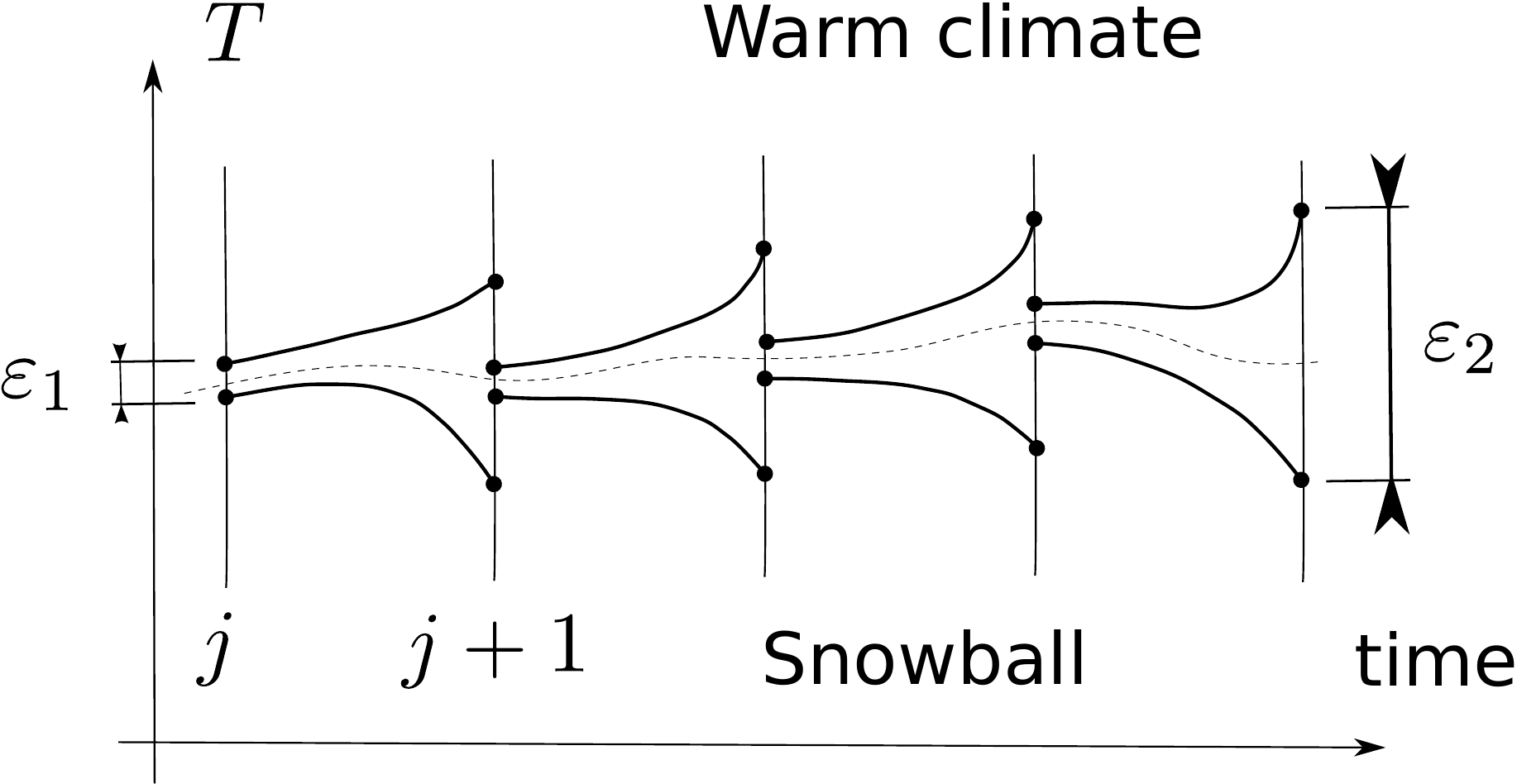}
        \caption{\label{edgetrackingscheme} 
        Scheme of the tracking method {\color{black}used} to construct the edge state in the climate model. We let the two initial conditions defined in Fig. \ref{fig:edge_tr_init} evolve until the two globally averaged surface temperatures differ by $\varepsilon_2$. We then repeat the bisection as in Fig. \ref{fig:edge_tr_init} until a steady state is obtained. See details in the text.}
    \end{center}
\end{figure}

\subsection{Algorithm for Tracking the Melancholia States}

We adapt the edge tracking algorithm so that in the initializing phase of the procedure the basin boundary between the warm and snowball states is closely bracketed by two initial conditions. See Fig. \ref{fig:edge_tr_init} for a cartoon representing the procedure. The initial ($j=0$) bracketing can be obtained as follows. One starts with two arbitrary points in phase space, $x_{W,i,j}$ and $x_{SB,i,j}$, $i=0$, $j=0$ (small black disks in the schematics), belonging to the two different basins of attraction. These points need not to belong to the attractors themselves, which are marked by W and SB in Fig. \ref{fig:edge_tr_init}  (red and blue objects). By iterative bisecting the basin boundary (thick green line - a simple geometrical object for convenience of visualization) is bracketed by points along a straight line in phase space a distance $\varepsilon_1$ apart. After each bisection, a  control simulation (thick black line) reveals on which side of the boundary the midpoint (indicated by the horizontal arrows) is situated. {\color{black}This is realized by checking the vicinity of which attractor (dashed black circles) the trajectory enters starting from the midpoint, possibly after a long waiting time.} It is easier to consider a suitably defined {\em scalar} indicator quantity, and measure $\varepsilon_1$ in this single dimension, and find scalar threshold values (indicated by horizontal dashed lines) that unmistakably indicate the outcome. 

The best indicator to be used for all phases of the edge tracking algorithm is the  globally averaged surface temperature $[T_s]$, similarly to what was discussed in \cite{BLL2014}. The basic reason for such a choice is simply the fact that at $0^{th}$ order the dynamics of the system can be approximated, as discussed above, by Eq. \ref{0DEBM}, where the energy landscape is shown in Fig. \ref{fig:double_well}. In the case of such a simple model, for initial conditions near the saddle, the ice-albedo feedback pushes the system towards the W (SB) attractor through a monotonic increase (decrease) of temperature. {\color{black}In case of the more complex model, monotonicity does not strictly hold, but its violation is restricted to small temperature scales: compare Fig. \ref{fig:bracketing} a) to Fig. \ref{fig:bracketing} b), and Fig. \ref{fig:bracketing} c) to Fig. \ref{fig:bracketing} d), respectively.}

After the initialization, we run the actual edge tracking algorithm; a cartoon is presented in Fig. \ref{edgetrackingscheme}. We launch the two simulations with initial conditions portrayed in Fig. \ref{fig:edge_tr_init} and stop them when the value of their globally averaged surface temperature is apart more than a given value $\varepsilon_2$. We then repeat the bisection procedure outlined in Fig. \ref{fig:edge_tr_init} and define  two nearby initial conditions  whose globally averaged surface temperature differs by less than $\varepsilon_1$. The two orbits originating from those two initial conditions end up in two different climates. But, since they are both initialized very close to the basin boundary, they remain near it for quite some time, quantified by the Lyapunov exponent corresponding to the dominating unstable dimension leading to the bistability.  In the limit of $\varepsilon_1,\varepsilon_2\rightarrow0$, we construct the dashed line representing a trajectory on the basin boundary. We repeat the procedure until a statistically steady state is realized, corresponding to having reached the edge state indicated with E in Fig. \ref{fig:edge_tr_init}. {\color{black}The procedure is then continued in order to be able to construct a trajectory long enough for reconstructing accurately enough the statistical properties of the edge state, and in particular for computing the maximum Lyapunov exponent (MLE).}

\section{Results}\label{results}

\subsection{Practical Implementation of the Edge Tracking Algorithm}
\begin{figure}[ht]
  \sidesubfloat[]{\includegraphics[width=0.45\textwidth]{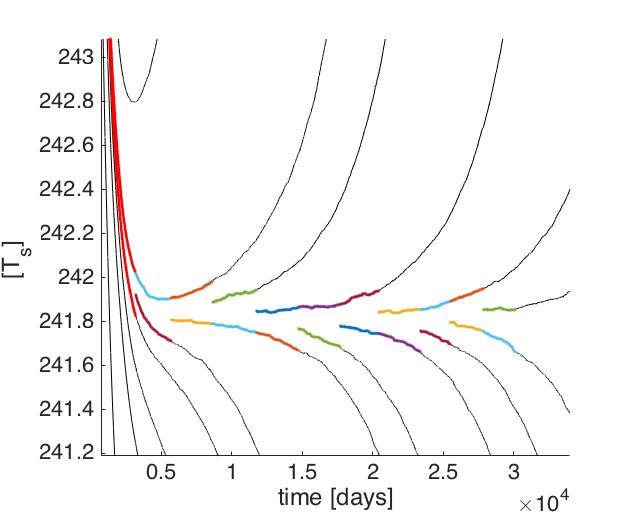}}  
  \sidesubfloat[]{\includegraphics[width=0.45\textwidth]{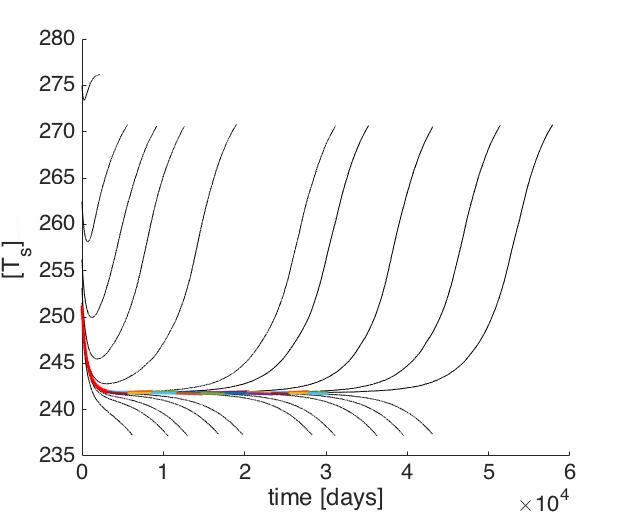}} \\ 
  \sidesubfloat[]{\includegraphics[width=0.45\textwidth]{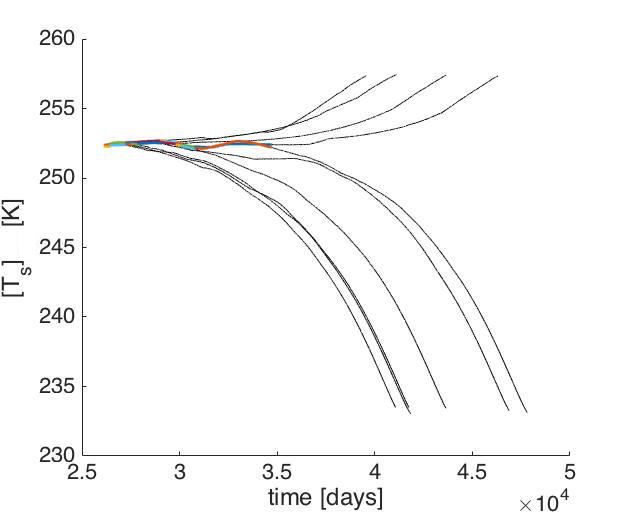}} 
  \sidesubfloat[]{\includegraphics[width=0.45\textwidth]{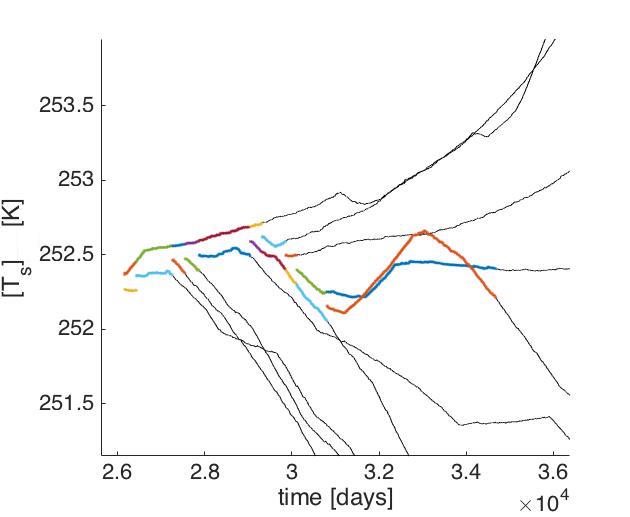}}
  \caption{How the edge tracking procedure works: bracketing and control trajectories in colour and black, respectively. We plot the time series of the globally averaged surface temperature $[T_s]$. a) $\mu=1.03$; b) Zoomed-in version of panel a); c) $\mu=0.9{\color{black}8}$; d) Zoomed-in version of panel c). Details in the text.\label{fig:bracketing}}
\end{figure}

Let's delve a bit into the details of how edge tracking works in two selected cases. Results are presented in Fig. \ref{fig:bracketing}, where we plot the time series of the globally averaged surface temperature $[T_s]$ for the bracketing and control trajectories. The reason why atmospheric variables are not included in the choice of the observable has to do with the fact that they have much larger fluctuations than the surface temperature, because of the smaller heat capacity of the atmosphere with respect to the surface. As discussed above, in order to construct an efficient algorithm along the lines of what depicted in Fig. \ref{edgetrackingscheme}, we want the short-term natural fluctuations of the observable to be much smaller than the difference of their climatological values in the two attractors.

After some testing, we have chosen $\varepsilon_1=0.1$ K and, additionally, we have selected $\varepsilon_2=2\varepsilon_1$, which implies that a single bisection is  taken per edge tracking $j-$cycle, as done in \cite{BLL2014}. Such a bracket size is much larger than the fast variability of $[T_s]$, so that we have little risk of being influenced by spurious trends. The control trajectories are continued up to a threshold value chosen to be close to the mean value for the stable warm or cold climate. 

In Fig. \ref{fig:bracketing}a) we show how the procedure works for $\mu=1.03$, with  \ref{fig:bracketing}b) providing a zoom: we are able to conclude that the procedure converges already in the $j=1$ cycle almost completely (bracketing trajectories shown in red), an edge state having $[T_s]\approx 241.8$ $K$ is identified, and we observe a (closely) monotonic (in fact, almost exponential) change of  $[T_s]$ as trajectories diverge from the edge state, pulled apart by the ice-albedo feedback.

Somewhat more interesting results are shown in Fig. \ref{fig:bracketing}c) and its zoomed-in version  \ref{fig:bracketing}d), where the edge tracking procedure is applied for  $\mu=0.98$. %, \textit{i.e.} pretty close to $\mu_{W/SB}\approx 0.97$, which defines  the tipping point at the low$-\mu$ range of the region of bistability. Comparing with Fig. \ref{fig:bracketing}a)-b) one observes the presence of slower and larger fluctuations. %: this low frequency variability can indeed associated to the fact that the edge state feel the presence of one attractor that is close to a critical transition, with the ensuing slow decay of correlations \cite{chekroun2014,Tantet2015a,Tantet2015b}. 
Figure \ref{fig:bracketing}d) clarifies that the time series of $[T_s]$ of trajectories going to different attractors can, in fact, cross, so that in general no trivial monotonic increase/decrease of the chosen indicator should be expected when a complex model is considered. In other terms, the cartoon provided in Fig. \ref{edgetrackingscheme} (and, \textit{a fortiori}, ultra-simplied points of view as expressed by Eq. \ref{0DEBM} and Fig. \ref{fig:double_well}) should be taken with a grain of salt. 

In the supplementary material, we include a movie showing the evolution of three orbits initialized near the Melancholia state realized for $\mu=1.0$ towards the W state, the SB state, plus one trajectory held on the edge state through the edge tracking algorithm described here. See the movie \texttt{three\_trajs\_atm\_temp.mp4} - URL: \texttt{https://youtu.be/mLYZiyzO8c4} (\texttt{three\_trajs\_surf\_temp.mp4} - URL: \texttt{https://youtu.be/OOYqUuG\_VUE}) for the evolution of the atmospheric (surface) temperature field.

%. The crosscorrelation of atmospheric variables belonging to the warm- and cold side trajectories is zero. As these uncorrelated signals perturb the equations for $T_{s,w}$ and $T_{s,c}$, the scalar time series of the global averages $[T_{s,w}]$ and $[T_{s,c}]$ may cross, even if the full fields do bracket the edge state. This does occur for $\mu=0.975$ (sample value/scenario \#2) seen in panel (c), which is a zoomed view of panel (a). When the atmospheric dynamics is less intense, like for scenario \#13 in panels (b),(d), the separation of $[T_{s,w}]$ and $[T_{s,c}]$ over time is more closely exponential.

%Using this procedure, we will reconstruct the Melancholia states for all range of parameters where multistability is found in the climate model. We can anticipate that the edge tracking method is successful and converges both in the cases where the edge state is a fixed point of a periodic orbit, and in the more complex case where the edge state lives on a strange set and features chaotic dynamics. As unexpected features we find a symmetry break, accompanying a bifurcation of the edge state from a fixed point to a periodic orbit, where a very slow wave is generated, such that the planet features also large longitudinal gradients of temperature and ice concentration, and a small parametric range where three steady states, and two edge states are found, thus going beyond the paradigm of bistability. 

\subsection{A Plethora of States beyond the Stable Climates}

\begin{figure}[ht] %[t!]
    \begin{center}
	%\scalebox{0.5}{\includegraphics{bif_ta_3}} 
	\includegraphics[width=0.85\textwidth]{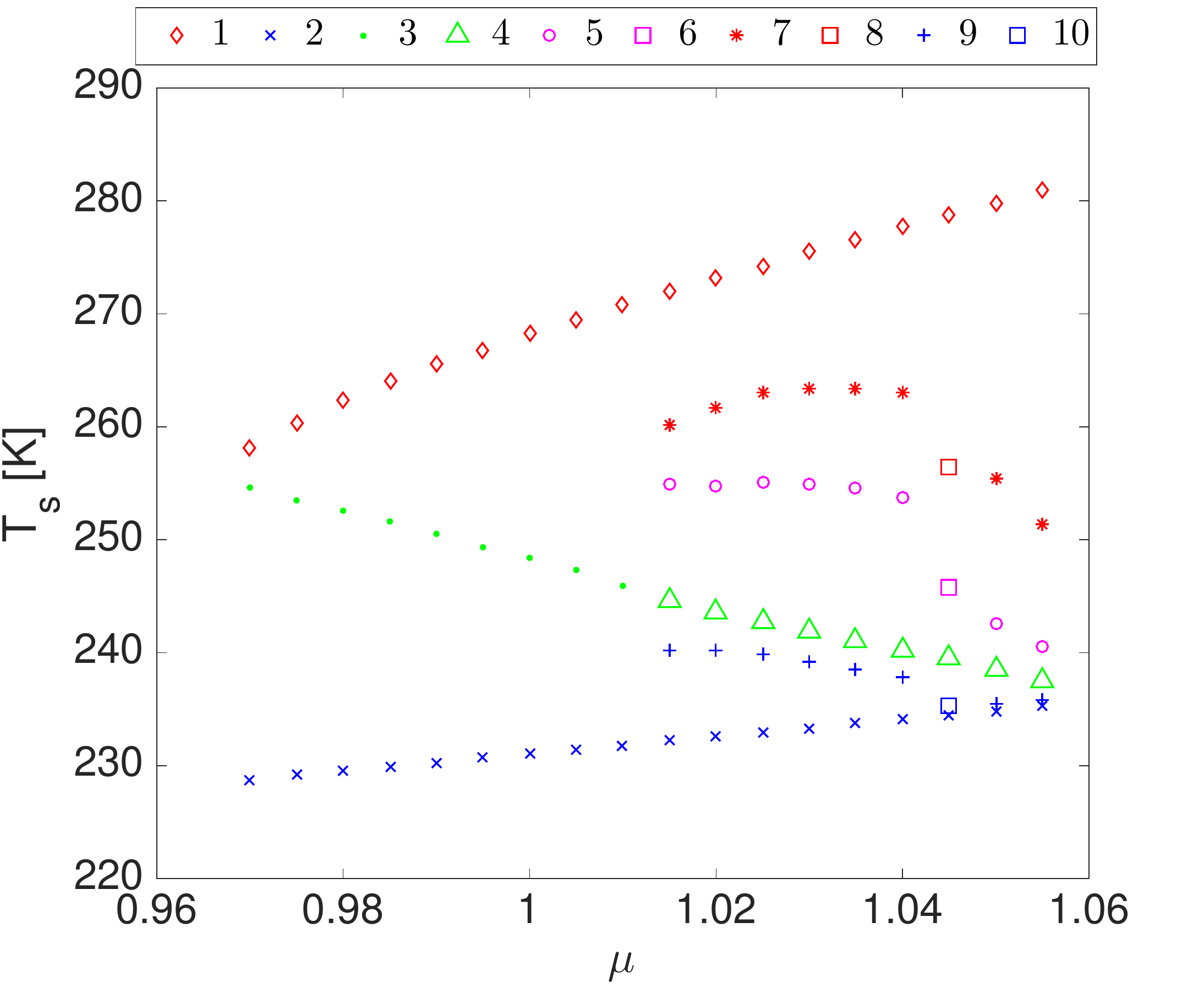}
	\captionsetup{singlelinecheck=off}
        \caption[my_option]{\label{fig:bif_tg_3} 
        Bifurcation diagram for the surface temperature. The markers in the legend are number-coded and correspond to qualitatively different states in the diagram. The markers $\#1$-$\#6$ refer to the globally averaged surface temperature $[T_s]$. For markers $\#7$-$\#10$ a different averaging for the temperature field is performed. See text for details.
%        \begin{enumerate}
%         \item W, $[\overline{T_s(t_1<t)}]_{x,y}$
%         \item C, $[\overline{T_s(t_1<t)}]_{x,y}$
%         \item E, $[\overline{T_s(t_1<t)}]_{x,y}$
%         \item P, $[\overline{T_s(t_1<t<t_2)}]_{x,y}$
%         \item E, $[\overline{T_s(t_3<t)}]_{x,y}$
%         \item M, $[\overline{T_s(t_3<t)}]_{x,y}$
%         \item E, $[\overline{T_s(y=y_w(t),t_3<t)}]_x$
%         \item M, $[\overline{T_s(y=y_w(t),t_3<t)}]_x$
%         \item E, $[\overline{T_s(y=y_m(t),t_3<t)}]_x$
%         \item M, $[\overline{T_s(y=y_m(t),t_3<t)}]_x$
%        \end{enumerate}
        }
    \end{center}
\end{figure}

In what follows we will show how much additional information one can draw from the climate model studied here beyond the usual bifurcation diagrams of the form shown in Fig. \ref{plasimsnowball}. We will be able to construct the Melancholia states, characterize their properties in terms of symmetry, variability, and degree of chaoticity, and prove that, near each tipping point, their properties converge to those of the stable climate that is in the process of losing stability. 

Figure \ref{fig:bif_tg_3} provides a first comprehensive summary of our results and should be compared with Fig. \ref{bodai14} discussed in \cite{BLL2014} and with Fig. \ref{edge}b) published in \cite{Schneider13022009}.   We portray the bifurcation diagram of our model, where $[T_s]$ for all the detected states is plotted vs the value of the control parameter $\mu$. A similar bifurcation diagram can be obtained by considering the globally averaged temperature of the lowest atmospheric layer $[T_a(\sigma=0.9)]$. The markers $\#1$ and $\#2$ indicated in the figure correspond to the W and SB attractors, and closely correspond to what is shown in Fig. \ref{plasimsnowball}. The marker $\#3$ corresponds to the Melancholia state sitting in-between the W and SB attractor and features decreasing values of $[T_s]$ with increasing values of $\mu$, similarly to what is shown in Fig. \ref{bodai14} for the simple model studied in \cite{BLL2014}. 

\begin{figure}[ht] %[t!]
    \begin{center}
            %\scalebox{0.5}{\includegraphics{bif_ta_3}} 
            \includegraphics[width=0.85\textwidth]{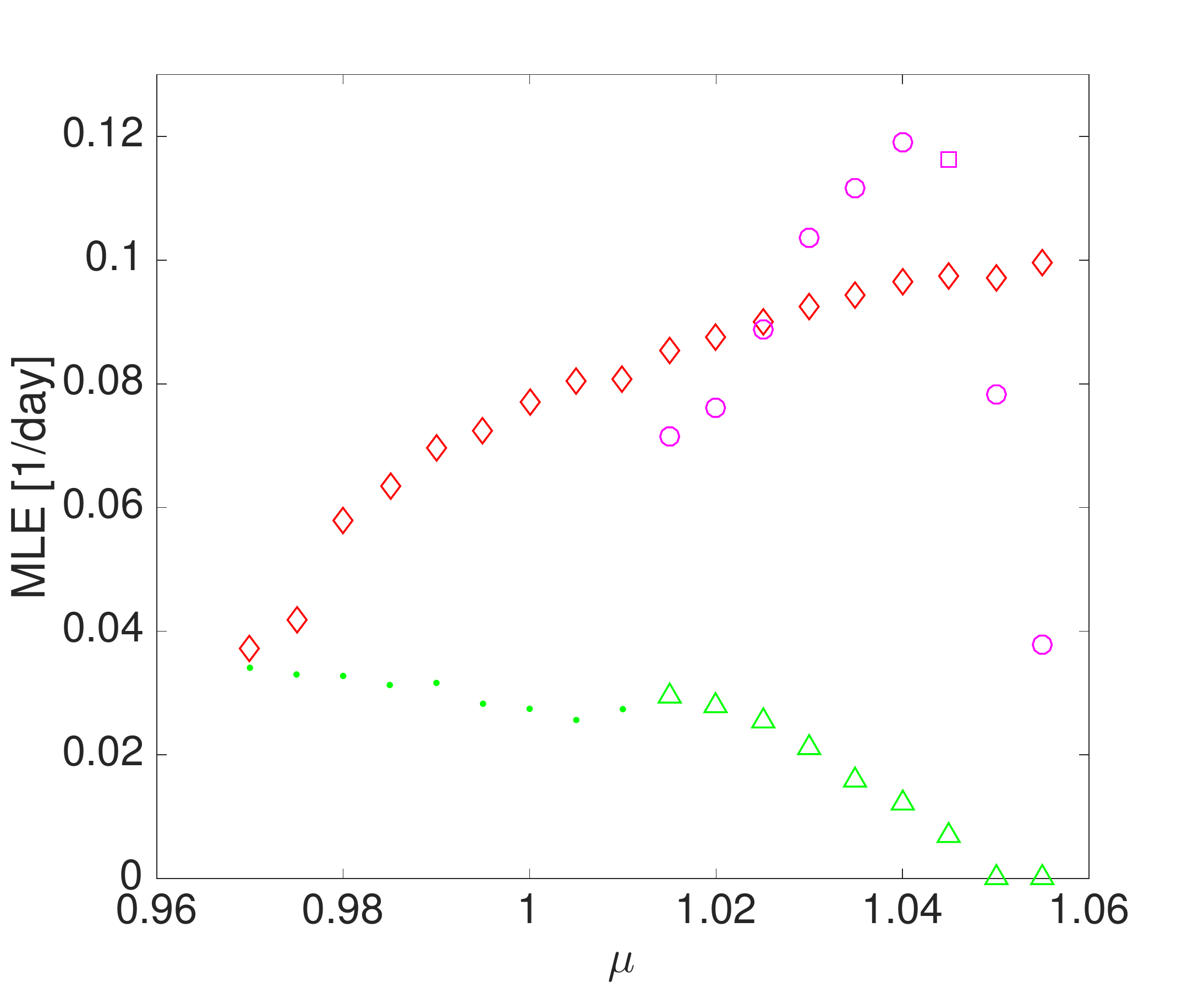}
        \caption{\label{fig:bif_MLE}
        Bifurcation diagram of the maximum Lyapunov exponents (MLE). Wheb the MLE is larger than zero we have a limited horizon of predictability  due to chaotic dynamics. Note that while the MLE can be rigorously defined for the states $\#1$, $\#3$, and $\#5$, a word of caution must be used in the case of states $\#4$. Same number coding as in Fig. \ref{fig:bif_tg_3}. Details in the text.}
    \end{center}
\end{figure}

Figure \ref{fig:bif_MLE} shows that  the state corresponding to the warm attractor - $\#1$ - features a positive MLE, \textit{i.e.}, it has a limited horizon of predictability, while the cold state - $\#2$ - is a fixed point, so that its MLE vanishes. {\color{black} What we obtain in terms of qualitative properties of the snowball state is partially unsatisfactory. Previous studies performed using comprehensive atmospheric models feature much weaker (yet present) atmospheric  variability in the snowball state than in the warm state; see e.g. \cite{Luchyst}. In our case the atmosphere is too \textit{quiet}, as a result of the very weak atmospheric meridional temperature gradient which is below (even if close to) the critical value needed for having sufficient baroclinicity supporting unstable motions. The main culprits for this behaviour we can indicate are the fact that a) the PUMA-GS has no continents and no orography, which makes it impossible to have geographically limited regions where higher baroclinicity can be present, thus leading to generation of cyclones (as preferentially happens in our planet in the westward portions of the Atlantic and Pacific storm tracks \cite{Holton}), and b) the simple ocean model we use is slightly too effective in transporting heat from the equator to the polar regions, thus reducing too much the meridional temperature gradient. Such - in our opinion, minor - limitation of our investigation could be overcome by repeating the analysis using the full PlaSim model with a  setup as in \cite{Luchyst}.} 

Most interestingly, we are able to \textit{identify chaoticity in  the Melancholia state (state $\#3$)}: we are here describing the properties of the dynamics restricted to the basin boundary between the W and SB attractors, so that the global instability due to the ice-albedo feedback is filtered out, and weather-related processes are instead retained. In other terms, we have that the Melancholia state features a dynamics of weather qualitatively analogous to what is found in the W state, including full life-cycles for mid-latitude disturbances and a non-trivial active Lorenz energy cycle. We present additional evidences of this in the movie included in the supplementary material.  See the evolution of the atmospheric temperature field in the edge state belonging to $\mu$=1.0 in the movie {\color{black}\texttt{three\_trajs\_atm\_temp.mp4} }- URL: \texttt{https://youtu.be/mLYZiyzO8c4}. 

\begin{figure}[ht] %[t!]
    \begin{center}
	%\scalebox{0.5}{\includegraphics{bif_ta_3}} 
	\includegraphics[width=0.85\textwidth]{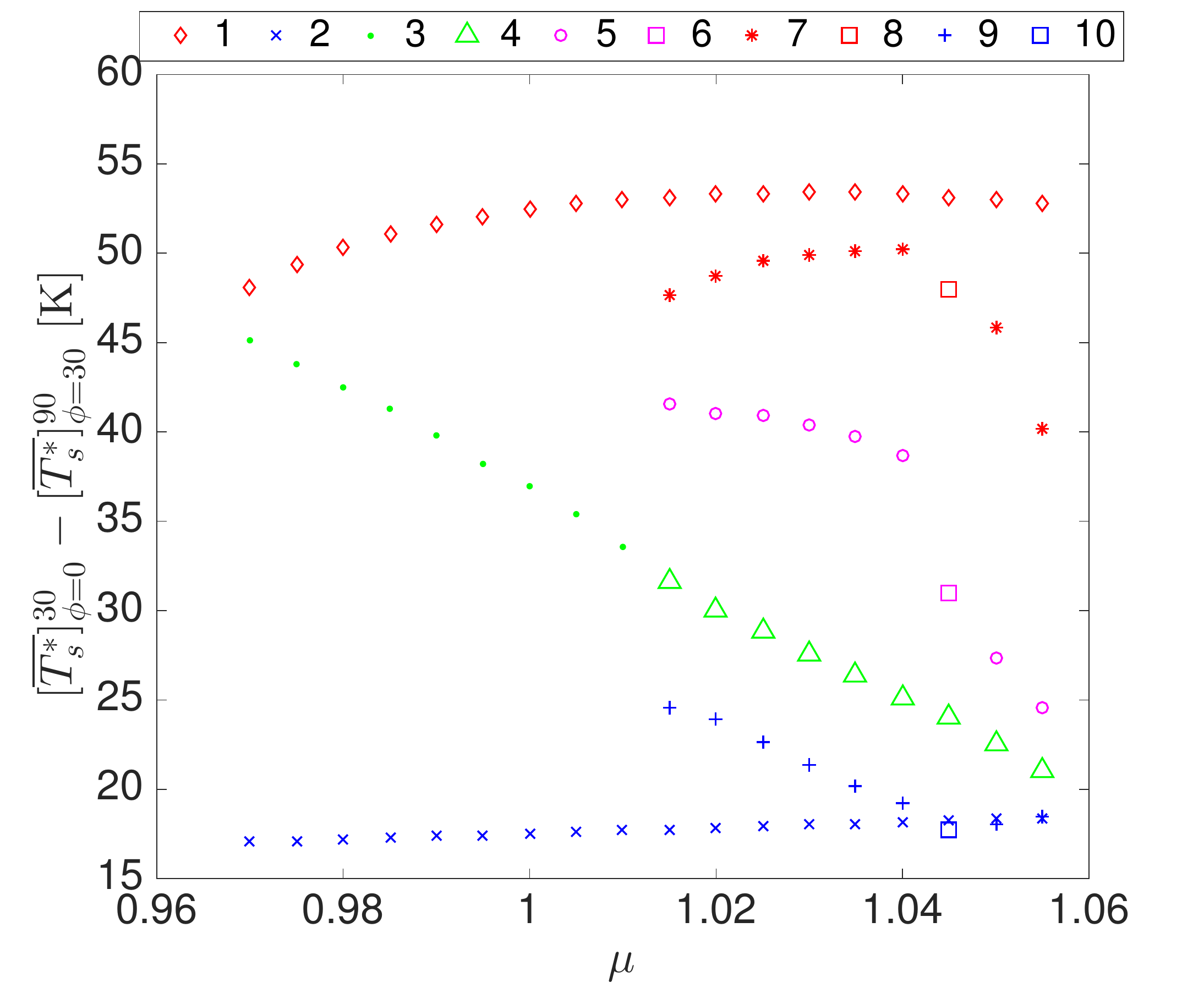}
	\captionsetup{singlelinecheck=off}
        \caption[my_option]{\label{fig:bif_tg_D2x_3} 
        Bifurcation diagram for a large scale meridional surface temperature gradient defined as the difference between low-latitude and high latitude `boxes', as done in~\cite{BLL2014}. Same number coding as in Fig. \ref{fig:bif_tg_3}. Details in the text.}
    \end{center}
\end{figure}

One finds - compare with Fig. \ref{fig:bif_tg_D2x_3} - that, broadly speaking (one cannot expect a one-to-one correspondence), the value of the MLE of states $\#1$, $\#2$, and $\#3$ is controlled by the intensity of the large scale meridional temperature gradient at surface, as suggested by the basics of baroclinic instability theory \cite{Holton}. We note that regions of strong meridional temperature gradient for the surface temperature correspond by and large to where the variability of the atmospheric temperature near surface is more pronounced, as a result of baroclinic disturbances (not shown).

Let's go back to Fig. \ref{fig:bif_tg_3}. As $\mu$ is increased between $1.010$ and $1.015$, a rather interesting phenomenon appears. The Melancholia state shown with marker $\#3$ goes through a symmetry breaking according to the following pattern. After an extremely long transient of the order of $100$ years or more (marker $\#4$), a large longitudinal modulation of the atmospheric and surface fields suddenly appears (see in the supplementary material the movies \texttt{edge\_symm\_break\_atm\_temp.mp4} - URL: \texttt{https://youtu.be/u4--tRnBBS8} and \texttt{edge\_symm\_break\_surf\_temp.mp4} - URL: \texttt{https://youtu.be/Q4YAbU9O15U}  {\color{black}corresponding to $\mu=1.025$}), so that about half of the planet warms up (marker $\#7$, showing the meridional average at the \textit{warmest longitude}, following the revolution of the field) and is in conditions relatively similar to those of the co-existing W state, while  the rest of the planet cools down and reaches conditions relatively similar to those of the co-existing SB state (marker $\#9$, showing the meridional average at the \textit{coldest longitude}). The longitudinal modulation impacts all latitudes and rotates with a constant and extremely low angular velocity, and one can define the average properties of the the Melancholia state (marker $\#5$) without the need for considering ultralong time scales (order of $10^3$ years). In order to define the properties of the system in the states $\#7$ and $\#9$ one needs to compute averages in a reference frame co-rotating with the temperature wave. {\color{black} Note that even the direction of rotation is not the same for all values of $\mu$.}

In the supplementary material (see the movie \texttt{edge\_symm\_break\_atm\_temp.mp4} - URL: \texttt{https://youtu.be/u4--tRnBBS8}) one can also see that the warm portion of the domain features a large temperature gradient and supports the growth of baroclinic disturances, which tend to decay as they reach the cold region, where the low temperature gradient does not provide suffiicient baroclinic forcing. In a loose sense, the Melancholia states realized as symmetry breaking of the transient state $\#4$ resemble some sort of chimera states \cite{panaggio}, yet  unstable ones.

An additional aspect of our results should be highlighted. For $\mu=1.045$ we have a fundamentally different phase portrait for our model, which seems to be specific to a small neighborhood of this value of $\mu$. We find that the system features three stable climates, which include, apart from the W and SB climate described elsewhere, the result of the bifurcation of the Melancholia state described by the markers $\#5$, $\#7$, and $\#9$ into a stable state (see markers $\#6$, $\#8$, and $\#10$, respectively), which displays similar features in terms of the presence of a semi-stationary pattern of strong longitudinal gradients of temperature. {\color{black}As far as we know, we are unaware of any other study reporting the existence of such an exotic climate state.}

One is naturally bound to expect the presence of additional Melancholia states sitting in-between the three stable climates found for this value of $\mu=1.045$. We do not pursue this investigation, because it would lead us to what we see at this stage as unnecessary complication. It is important to note that it would have been extremely unlikely to find such an additional \textit{attracting climatic steady state}, had we not used the edge tracking algorithm, because the width of parametric window where it exists is small, and rather limited is also the size of its basin of attraction.  

The presence of a large time scale separation between the rotation and all the other dynamical processes allows for constructing the statistics embodied in the markers $\#7$ and $\#9$, which are computed following the slow rotating wave, in order to guarantee homogeneity in the results.  %Similarly, the presence of an extremely long transient allows for defining  the properties of the transient state described by marker $\#4$. We note that the properties of the transient Melancholia state  $\#4$  join on regularly with the actual Melancholia state portrayed with the marker {\color{black}$\#3$}. 

Additionally, for  $\mu\approx\mu_{W\rightarrow SB}$, the properties of the Melancholia state get very close to those of the W state, in agreement with the fact that at the tipping point the edge state and the W attractor intersect, leading to the loss of stability. For $\mu\approx\mu_{SB\rightarrow W}$, we have that the properties of the Melancholia state (marker $\#5$), transient Melancholia state (marker $\#4$), and SB state tend all to converge, as the tipping point is reached.

By considering the argument of time scale separation, it is also possible to construct a (pseudo-)MLE for state $\#4$ (we cannot obviously take an infinite time horizon to compute it, because of the symmetry breaking process in action), whose value joins on with continuity with what is found for the Melancholia state denoted by marker $\#3$, as shown in Fig. \ref{fig:bif_MLE}. It is also possible to compute the MLE for state $\#5$. %, even if some lengthy calculations are needed give the extreme heterogeneity of the attractor. 
One finds that the values of the (pseudo-)MLE obtained for state $\#4$ follow closely those of the actual Melancholia state $\#3$ (and obeys the previously discussed relationship between value of the MLE and the strength of the meridional temperature gradient), while the MLE values  obtained for state $\#5$ are much larger (even larger than those realized in state $\#1$), despite the presence of a weaker average meridional temperature gradient; see Fig. \ref{fig:bif_tg_D2x_3}. This can be explained by considering that in state $\#5$, in addition to the presence of a meridional temperature gradient, the system experiences  a large longitudinal temperature gradient too, which also supports the growth of instabilities, separating the warm and cold portions of the planet (states $\#7$ and $\#9$, respectively). The strongly nonlinear dynamics of cyclones when transiting from the relatively cold to the relatively warm regions (see in the supplementary material the movie \texttt{edge\_symm\_break\_atm\_temp.mp4} - URL: \texttt{https://youtu.be/u4--tRnBBS8}) is a manifestation of this effect.

\begin{figure}[ht]
  \sidesubfloat[]{\includegraphics[width=0.44\textwidth]{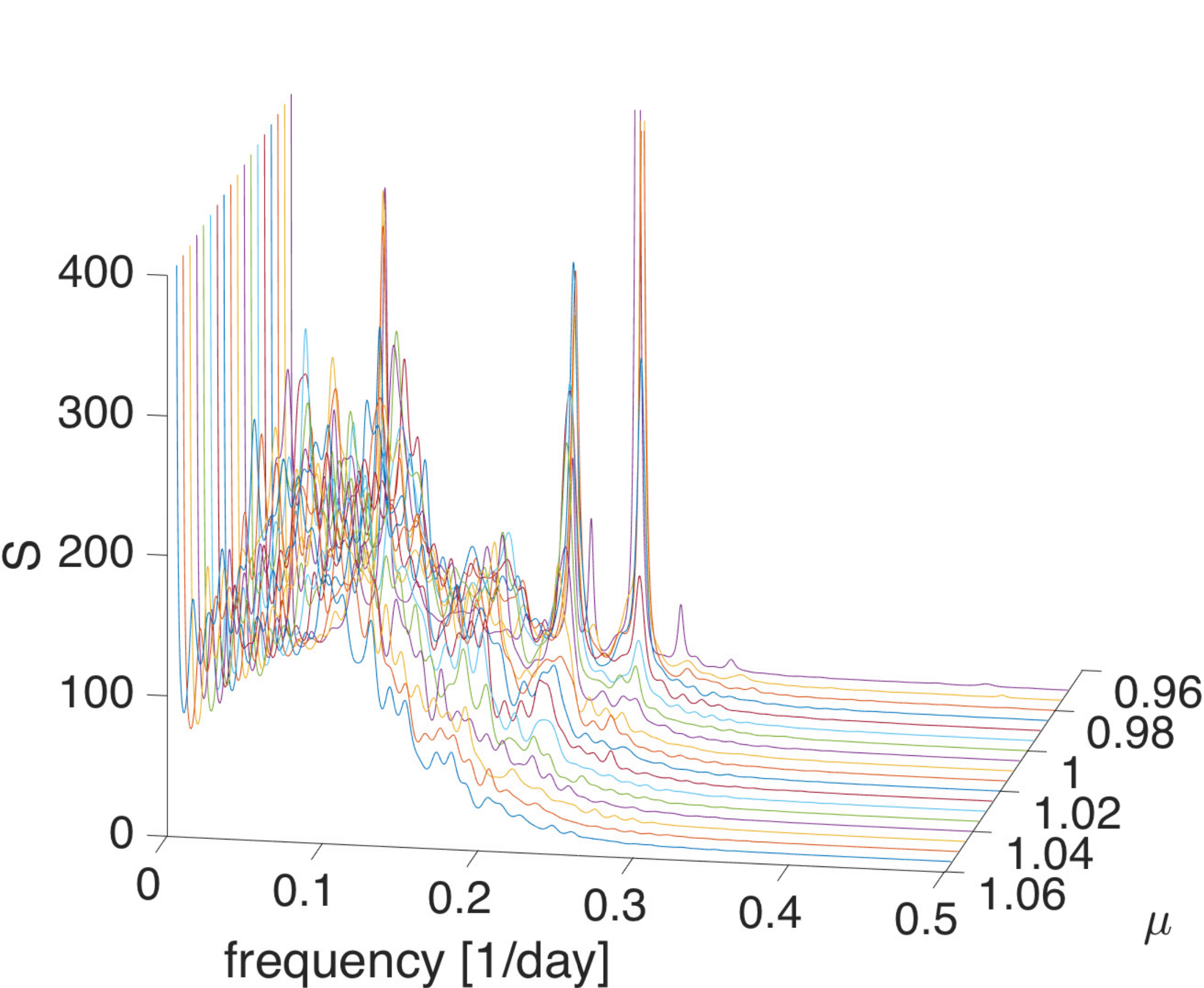}\label{fig:pow_spectra_sub1}}\ %quad
  \sidesubfloat[]{\includegraphics[width=0.44\textwidth]{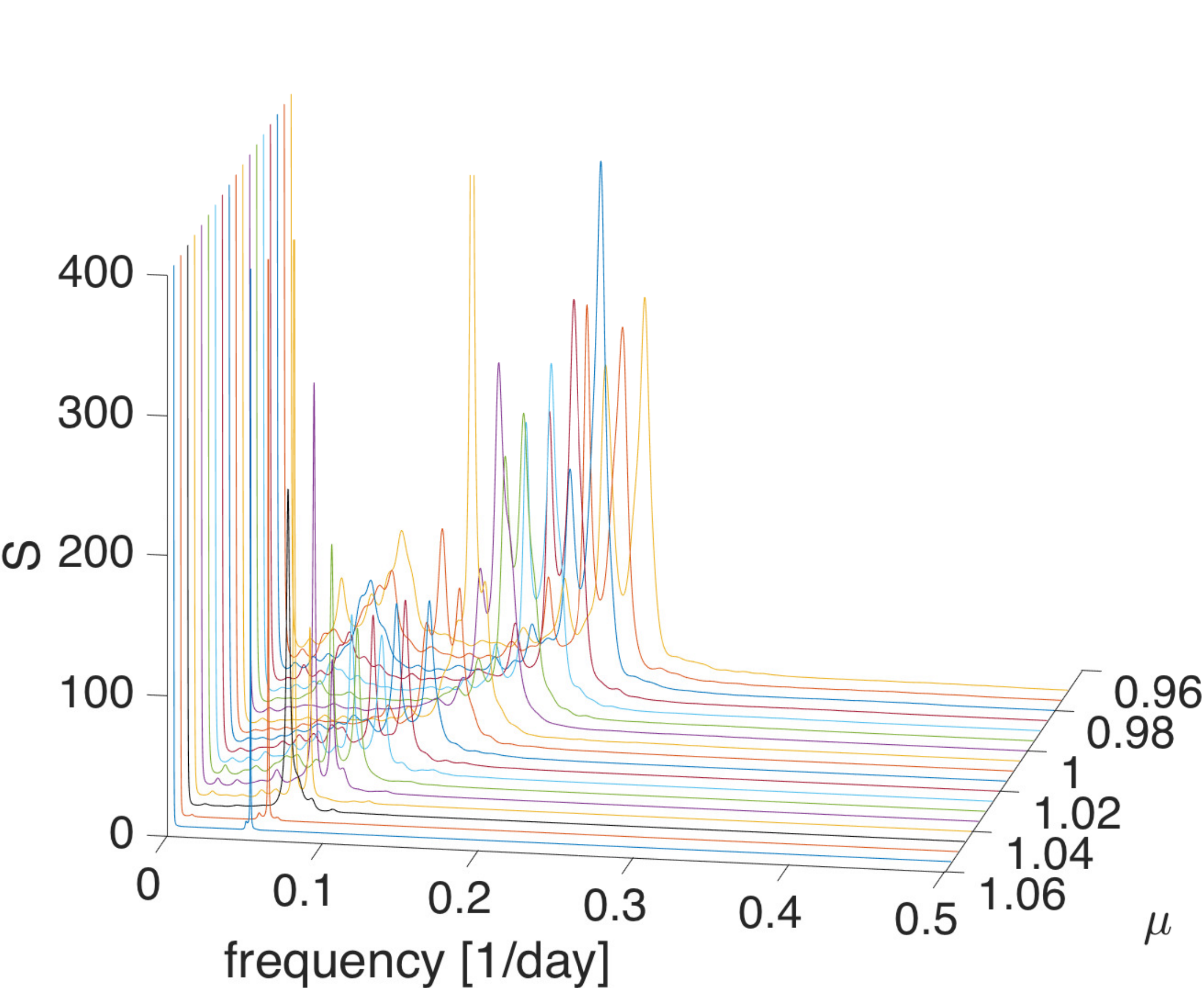}\label{fig:pow_spectra_sub2}} \\
  \sidesubfloat[]{\includegraphics[width=0.44\textwidth]{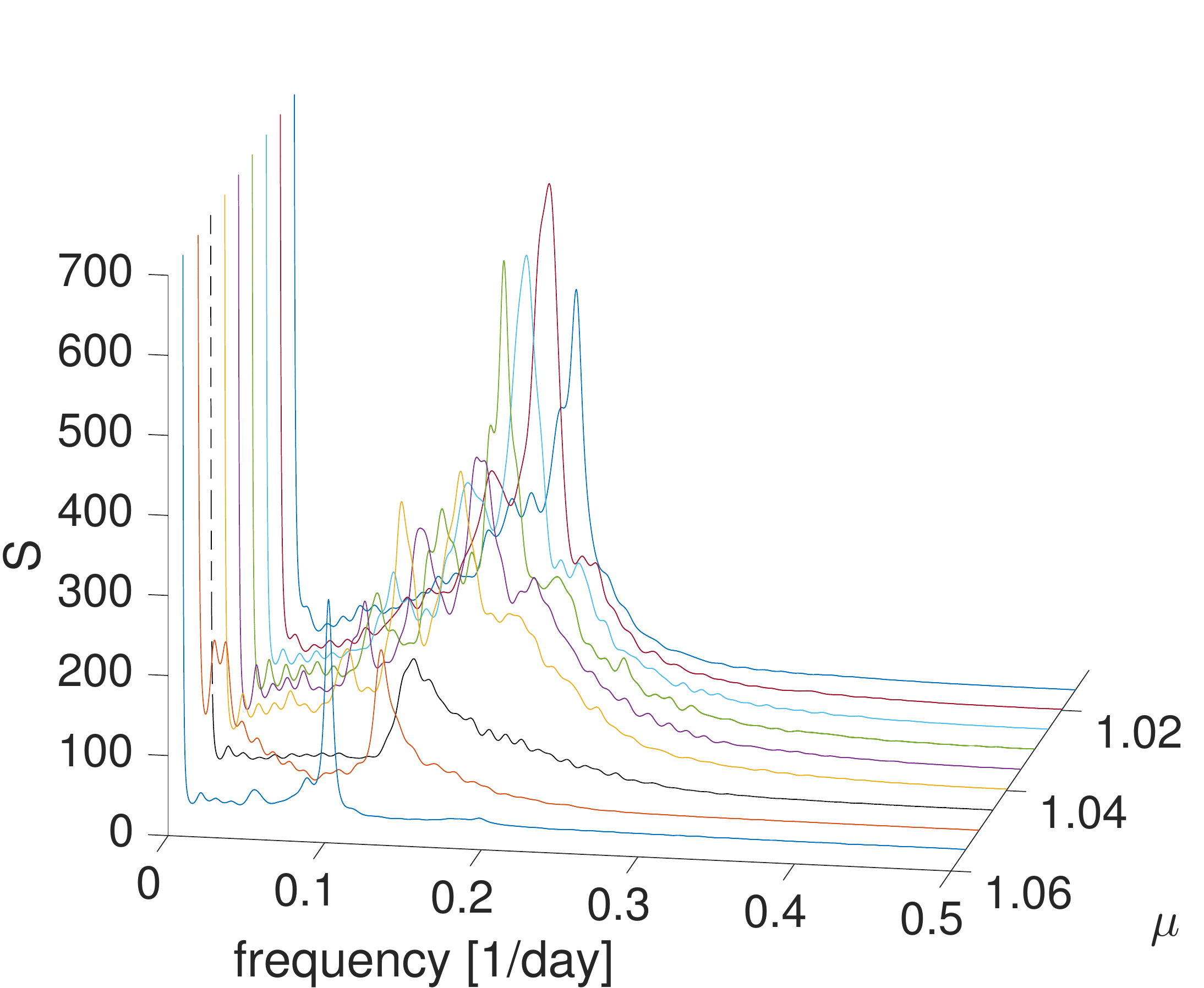}\label{fig:pow_spectra_sub3}}\ %quad
  \sidesubfloat[]{\includegraphics[width=0.44\textwidth]{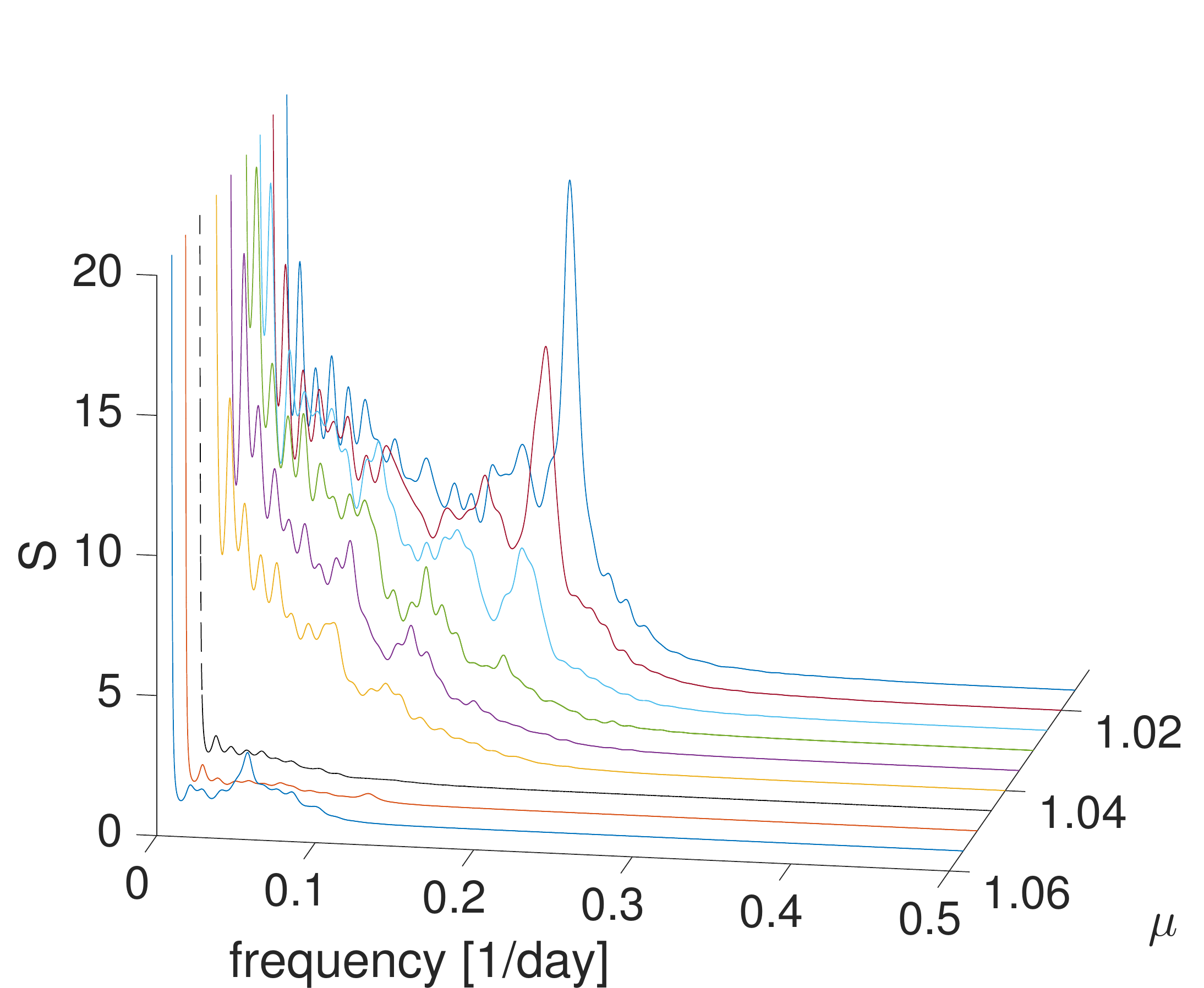}\label{fig:pow_spectra_sub4}} 
  \caption{Power spectra of $T_a(\sigma=0.9)$ for a grid point situated at $\approx30^{\circ}$ $N$ latitude.The panels describe the properties of states coded by numbered markers in Fig. \ref{fig:bif_tg_3}; \ref{fig:pow_spectra_sub1}: state $\#1$; \ref{fig:pow_spectra_sub2}: states $\#3$, and $\#4$; \ref{fig:pow_spectra_sub3}: state $\#7$ and state $\#8$ (black line); \ref{fig:pow_spectra_sub4}: state $\#9$ and state $\#10$ (black line). Note the different ranges of the total power $S$ measured on the vertical axis. Details in the text.\label{powerspe}}%  Note that for better visibility the direction of the axis of $\mu$ in panels \ref{fig:pow_spectra_sub1} and \ref{fig:pow_spectra_sub2} is the reverse of those in \ref{fig:pow_spectra_sub3} and  \ref{fig:pow_spectra_sub4}. We can observe a decreasing of characteristic frequencies with increasing $\mu$; this can be observed also in the video ***.mp4. }\label{fig:pow_spectra}
\end{figure}

A further characterization of the dynamics of the model can be obtained by looking at the power spectrum of an atmospheric variable, keeping  in mind that a broadband spectrum is a signature of chaotic dynamics. The power spectra for the  $T_a(\sigma=0.9)$ in a grid point situated at $30^\circ$ $N$ are presented in Fig. \ref{powerspe}. We have that, in agreement with what is reported in Fig. \ref{fig:bif_MLE}, the states corresponding to the W attractor feature a broadband spectrum for all values of $\mu$, while no variability is present for the SB states (not shown). 

As for states $\#3$ and $\#4$, we have that for increasing $\mu$ the total intensity $S$ and the breadth of the spectrum decreases as well as featuring an overall shift to lower frequencies, as a result of the decrease of the meridional temperature gradient (see Fig. \ref{fig:bif_tg_D2x_3}), and, by thermal wind relation \cite{Holton}, of the intensity of the jet, which defines the time scales of the mid-latitude. For very large values of $\mu$, the system reaches a quasi-periodic behaviour, where one or few frequencies are present. 

When constructing the power spectra of states $\#7$-$\#10$ (note that no information on these states could be obtained by looking at the MLE), we make sure to discard the effect of the longitudinally slowly moving temperature  wave. The power spectra of the cold regions of the longitudinally-modulated climate, corresponding to states $\#9$ and $\#10$, are extremely weak, corresponding to the fact that the variability is only inherited from the active dynamics taking place in the warm regions corresponding to the states $\#7$ and $\#8$. The variability is intense with broad spectrum for states $\#7$ when $\mu<1.04$, in agreement with the fact that a large horizontal temperature gradient is found across the mid-latitudes.

{\color{black}\subsection{The Geometry of the Basin Boundary}\label{boundarygeometry}
A very nontrivial aspect of the problem we are studying is the characterization of the geometrical properties of the boundary between the basins of attraction of the stable climates. The cartoons shown in Figs. \ref{fig:edge_tr_init}-\ref{edgetrackingscheme} suggest that the two basins of attraction are separated by a simple manifold of co-dimension one. Clearly, the existence of two separate \textit{true} attractors (as opposed to the case of transient turbulence studied by Eckhardt and collaborators) implies that the co-dimension of the boundary can be at most one, because otherwise one would automatically have holes allowing for transitions between the two basins of attraction, in contrast with their property of being invariant. Nonetheless, nothing prevents in principle the boundary from being a more complex geometrical object of co-dimension smaller than one, such that the intersection between such a boundary and a line connecting the two competing attractors has  a non-trivial Cantor-like structure. 

\begin{figure}[ht]
    \begin{center}
	%\scalebox{0.5}{\includegraphics{bif_ta_3}} 
	a)\includegraphics[angle=0,width=0.45\textwidth]{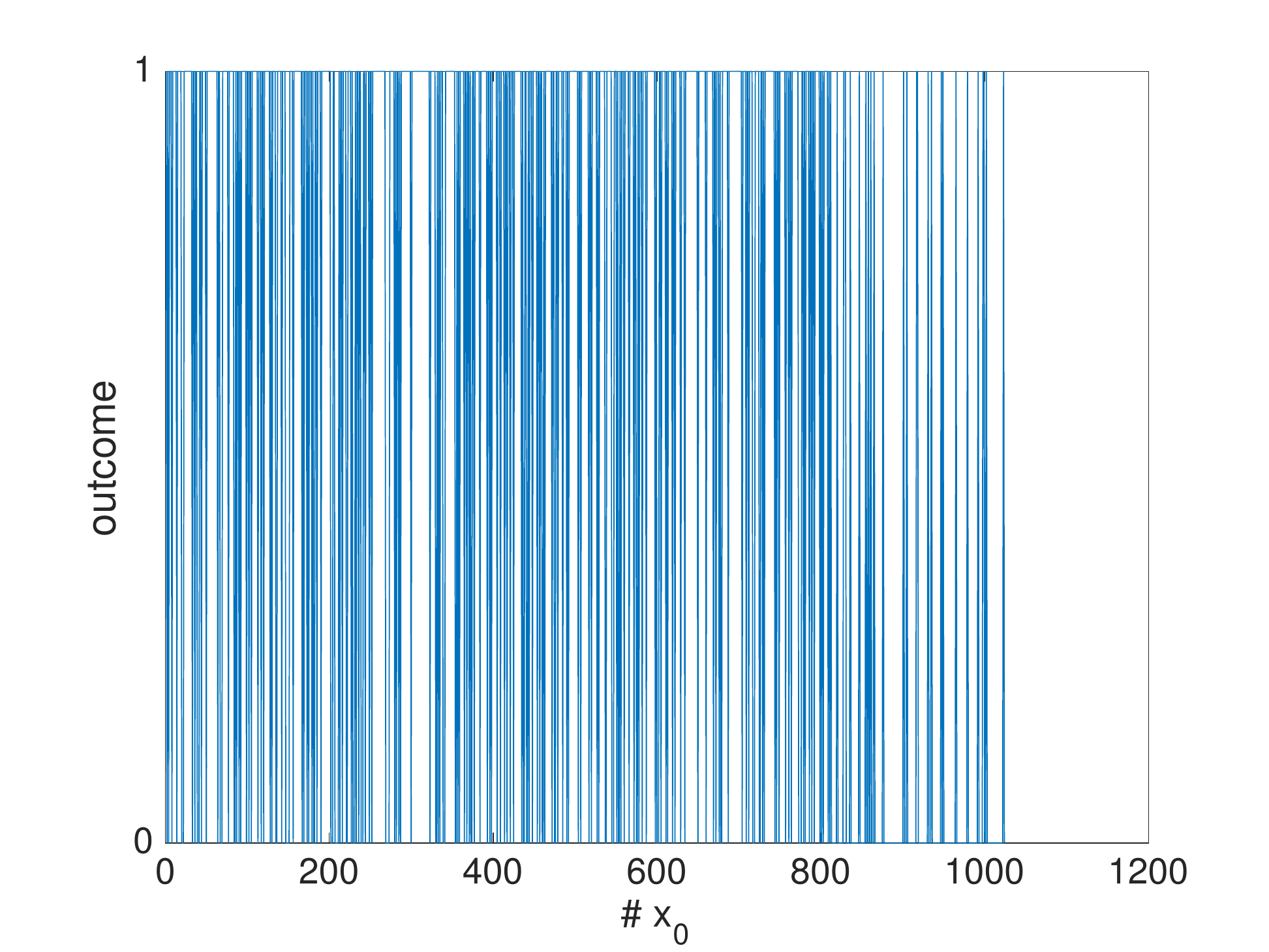}
	b)\includegraphics[angle=0,width=0.45\textwidth]{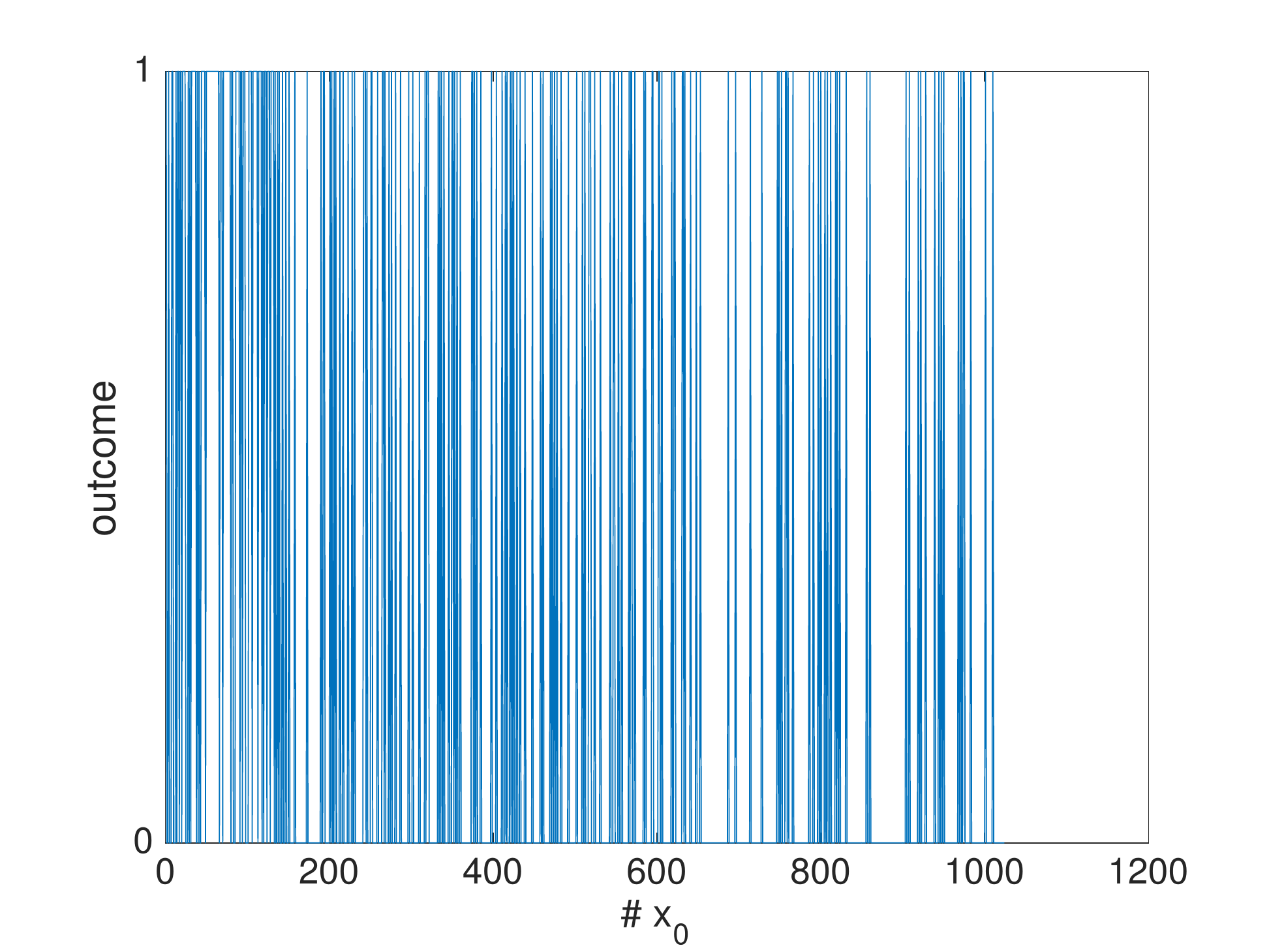}
	        \caption{a) Outcome of the the forward integration of 1024 very similar initial conditions,  populating a straight line in phase space that traverses the basin boundary. The difference between the globally averaged surface temperature of the 1$^{st}$ and of the $1024^{th}$ is 0.5 $K$. Outcome 1 indicates that the trajectory ends in the warm state, and outcome 0 indicates that the trajectory ends in the snowball state. b) Same as a), but for a different set of initial conditions. Details in the text. 
	        \label{boundaryoutcomes} }
    \end{center}
\end{figure}

This issue has been studied in simple mathematical models in, e.g.,  \cite{Grebogi83,Kaplan84,McDonald85,Grebogi87,Vollmer09}, and has practical relevance as the presence of a fractal boundary makes the prediction of the final state given the knowledge of the initial conditions of the system (with finite precision)  a very non-trivial task. In other terms, it controls  what Lorenz called the \textit{predictability of the second kind} \cite{Lor67,Peixoto:1992}. 

As discussed in, e.g., \cite{Tel2006}, the presence of a chaotic edge state is a necessary (but not sufficient) condition for the existence of a geometrically complex basin boundary. In fact, the geometrical complexity of the boundary requires that that the MLE characterizing the separation of trajectories inside the basin boundary is larger than the Lyapunov exponent describing the separation of trajectories in the unstable direction (and leading to the orbits ending up in either attractor) \cite{Grebogi83,McDonald85,Grebogi87,Vollmer09}.

Following a suggestion by A. Pikovsky during a public presentation of some preliminary results then included in the present paper, we have decided to investigate the geometrical properties of the basin boundary in which the Melancholia states are embedded. We have considered the case $\mu=0.98$, for which we have already found a chaotic Melancholia state, and have considered two nearby initial conditions belonging, respectively, to the basin of attraction of the warm and of the snowball state. These two initial conditions are represented by the first bracket seen in Fig. \ref{fig:bracketing}d). The two initial conditions are very similar in terms of dynamical and thermodynamical properties and differ by only $0.5$ $K$ in terms of globally averaged surface temperature. We have then considered additional 1022 initial conditions obtained by equispaced convex linear interpolations of the two original initial conditions, which results in the fact that their corresponding globally averaged surface temperatures are ordered and equispaced, so that the globally averaged surface temperature of two neighbouring initial conditions differs by about $5\times10^{-4}$ $K$.

We have then integrated forward each of these initial conditions and have labelled with 1 the orbits ending in the warm state and 0 the orbits ending in the snowball state. The results are shown in Fig. \ref{boundaryoutcomes}a). Against intuition, one finds an extremely complex geometric structure across the boundary. Very often, nearby initial conditions belong to different basins of attraction. Similar results are obtained by repeating the investigation in another region of the phase space: see Fig. \ref{boundaryoutcomes}b), where we follow the procedure detailed above starting from the fifth bracket in Fig. \ref{fig:bracketing}d). Clearly, at finite precision one cannot in principle exclude the possibility that the basin boundary is a regular - yet very tightly folded - manifold, rather than being a strange geometrical object. Further analysis is shown in Fig. \ref{boundarydimension}, where we have computed the information dimension of the  intersection set between the segment including the 1024 initial conditions above and the basin boundary  using the data shown in Fig.\ref{boundaryoutcomes}a). Entirely compatible results are obtained using, instead, the data in Fig.\ref{boundaryoutcomes}b). We find an information dimension of about 0.98, which indicates that - at least within a certain region - the boundary has almost full dimension.

\begin{figure}[ht]
    \begin{center}
	%\scalebox{0.5}{\includegraphics{bif_ta_3}} 
	\includegraphics[angle=0,width=0.7\textwidth]{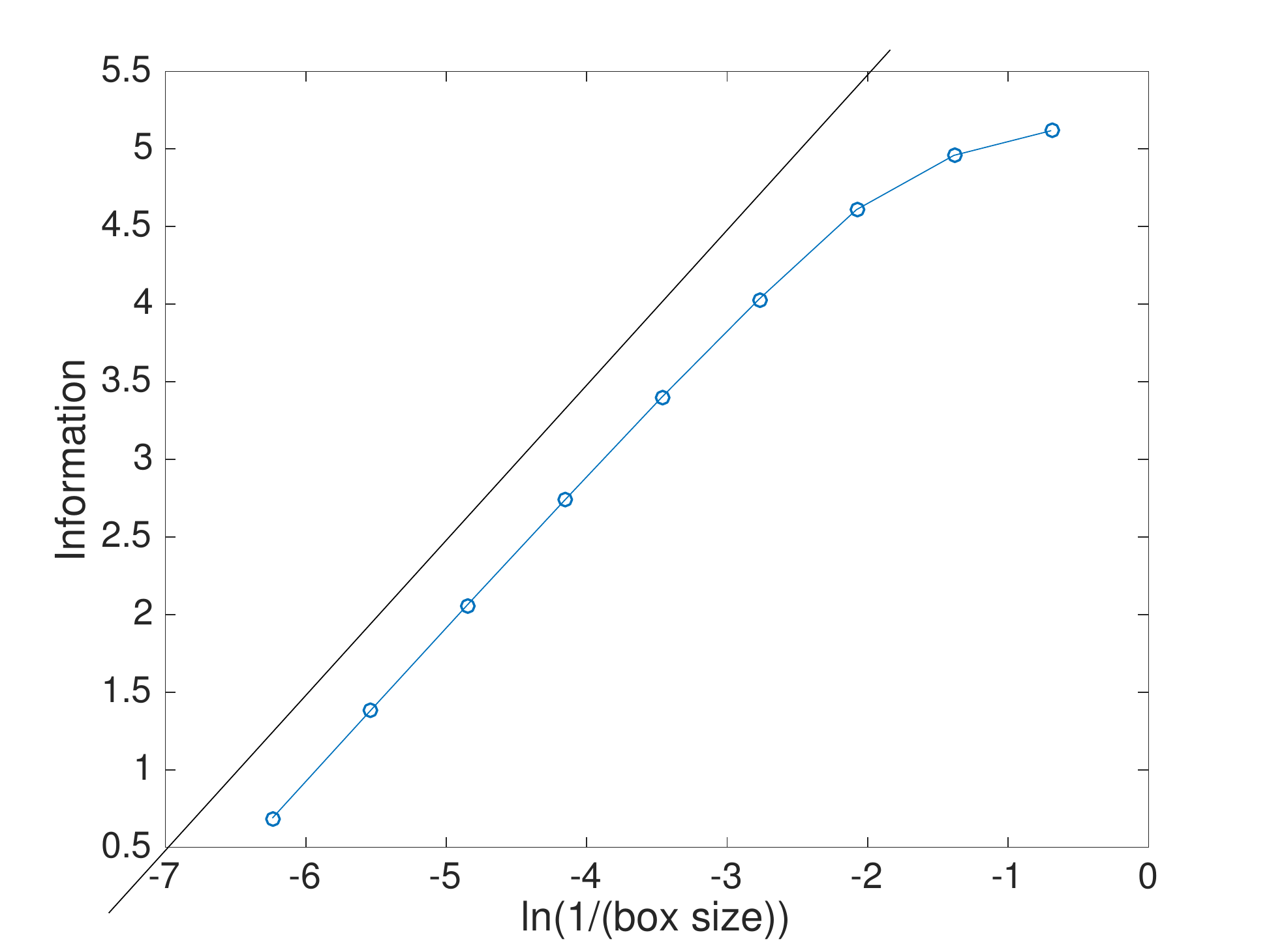}
	        \caption{Computation of the information dimension along the transversal direction of the basin boundary. The slope of the straight line is about 0.98.\label{boundarydimension} }
    \end{center}
\end{figure}

As a result, our ability to predict the final state of the system (warm or snowball) given the initial conditions (which can be measured using the uncertainty exponent \cite{Ott2002}) is extremely low. Going back to the mathematical conditions discussed in \cite{Grebogi83,McDonald85,Grebogi87,Vollmer09}, we must have that the instability inside the basin boundary is (much) stronger than the instability in the unstable direction. And here physical intuition comes to our help: the instability occurring inside the basin boundary is associated to weather processes, and corresponds to the baroclinic processes (having characteristic time scales of tens of days) responsible for the generation of eddies in the Melancholia state, see Fig. \ref{fig:bif_MLE}. Instead, the instability in the transversal direction results from the ice-albedo feedback, and acts on time scales of years \cite{Ghil,BLL2014}. Concluding, the strangeness of the basin boundary comes from the large time scale separation between weather and climatic unstable processes.

A natural question one may ask is why, despite such a geometrical complexity, the edge tracking algorithm still works well. The reason is that the bisection algorithm removes the motions in the unstable direction associated to the ice-albedo feedback. Therefore, the constructed pseudo-trajectory along the basin boundary is constrained to explore the saddle set with no regard to its geometry and to be eventually attracted to the edge state.}%  trajectory jump between and across different foliations of the basin boundary along on the transversal direction, but this causes no problems because of the attractiveness of the edge state.}
 
\section{Summary and Conclusions}\label{concl}
In this paper we have attempted to provide a review of some ideas concerning the study of critical transitions in the context of climate science. We have first briefly recapitulated under which conditions one can expect a regular response of the climate system to forcings, and have clarified that the response theory introduced by Ruelle gives the mathematical framework for addressing comprehensively the problem of predicting how the statistical properties of the climate system change as a result of (possibly time-dependent) perturbations to its dynamics. We have then clarified that the lack of regularity in the response of the system to perturbations, associated to a slowing-down in the rate at which the system decorrelates, flags being near to conditions where critical transitions take place.

In order to have a more complete picture of climate dynamics, one needs to take a global point of view in the study of the phase space of the climate system and consider the possibility of the existence of multiple steady states. We have then  briefly discussed the literature concerning the existence of multistability in the Earth system, with a special focus on the key scientific issue regarding the co-existence of snowball and warm climate conditions given the current astrophysical and astronomical conditions. In this regard, we have presented a hierarchy of conceptual to more realistic mathematical models commonly used for studying multistability in a geophysical (and not only) context and have underlined virtues and limitations of using a stochastic dynamical point of view. 

While typically one wants to study the properties of the co-existing climate states and of, \textit{e.g.}, the noise-induced probability of jumping from one to the other, the question we find most interesting and indeed novel regarding the specific climatic problem we focus on is to define what - dynamically - lies in-between the multiple (say, two) co-existing steady states of the climate system. In other terms: is there a special dynamical configuration that generalizes the saddle point in an energy landscape and acts as gate facilitating the transitions between the two attractors?

The edge state introduced by Eckhardt and collaborators for studying turbulent fluid systems has {\color{black}proven} rather useful for our study. The edge state is the attracting set embedded in and relative to the  boundary between the two basins of attraction of the co-existing attractors. The edge state attracts, in fact, almost all the initial conditions belonging to the boundary of the basins of attraction. The edge state cannot be found by direct numerical integration, but can be constructed through a control algorithm - the edge tracking algorithm - that allows for bracketing nearby trajectories evolving on either side of the basins boundary.

In recent years we have followed a scientific programme aiming at a comprehensive view on climate response that has led us in the last years to study - using the same climate model - a) the macroscopic thermodynamical properties of the climate system associated to the critical transitions responsible for the snowball/warm bistability; b) the smooth climate response to perturbations using the Ruelle response theory; and c) the breakdown of the smooth response near the critical transitions due to slow decay of correlations using a the transfer operator method in a reduced space. 

In this paper we have aimed at completing the puzzle by taking a global point of view with the analysis of dynamical regimes that do not correspond to attracting steady states of the climate, but rather constitute the edge states sitting in-between co-existing climate states. We refer to these states as Melancholia states. We have decided to stick to the point of view of statistical mechanics that proposes to investigate complex systems using a purely deterministic dynamics, with the hope of being able to understand more deeply the effect of adding noise. 

By adapting the edge tracking algorithm to a climate model that is somewhat simpler than what was used in previous investigations (yet Earth-like), we have been able to identify a plethora of climate states and study the properties of the Melancholia states existing in the region of bistability where snowball and warm climate attractors coexist. 

Our control parameter is $\mu$, the ratio between the considered and the present-day solar {\color{black}constant}. The tipping points are in fact reached when one of the climate attractors collides with the Melancholia state. We provide ample diagnostics for characterizing such states, including some movies that are available as supplementary material. The Melancholia states are obviously unstable with respect to the ice-albedo feedback, and, once such an instability is kept under control, they exhibit in some parametric range the kind of variability usually associated to regular climate conditions. The Melancholia states feature, \textit{e.g.} mid-latitude cyclones growing as a result of baroclinic instability fuelled by the meridional temperature gradient, and have a chaotic behaviour associated to a limited horizon of predictability. 

As $\mu$ is increased from the value corresponding to the tipping point from the warm to the snowball state, one finds that the Melancholia states undergo a symmetry breaking bifurcation, after which they are characterized by strongly zonally non-symmetric conditions, in the form of a slowly rotating temperature anomaly wave. This regime is a sort of mix of the co-existing warm and snowball attractors, and feature a complex dynamics at the interface between the cold and warm regions. In a small window of values of $\mu$, one finds three stable climate attractors, but this regime is not further explored in this paper. After such a window, the asymmetric Melancholia state is recovered and converges to the cold attractor at the tipping point associated to the transition from the snowball to the warm state. 

{\color{black} The Melancholia states discussed in this paper correspond to climatic regimes that \textit{cannot} be realized in nature nor can be obtained by a forward integration of a numerical model.  We would like to suggest two arguments supporting the idea that the Melancholia states have, nonetheless, great physical relevance. 

First, they can be used to flag  \textit{potentially dangerous} forcings acting on systems originally living in one of the coexisting attractors. Obviously, forced transitions between the two basins of attraction can take place also bypassing the Melancholia state, but, surely, if the forced system resembles more and more a Melancholia state, the risk of catastrophic changes taking place if the forcing is not stopped is very real. Clearly, the risk is much higher when the unperturbed system lives in an attractor close to the Melancholia state - so in a parametric range near the critical transition.  

Second, as we have discussed in the latter part of our paper, Melancholia states might morph into actual stable yet very nontrivial stable climate states for small changes in the value of the parameters. These exotic climate states might be extremely hard to discover through direct numerical integration, given the limited size of their basin of attraction and the small parametric window where they actually exist.}

%, nor to guess \textit{a-priori} its existence. 

%we underline that the methodology proposed here has allowed us to find \textit{also} a new attracting climatic steady state characterized by a rather fascinating dynamical regime. One can safely argue that it would have unlikely to discover such a climate through direct numerical integration, 

The results presented provide a first example of a successful reconstruction of the global properties of a climate model possessing multiple steady states and complex regimes of motion, and hope to provide a positive stimulation in the direction of having a more thorough understanding  of the properties of the climate system. It seems then relevant to extend the study of Melancholia state for the Earth and for other planets, as they may be key to understanding the evolution and the statistics of observed planetary atmospheres. 

{\color{black}Additionally, we have shown that the basin boundary between the basins of attraction has a  complex geometrical structure, as a result of the very different time scales associated to weather-like (baroclinic processes) and climatic instabilities (ice-albedo feedback). Obtaining a rather detailed geometrical description in such a high-dimensional system as the one analysed here seems quite an achievement. The presence of a \textit{thick} basin boundary with a complex geometry makes the prediction of the final state given a finite precision knowledge of the initial conditions extremely challenging, thus indicating a reduced predictability of the second kind.}

As a future investigation, we aim at relating the dynamical properties of the edge state to the statistics of noise-induced transitions between competing attractors, starting from simple up to more complex multistable models. {\color{black}As the edge state is \textit{the} gate controlling the noise-induced transitions between the two co-existing attractors, its dynamical properties are crucial in assessing how likely a random forcing might lead to a catastrophic transition to another basin of attraction. Therefore, one could hope to derive estimates or semi-quantitative statements relating the properties of the edge state to the escape rates one can study via the Freidlin-Wentzell theory. Additionally, we wish to analyse in detail the geometry of the basin boundary and investigate the relationship between the escape rate and  Lyapunov exponent along the unstable direction, along the lines indicated in \cite{Tel2006}. 

Additionally, we remark that the existence of more than two stable states leads to the existence of a potentially more complex partition of the phase space in different basins of attraction, and of an ensuing more complex population of edge states sitting in-between different pairs of attractors. 

In terms of specific geophysical relevance, it seems relevant to go deeper in the properties of the Melancholia states found in this study and analyse in greater detail a) more specific physical reasons  behind the presence of such a complex basin boundary (e.g. the role of localised instabilities in the physical space) ; b) the physical mechanisms behind the symmetry break found in the upper range of bistability, and, c) indeed, the apparent exchange of stability that morphs a Melancholia state into a new stable climate state and \textit{viceversa} for small modulations of the solar constant. 

The properties and relevance of the newly found stable climate state characterized by the break-up of the zonal symmetry and by intense horizonal temperature gradients seem indeed worth investigating. As far as we are aware, no previous studies have found any qualitatively analogous stable state.

A  minor pitfall of the model used in this study is that it features a stable atmospheric circulation with no fluctuations in the snowball states. Instead, the snowball states obtained using more comprehensive models have weak yet non-vanishing atmospheric variability, as baroclinic instability is active in the atmosphere, despite the low meridional temperature gradient. Hence, repeating our analysis with a climate model like the one used in \cite{Luchyst} seems a useful future exercise.

As discussed earlier in the paper, a more sophisticated analysis of the mechanisms behind the multistability of the Earth climate using models able to resolve accurately the oceanic time scales suggests the possibility of an additional climate state (the slushball state) co-existing with the warm and the snowball state. A complete analysis of the Melancholia states situated in-between the possible climate states would definitely provide a more complete picture of the global dynamical properties of the Earth system and possibly lead to identifying additional, nontrivial  stable climates  of relevance for paleoclimate and for the study of exoplanetary systems. 

}

\subsection*{Acknowledgements}
The authors wish to express gratitude to B. Eckhardt for multiple intellectual stimulations on the theory and intuitive aspects of the edge state and on the details of the edge tracking method{\color{black}; to A. Pikovsky for  pointing out the importance of looking at the geometry of the basin boundary; to C. Grebogi, A. Politi, and J. Yorke for various very inspiring discussions; and to one anonymous reviewer; to D. Abbott for providing insightful comments on the manuscript; and to L. von Trier for directing and producing the movie Melancholia}  The authors also wish to thank F. Lunkeit and E. Kirk for the great support on many computational aspects of this paper. The authors acknowledge the support of the FP7-ERC StG Grant NAMASTE, of the Horizon2020 Project CRESCENDO, and of the DFG Cluster of Excellence CliSAP. VL acknowledges the support of the DFG project MERCI. VL wishes to thank GGG for many things, which include forcing him to watch Melancholia. 
\vspace{10 mm}
\subsection*{References}
\providecommand{\newblock}{}


\begin{thebibliography}{100}
\expandafter\ifx\csname url\endcsname\relax
  \def\url#1{{\tt #1}}\fi
\expandafter\ifx\csname urlprefix\endcsname\relax\def\urlprefix{URL }\fi
\providecommand{\eprint}[2][]{\url{#2}}

\bibitem{ghil_topics_1987}
Ghil M and Childress S 1987 {\em Topics in geophysical fluid dynamics:
  atmospheric dynamics, dynamo theory, and climate dynamics\/} (Heildelberg:
  Springer)

\bibitem{Chekroun2011}
Chekroun M~D, Simonnet E and Ghil M 2011 {\em Physica D: Nonlinear Phenomena\/}
  {\bf 240} 1685 -- 1700 ISSN 0167-2789
  \urlprefix\url{http://www.sciencedirect.com/science/article/pii/S016727891100145X}

\bibitem{LBHRPW14}
Lucarini V, Blender R, Herbert C, Ragone F, Pascale S and Wouters J 2014 {\em
  Reviews of Geophysics\/} {\bf 52} 809--859 ISSN 1944-9208
  \urlprefix\url{http://dx.doi.org/10.1002/2013RG000446}

\bibitem{LLR16}
Lucarini V, Ragone F and Lunkeit F 2016 {\em Journal of Statistical Physics\/}
  1--29 ISSN 1572-9613
  \urlprefix\url{http://dx.doi.org/10.1007/s10955-016-1506-z}

\bibitem{Ghil:2013}
Ghil M 2013 {\em J. Atmos. Sci.\/} {\bf 33} 3--20

\bibitem{Ghil2015}
Ghil M 2015 {A mathematical theory of climate sensitivity or, How to deal with
  both anthropogenic forcing and natural variability?} {\em {Climate Change :
  Multidecadal and Beyond}\/} ed P C~C, M G, M L and M W~J (Kluwer) pp 31--51

\bibitem{Lor67}
Lorenz E 1967 The nature and theory of the general circulation of the
  atmosphere vol. 218.TP.115, World Meteorological Organization

\bibitem{Peixoto:1992}
Peixoto J~P and Oort A~H 1992 {\em Physics of climate\/} (New York: American
  Institute of Physics)

\bibitem{Kleidon05}
Kleidon A and Lorenz R (eds) 2005 {\em Non-equilibrium thermodynamics and the
  production of entropy\/} (Berlin: Springer)

\bibitem{Lucarini09PRE}
Lucarini V 2009 {\em Phys. Rev. E\/} {\bf 80}

\bibitem{G14}
Gallavotti G 2014 {\em Nonequilibrium and irreversibility\/} (Springer, New
  York)

\bibitem{IPCC01}
{Intergovernmental Panel on Climate Change [Eds: J Houghton et al]} 2001 {\em
  {IPCC Third Assessment Report: Working Group I Report "The Physical Science
  Basis"}\/} (Cambridge University Press)

\bibitem{IPCC07}
{Intergovernmental Panel on Climate Change [Eds: S Solomon et al]} 2007 {\em
  {Climate Change 2007 - The Physical Science Basis: Working Group I
  Contribution to the Fourth Assessment Report of the IPCC}\/} (Cambridge, UK
  and New York, NY, USA: Cambridge University Press) ISBN 0521880092
  \urlprefix\url{http://www.worldcat.org/isbn/0521880092}

\bibitem{IPCC13}
{Intergovernmental Panel on Climate Change [Eds: T Stocker et al]} 2014 {\em
  Climate Change 2013 - The Physical Science Basis IPCC Working Group I
  Contribution to AR5\/} (Cambridge: Cambridge University Press) ISBN
  9781107415324 \urlprefix\url{http://dx.doi.org/10.1017/cbo9781107415324}

\bibitem{RLL16}
Ragone F, Lucarini V and Lunkeit F 2015 {\em Climate Dynamics\/} {\bf 46}
  1459--1471 ISSN 1432-0894
  \urlprefix\url{http://dx.doi.org/10.1007/s00382-015-2657-3}

\bibitem{Grits_Luca2016}
{Gritsun} A and {Lucarini} V 2016 {\em ArXiv e-prints\/} (\textit{Preprint}
  \eprint{1604.04386})

\bibitem{Tantet2015b}
{Tantet} A, {Lucarini} V, {Lunkeit} F and {Dijkstra} H~A 2015 {\em ArXiv
  e-prints\/} (\textit{Preprint} \eprint{1507.02228})

\bibitem{Luchyst}
Lucarini V, Fraedrich K and Lunkeit F 2010 {\em Q. J. Royal Met. Soc.\/} {\bf
  136} 2--11

\bibitem{Lucarini2013a}
Lucarini V, Pascale S, Boschi V, Kirk E and Iro N 2013 {\em Astronomische
  Nachrichten\/} {\bf 334} 576--588 ISSN 00046337 (\textit{Preprint}
  \eprint{arXiv:1303.5937v1})

\bibitem{Boschi}
Boschi R, Lucarini V and Pascale S 2013 {\em Icarus\/} {\bf 226} 1724--1742

\bibitem{Lenton2008}
Lenton T~M, Held H, Kriegler E, Hall J~W, Lucht W, Rahmstorf S and Schellnhuber
  H~J 2008 {\em Proceedings of the National Academy of Sciences\/} {\bf 105}
  1786--1793 (\textit{Preprint}
  \eprint{http://www.pnas.org/content/105/6/1786.full.pdf})
  \urlprefix\url{http://www.pnas.org/content/105/6/1786.abstract}

\bibitem{Scheffer12}
Scheffer M, Carpenter S~R, Lenton T~M, Bascompte J, Brock W, Dakos V, van~de
  Koppel J, van~de Leemput I~A, Levin S~A, van Nes E~H, Pascual M and
  Vandermeer J 2012 {\em Science\/} {\bf 338} 344--348 ISSN 0036-8075
  (\textit{Preprint}
  \eprint{http://science.sciencemag.org/content/338/6105/344.full.pdf})
  \urlprefix\url{http://science.sciencemag.org/content/338/6105/344}

\bibitem{Budyko}
Budyko M 1969 {\em Tellus\/} {\bf 21} 611--619

\bibitem{Sellers}
Sellers W 1969 {\em J. Appl. Meteorol.\/} {\bf 8} 392--400

\bibitem{Ghil}
Ghil M 1976 {\em Journal of Atmospheric Sciences\/} {\bf 33} 3

\bibitem{Hoffman98}
Hoffman P, Kaufman A, Halverson G and Schrag D 1998 {\em Science\/}  1342--1346

\bibitem{saltzman_dynamical}
Saltzman B 2001 {\em Dynamical Paleoclimatology\/} (Academic Press New York)

\bibitem{Pierrehumbert}
Pierrehumbert R~T, Abbot D, Voigt A and Koll D 2011 {\em Annual Review of Earth
  and Planetary Sciences\/} {\bf 39} 417

\bibitem{hasselmann_stochastic_1976}
Hasselmann K 1976 {\em Tellus\/} {\bf 28} 473--485

\bibitem{arnold_hasselmanns_2001}
Arnold L 2001 {\em Stochastic climate models\/} {\bf 49} 141--158

\bibitem{T09}
Touchette H 2009 {\em Physics Reports\/} {\bf 478} 1--69

\bibitem{FW98}
Freidlin M~I and Wentzell A~D 1998 {\em Random perturbations of dynamical
  systems\/} (New York: Springer)

\bibitem{Bouchet2014}
Bouchet F, Laurie J and Zaboronski O 2014 {\em Journal of Statistical
  Physics\/} {\bf 156} 1066--1092 ISSN 1572-9613
  \urlprefix\url{http://dx.doi.org/10.1007/s10955-014-1052-5}

\bibitem{Laurie2015}
Laurie J and Bouchet F 2015 {\em New Journal of Physics\/} {\bf 17} 015009
  \urlprefix\url{http://stacks.iop.org/1367-2630/17/i=1/a=015009}

\bibitem{PhysRevLett.96.174101}
Skufca J~D, Yorke J~A and Eckhardt B 2006 {\em Phys. Rev. Lett.\/} {\bf 96}(17)
  174101 \urlprefix\url{http://link.aps.org/doi/10.1103/PhysRevLett.96.174101}

\bibitem{Schneider13022009}
Schneider T~M and Eckhardt B 2009 {\em Phil. Trans. R. Soc. A\/} {\bf 367}
  577--587 (\textit{Preprint}
  \eprint{http://rsta.royalsocietypublishing.org/content/367/1888/577.full.pdf+html})
  \urlprefix\url{http://rsta.royalsocietypublishing.org/content/367/1888/577.abstract}

\bibitem{Vollmer09}
Vollmer J, Schneider T~M and Eckhardt B 2009 {\em New Journal of Physics\/}
  {\bf 11} 013040
  \urlprefix\url{http://stacks.iop.org/1367-2630/11/i=1/a=013040}

\bibitem{BLL2014}
B{\'o}dai T, Lucarini V, Lunkeit F and Boschi R 2014 {\em Climate Dynamics\/}
  {\bf 44} 3361--3381 ISSN 1432-0894
  \urlprefix\url{http://dx.doi.org/10.1007/s00382-014-2206-5}

\bibitem{puma}
Fraedrich K, Kirk E and Lunkeit F 1998 {PUMA: Portable University Model of the
  Atmosphere} Tech. rep. Deutsches Klimarechenzentrum Hamburg

\bibitem{Grebogi83}
Grebogi C, Ott E and Yorke J~A 1983 {\em Phys. Rev. Lett.\/} {\bf 50}(13)
  935--938 \urlprefix\url{http://link.aps.org/doi/10.1103/PhysRevLett.50.935}

\bibitem{Kaplan84}
Kaplan J~L, Mallet-Paret J and Yorke J~A 1984 {\em Ergodic Theory and Dynamical
  Systems\/} {\bf 4} 261--281

\bibitem{McDonald85}
McDonald S~W, Grebogi C, Ott E and Yorke J~A 1985 {\em Physica D: Nonlinear
  Phenomena\/} {\bf 17} 125 -- 153 ISSN 0167-2789
  \urlprefix\url{http://www.sciencedirect.com/science/article/pii/0167278985900016}

\bibitem{Grebogi87}
Grebogi C, Ott E and Yorke J~A 1987 {\em Science\/} {\bf 238} 632--638 ISSN
  0036-8075 (\textit{Preprint}
  \eprint{http://science.sciencemag.org/content/238/4827/632.full.pdf})
  \urlprefix\url{http://science.sciencemag.org/content/238/4827/632}

\bibitem{K57}
Kubo R 1957 {\em Journal of the Physical Society of Japan\/} {\bf 12} 570--586
  \urlprefix\url{http://jpsj.ipap.jp/link?JPSJ/12/570/}

\bibitem{kubo_statistical_1988}
Kubo R, Toda M and Hashitsume N 1988 {\em Statistical physics {II:}
  nonequilibrium Statistical Mechanics\/} (Heidelberg: Springer)

\bibitem{marconi2008}
Marconi U~M~B, Puglisi A, Rondoni L and Vulpiani A 2008 {\em Phys. Rep.\/} {\bf
  461} 111

\bibitem{LC12}
Lucarini V and Colangeli M 2012 {\em Journal of Statistical Mechanics: Theory
  and Experiment\/} {\bf 2012} P05013
  \urlprefix\url{http://stacks.iop.org/1742-5468/2012/i=05/a=P05013}

\bibitem{R97}
Ruelle D 1997 {\em Communications in Mathematical Physics\/} {\bf 187} 227--241

\bibitem{ruelle_smooth_1999}
Ruelle D 1999 {\em Journal of Statistical Physics\/} {\bf 95} 393Ð468

\bibitem{ruelle_review_2009}
Ruelle D 2009 {\em Nonlinearity\/} {\bf 22} 855--870

\bibitem{B00}
Baladi V 2000 {\em Positive Transfer Operators and Decay of Correlations\/}
  (Singapore: World Scientific)

\bibitem{BL07}
Butterley O and Liverani C 2007 {\em Journal of Modern Dynamics\/} {\bf 1}
  301--322 ISSN 1930-5311
  \urlprefix\url{http://aimsciences.org/journals/displayArticlesnew.jsp?paperID=2192}

\bibitem{B08}
Baladi V 2008 {\em Nonlinearity\/} {\bf 21} T81
  \urlprefix\url{http://stacks.iop.org/0951-7715/21/i=6/a=T01}

\bibitem{B14b}
{Baladi} V 2014 {\em ArXiv e-prints\/} (\textit{Preprint} \eprint{1408.2937})

\bibitem{BKL16}
{Baladi} V, {Kuna} T and {Lucarini} V 2016 {\em ArXiv e-prints\/}
  (\textit{Preprint} \eprint{1603.09690})

\bibitem{GC95}
Gallavotti G and Cohen E~G~D 1995 {\em Journal of Statistical Physics\/} {\bf
  80} 931--970

\bibitem{lucarini08}
Lucarini V 2008 {\em Journal of Statistical Physics\/} {\bf 131}(3) 543--558
  10.1007/s10955-008-9498-y

\bibitem{L09}
Lucarini V 2009 {\em Journal of Statistical Physics\/} {\bf 134}(2) 381--400
  10.1007/s10955-008-9675-z

\bibitem{lucarini2011}
Lucarini V and Sarno S 2011 {\em Nonlin. Processes Geophys\/} {\bf 18} 7--28

\bibitem{lorenz79}
Lorenz E 1979 {\em J. Atmos. Sci\/} {\bf 36} 1367--1376

\bibitem{AM07a}
Abramov R~V and Majda A~J 2007 {\em Nonlinearity\/} {\bf 20} 2793--2821

\bibitem{zwanzig_memory_1961}
Zwanzig R 1961 {\em Physical Review\/} {\bf 124} 983--992

\bibitem{mori_transport_1965}
Mori H 1965 {\em Progress of Theoretical Physics\/} {\bf 33} 423--455

\bibitem{gritsun_climate_2007}
Gritsun A and Branstator G 2007 {\em Journal of the Atmospheric Sciences\/}
  {\bf 64} 2558--2575

\bibitem{GBM}
Gritsun A, Branstator G and Majda A 2008 {\em Journal of the Atmospheric
  Sciences\/} {\bf 65} 2824--2841 (\textit{Preprint}
  \eprint{http://dx.doi.org/10.1175/2007JAS2496.1})
  \urlprefix\url{http://dx.doi.org/10.1175/2007JAS2496.1}

\bibitem{cooper_climate_2011}
Cooper F~C and Haynes P~H 2011 {\em Journal of the Atmospheric Sciences\/} {\bf
  68} 937--953

\bibitem{EHL04}
Eyink G~L, Haine T~W~N and Lea D~J 2004 {\em Nonlinearity\/} {\bf 17} 1867
  \urlprefix\url{http://stacks.iop.org/0951-7715/17/i=5/a=016}

\bibitem{W13}
Wang Q 2013 {\em Journal of Computational Physics\/} {\bf 235} 1 -- 13 ISSN
  0021-9991
  \urlprefix\url{http://www.sciencedirect.com/science/article/pii/S0021999112005360}

\bibitem{GPTCLP07}
Ginelli F, Poggi P, Turchi A, Chat\'e H, Livi R and Politi A 2007 {\em Phys.
  Rev. Lett.\/} {\bf 99}(13) 130601
  \urlprefix\url{http://link.aps.org/doi/10.1103/PhysRevLett.99.130601}

\bibitem{svita88}
Cvitanov\'ic P 1988 {\em Phys. Rev. Lett.\/} {\bf 61}(24) 2729--2732
  \urlprefix\url{http://link.aps.org/doi/10.1103/PhysRevLett.61.2729}

\bibitem{CE91}
Cvitanov\'ic P and Eckhardt B 1991 {\em Journal of Physics A: Mathematical and
  General\/} {\bf 24} L237
  \urlprefix\url{http://stacks.iop.org/0305-4470/24/i=5/a=005}

\bibitem{GuckHolmes83}
Guckenheimer J and Holmes P 1983 {\em Nonlinear Oscillations, Dynamical
  Systems, and Bifurcations of Vector Fields\/} (Springer, New York)

\bibitem{Arnold}
ArnolÕd V 1992 {\em Catastrophe theory, 3d edition\/} (Berlin: Springer)

\bibitem{dijkstra2013}
Dijkstra H 2013 {\em Nonlinear Climate Dynamics\/} (Cambridge: Cambridge
  University Press)

\bibitem{Ashwin12}
Ashwin P, Wieczorek S, Vitolo R and Cox P 2012 {\em Philosophical Transactions
  of the Royal Society of London A: Mathematical, Physical and Engineering
  Sciences\/} {\bf 370} 1166--1184 ISSN 1364-503X (\textit{Preprint}
  \eprint{http://rsta.royalsocietypublishing.org/content/370/1962/1166.full.pdf})
  \urlprefix\url{http://rsta.royalsocietypublishing.org/content/370/1962/1166}

\bibitem{Lucarini05}
Lucarini V, Calmanti S and Artale V 2005 {\em Climate Dynamics\/} {\bf 24}
  253--262 ISSN 1432-0894
  \urlprefix\url{http://dx.doi.org/10.1007/s00382-004-0484-z}

\bibitem{Lucarini07}
Lucarini V, Calmanti S and Artale V 2007 {\em Russian Journal of Mathematical
  Physics\/} {\bf 14} 224--231 ISSN 1555-6638
  \urlprefix\url{http://dx.doi.org/10.1134/S1061920807020124}

\bibitem{rahmstorf}
Rahmstorf S, Crucifix M, Ganopolski A, Goosse H, Kamenkovich I, Knutti R,
  Lohmann G, Marsh R, Mysak L~A, Wang Z and Weaver A~J 2005 {\em Geophys. Res.
  Lett.\/} {\bf 32}

\bibitem{Ott2002}
Ott E 2002 {\em Chaos in Dynamical Systems\/} (New York: Cambridge University
  Press)

\bibitem{Pikovsky2001}
Pikovsky A, Rosenblum M and Kurths J 2001 {\em {Synchronization: A Universal
  Concept in Nonlinear Sciences}\/} (Cambridge: Cambridge University Press)

\bibitem{chekroun2014}
Chekroun M~D, Neelin D~J, Kondrashov D, C M~J and Ghil M 2014 {\em Proceedings
  of the National Academy of Sciences\/} {\bf 111} 1684--1690

\bibitem{Tantet2015a}
Tantet A, van~der Burgt F~R and Dijkstra H~A 2015 {\em Chaos: An
  Interdisciplinary Journal of Nonlinear Science\/} {\bf 25} 036406

\bibitem{butterley2007}
Butterley O and Liverani C 2007 {\em J. Mod. Dyn.\/} {\bf 1} 301--322

\bibitem{Hoffman}
Hoffman P~F, Kaufman A~J, Halverson G~P and Schrag D~P 2002 {\em Science\/}
  {\bf 281} 1342

\bibitem{HoffmanSchrag}
Hoffman P~F and Schrag D~P 2002 {\em Terra Nova\/} {\bf 14} 129

\bibitem{Lewis07}
Lewis J~P, Weaver A~J and Eby M 2007 {\em Journal of Geophysical Research:
  Oceans\/} {\bf 112} n/a--n/a ISSN 2156-2202 c11014
  \urlprefix\url{http://dx.doi.org/10.1029/2006JC004037}

\bibitem{Abbott}
Abbot D~S, Voigt A and Koll D 2011 {\em Journal of Geophysical Research:
  Atmospheres\/} {\bf 116} n/a--n/a ISSN 2156-2202 d18103
  \urlprefix\url{http://dx.doi.org/10.1029/2011JD015927}

\bibitem{Rose2015}
Rose B~E~J 2015 {\em Journal of Geophysical Research: Atmospheres\/} {\bf 120}
  1404--1423 ISSN 2169-8996 2014JD022659
  \urlprefix\url{http://dx.doi.org/10.1002/2014JD022659}

\bibitem{north81}
North G~R, Cahalan R~F and Coakley J~A 1981 {\em Reviews of Geophysics\/} {\bf
  19} 91--121 ISSN 1944-9208
  \urlprefix\url{http://dx.doi.org/10.1029/RG019i001p00091}

\bibitem{Holton}
Holton J 2004 {\em An introduction to Dynamic Meteorology\/} (Accademic Press)

\bibitem{plasim}
Fraedrich K, Jansen H, Kirk E, Luksch U and Lunkeit F 2005 {\em Meteorologische
  Zeitschrift\/} {\bf 14} 299--304

\bibitem{MC11}
Monahan A~H and Culina J 2011 {\em Journal of Climate\/} {\bf 24} 3068--3088
  \urlprefix\url{http://dx.doi.org/10.1175/2011JCLI3641.1}

\bibitem{arnold1988}
Arnold L 1988 {\em {Random Dynamical Systems}\/} (New York: Springer)

\bibitem{K92}
Kifer Y 1992 {\em Inventiones mathematicae\/} {\bf 110} 337--370 ISSN 0020-9910
  \urlprefix\url{http://dx.doi.org/10.1007/BF01231336}

\bibitem{Kuehn11}
Kuehn C 2011 {\em Physica D: Nonlinear Phenomena\/} {\bf 240} 1020 -- 1035 ISSN
  0167-2789
  \urlprefix\url{http://www.sciencedirect.com/science/article/pii/S0167278911000443}

\bibitem{K40}
Kramers H~A 1940 {\em Physica\/}  284--304

\bibitem{HTB90}
H{\"a}nggi P, Talkner P and Borkovec M 1990 {\em Reviews of Modern Physics\/}
  {\bf 62} 251

\bibitem{LFW12}
Lucarini V, Faranda D and Wouters J 2012 {\em J. Stat. Phys.\/} {\bf 147}
  63--73

\bibitem{Grafke2015}
Grafke T, Grauer R and Schäfer T 2015 {\em Journal of Physics A: Mathematical
  and Theoretical\/} {\bf 48} 333001
  \urlprefix\url{http://stacks.iop.org/1751-8121/48/i=33/a=333001}

\bibitem{WL12}
Wouters J and Lucarini V 2012 {\em Journal of Statistical Mechanics: Theory and
  Experiment\/} {\bf 2012} P03003

\bibitem{WL13}
Wouters J and Lucarini V 2013 {\em Journal of Statistical Physics\/} {\bf 151}
  ISSN 0022-4715, 1572-9613
  \urlprefix\url{http://link.springer.com/10.1007/s10955-013-0726-8}

\bibitem{CLW15a}
Chekroun M, Liu H and Wang S 2015 {\em Approximation of Stochastic Invariant
  Manifolds: Stochastic Manifolds for Nonlinear SPDEs I\/} SpringerBriefs in
  Mathematics (Springer International Publishing)
  \urlprefix\url{https://books.google.com/books?id=1-7nBQAAQBAJ}

\bibitem{CLW15b}
Chekroun M, Liu H and Wang S 2015 {\em Approximation of Stochastic Invariant
  Manifolds: Stochastic Manifolds for Nonlinear SPDEs II\/} SpringerBriefs in
  Mathematics (Springer International Publishing)

\bibitem{Dit99a}
Ditlevsen P~D 1999 {\em Phys. Rev. E\/} {\bf 60}(1) 172--179
  \urlprefix\url{http://link.aps.org/doi/10.1103/PhysRevE.60.172}

\bibitem{IP06}
{Imkeller, P and Pavlyukevich, I} 2006 {\em Stochastic Processes Appl.\/} {\bf
  116} 611--642

\bibitem{Dit99b}
Ditlevsen P~D 1999 {\em Geophysical Research Letters\/} {\bf 26} 1441--1444
  ISSN 1944-8007 \urlprefix\url{http://dx.doi.org/10.1029/1999GL900252}

\bibitem{Dit10}
Ditlevsen P~D and Johnsen S~J 2010 {\em Geophysical Research Letters\/} {\bf
  37} n/a--n/a ISSN 1944-8007 l19703
  \urlprefix\url{http://dx.doi.org/10.1029/2010GL044486}

\bibitem{GIHP11}
Gairing J, Imkeller P, Hein C and Pavlyukevich I 2011 {\em Mathematisches
  Forschungsinstitut Oberwolfach\/} {\bf Report N. 40/2011}

\bibitem{PhysRevE.78.037301}
Schneider T~M, Gibson J~F, Lagha M, De~Lillo F and Eckhardt B 2008 {\em Phys.
  Rev. E\/} {\bf 78}(3) 037301
  \urlprefix\url{http://link.aps.org/doi/10.1103/PhysRevE.78.037301}

\bibitem{Grossmann2000}
Grossmann S 2000 {\em Rev. Mod. Phys.\/} {\bf 72}(2) 603--618
  \urlprefix\url{http://link.aps.org/doi/10.1103/RevModPhys.72.603}

\bibitem{Hof2006}
Hof B, Westerweel J, Schneider T~M and Eckhardt B 2006 {\em Nature\/} {\bf 443}
  59--62 \urlprefix\url{http://dx.doi.org/10.1038/nature05089}

\bibitem{Hof2008}
Hof B, de~Lozar A, Kuik D~J and Westerweel J 2008 {\em Phys. Rev. Lett.\/} {\bf
  101}(21) 214501
  \urlprefix\url{http://link.aps.org/doi/10.1103/PhysRevLett.101.214501}

\bibitem{LucACP}
Lucarini V, Fraedrich K and Lunkeit F 2010 {\em Atmos. Chem. Phys\/} {\bf 10}
  9729--9737

\bibitem{LucariniRagone}
Lucarini V and Ragone F 2011 {\em Rev. Geophys.\/} {\bf 49} RG1001

\bibitem{panaggio}
Panaggio M~J and Abrams D~M 2015 {\em Nonlinearity\/} {\bf 28} R67
  \urlprefix\url{http://stacks.iop.org/0951-7715/28/i=3/a=R67}

\bibitem{Tel2006}
T\'{e}l T and Gruiz M 2006 {\em {Chaotic Dynamics}\/} (Cambridge: Cambridge
  University Press)

\end{thebibliography}
\end{document}